# Atomic Quantum Technologies for Quantum Matter and Fundamental Physics Applications

Jorge Yago Malo,[1,2,*] Luca Lepori,[3,4,5,†] Laura Gentini,[1,‡] and Maria Luisa Chiofalo[1,2,§]

[1] *Department of Physics, University of Pisa, Largo Bruno Pontecorvo 3, 56126 Pisa, Italy*
[2] *INFN, Pisa section, Largo Bruno Pontecorvo 3, 56126 Pisa, Italy*
[3] *Department of Mathematical, Physical, and Computer Science, University of Parma , Parco Area delle Scienze 53/A, 43124 Parma, Italy.*
[4] *INFN, group of Parma, Parco Area delle Scienze 7/A, 43124, Parma, Italy.*
[5] *QSTAR and INO-CNR and LENS, Largo Enrico Fermi 2, 50125 Firenze, Italy.*

Physics is living an era of unprecedented cross-fertilization among the different areas of science. In this perspective review, we discuss the manifold impact that state-of-the-art ultracold-atom quantum technologies can have in fundamental and applied science through the development of platforms for quantum simulation, computation, metrology and sensing. We illustrate how the engineering of table-top experiments with atom technologies is engendering applications to understand problems in condensed matter and fundamental physics, cosmology and astrophysics, unveil foundational aspects of quantum mechanics, and advance quantum chemistry and the emerging field of quantum biology. In this journey, we take the perspective of two main approaches, i.e. creating quantum analogues and building quantum simulators, highlighting that independently of the ultimate goal of a universal quantum computer to be met, the remarkable transformative effects of these achievements remain unchanged. We wish to convey three main messages. First, this atom-based quantum technology enterprise is signing a new era in the way quantum technologies are used for fundamental science, even beyond the advancement of knowledge, which is characterised by truly cross-disciplinary research, extended interplay between theoretical and experimental thinking, and intersectoral approach. Second, quantum many-body physics is unavoidably taking center stage in frontier's science. Third, quantum science and technology progress will have *capillary* impact on society, meaning this effect is not confined to isolated or highly specialized areas of knowledge, but is expected to reach and have a pervasive influence on a broad range of society aspects: while this happens, the adoption of a responsible research and innovation approach to quantum technologies is mandatory, to accompany citizens in building awareness and future scaffolding. Following on all the above reflections, this perspective review is thus aimed at scientists active or interested in interdisciplinary research, providing the reader with an overview of the current status of these wide fields of research where ultracold-atomic platforms play a vital role in their description and simulation.

## I. INTRODUCTION

The controllable manipulation and characterization of quantum matter have a wide range of potential applications. Over the years, the horizon of these applications has been moved far beyond that originally motivating the development of quantum simulators, like for example the simulation of yet non-well understood condensed-matter phenomena that were hard to investigate in their native media. Since then, applications have bloomed in many directions, including the creation of quantum simulators for fundamental physics and cosmology [1], the design of devices for precision measurements in quantum sensing and communications [2], or the development of high-fidelity qubits for quantum computers [3]. In particular, quantum computers constitute the application that could have the farthest-reaching impacts: despite them being still in their early development, they could enable the design of molecules for medicine, biology, and agronomy, as well as the creation of energetically efficient materials, and simulation of complex systems for applications in finance or artificial intelligence [3]. If the first quantum revolution has led to the development of powerful miniaturized devices and influenced culture, art, and philosophy in the 20th century, the second revolution could deeply transform economy, labor market, and day-to-day use of devices with novel capabilities.

Quantum many-body physics is a crucial area of study that underpins quantum technologies (QT) in an interdisciplinary, transversal, and global manner, given that its underlying theory represents the scaffolding for a number of technological platforms and experiments. Quantum many-body physics draws from atomic physics to model and compute microscopic interactions, while it relies on the unifying and foundational concepts of broken symmetry and conservation laws [4] to understand the emergence of different quantum states of matter. In fact, while system symmetries can be possibly reduced by tuning a number of knobs, different (quantum) states of matter set in and - irrespective of details - the well-known conservation laws are preserved with the many-particles traits ensured by thermodynamic and statistical mechanics. It is therefore not surprising that the corresponding concepts and methods of many-particle physics are shared with, and often essential to, high-energy physics problems like the quark-gluons matter in the Quantum Chromo-Dynamics (QCD) phase diagram, nuclear physics quests like the equation of state of neutron stars [5], and even biological problems [6–8]. In fact, quantum simulation technologies, in both their theoretical and experimental joined development, open a wide research space where universe-related problems can be investigated to some extent in the four squared meters size of the optical table in a QT laboratory.

---

*correspondence at: jorge.yago@unipi.it
†correspondence at: luca.lepori@unipr.it
‡correspondence at: laura.gentini@df.unipi.it
§correspondence at: marilu.chiofalo@unipi.it

So, what is a quantum simulator and how quantum technologies are making this concept accessible? Quantum simulators, as proposed by Richard Feynman [9], involve the mapping of a quantum system that is to be understood by means of an experiment or simulation performed with another highly controllable quantum system. In order to achieve this task, two approaches can generally be addressed: creating quantum analogues and building quantum simulators. The former involves the design of an analogy between two (quantum) systems, to explore common features and learn through associative creativity. The latter involves the encoding of specific microscopic system Hamiltonians in a quantum platform that is highly controllable, in what can be seen as a specific type of quantum computation. The boundary between the two methods is often blurred, since paradigmatic model Hamiltonians, as the Ising or the Hubbard model just to mention two of them, can be used to either create an analog simulation or to code microscopic quantum problems. Different platforms for quantum technologies are being developed in parallel, involving scientific research institutions and companies [10–12], as for example IonQ [13] and Alpine Quantum Computing [14] for trapped ions plaforms; Quera [15], Planqc [16] and Pasqal [17] for neutral atoms ones; or Atom Computing for nuclear spin qubit realization [18]; while instead some companies like Zapata Computing [19] focus on cross-platform applications like Generative AI. In fact, this field of research represents one of the most intersectoral worldwide technological efforts.

The state-of-the-art science that can be addressed with cold and ultracold-atom based technology platforms, is the focus of the present review.

Atom-based quantum technologies have proven to be useful in simulating quantum-matter systems that would otherwise be inaccessible or difficult to control, working as a rich test-bed for creating quantum analogies, quantum simulators but also as quantum sensors and computers, albeit with varying levels of efficacy [20]. Numerous paradigmatic implementations of condensed-matter physics problems include the crossover from a Bardeen-Cooper-Schrieffer (BCS)-type [21] superconductivity/superfluidity to a Bose-Einstein condensation (BEC) of composite bosons, relevant for high-temperature superconductivity [22–24], the transition from superfluidity to Mott insulator states [25, 26], systems frustrated by different incommensurabilities [27, 28], quantum simulators for fermionic systems [29–31], topological phases [32], or the entirely new concept of many-body localization [33, 34]. These latter implementations involve the breaking of the ergodic thermal hypothesis with the exploration of quantum phase transitions in open, driven-dissipative, out-of-equilibrium quantum systems, in turn connecting with the idea of dissipation engineering [35, 36]. These non-exhaustive list of condensed-matter physics applications highlights the power of QTs, that can be transferred to diverse applied fields, as discussed in this review.

Indeed, atom-based quantum technologies also hold great promise for addressing problems in fundamental physics. Here, our understanding relies on four fundamental theories, Quantum mechanics and the Standard Model of particle physics, General relativity and the Standard model of cosmology. Despite the successes of these theories in the corresponding domains, a unified description has yet to be found that can encompass all phenomena across the more than 40 orders of magnitude from the atomic to the astrophysics and cosmology length scales. Fundamental questions remain, including the behavior of General relativity at the atomic level and the possible existence of additional forces, the nature of Dark matter and, at the quantum scale, the measurement problem and the quantum to classical crossover. The reconciliation of Quantum Mechanics with General Relativity stands as an unresolved problem, which may require modifications to one or both theories.

For instance, as we discuss in this review, the BCS-BEC crossover is relevant to the low-temperature region of the QCD phase diagram from color-superconductivity to quark confinement in the baryon formation [5], as well as the equation of state for quark matter and for neutrons in neutron stars, where adimensional values for the scattering length and effective range of the interactions are similar to those in certain atomic gases [5]. Additionally, gravity is being explored as an effective theory in superfluids [37, 38], including the observation of Hawking radiation and temperature in acoustic black holes [39, 40], and highlighting the geometric nature of these concepts [41]. These are some relevant examples of the role that quantum simulation can play in the understanding of these theories. However, the applicability of QT does not end there.

In recent years in fact, quantum metrology has emerged as a new paradigm [2] for testing general relativity and for accurately measuring fundamental constants [42–47]. This field has seen a variety of applications, including tests of the equivalence principle using atomic interferometry [48], measuring gravity at micrometer distances with an atomic pendulum [49], and detecting gravitational waves at intermediate frequencies [50], that are not accessible to LIGO-VIRGO. The use of non-classical states of matter, such as squeezed or many-body entangled states, has also been proposed as a way to reduce quantum uncertainty in interferometers [2]. Atomic clocks have also seen significant progress in recent years, with best-performing clocks failing only half a second in the age of the universe [51]. With the ability to keep cold molecules coherent over hundreds of microns, researchers are beginning to envision table-top experiments that could explore the fabric of spacetime [51]. Quantum metrology has also found applications in testing the foundations of quantum mechanics at the border with general relativity [52], and in exploring the hypothesis of ultra-light scalar-bosons for dark matter [50, 53].

While quantum technologies were born within atomic-molecular-optics physics and condensed matter, they have now progressively become a modern research domain relevant for disciplines from computer science to engineering and life sciences, and have as well attracted worldwide efforts aimed at qualified outreach tools and responsible research and innovation concerns. These active fields of research are linked in the present review. In particular, quantum technologies form the foundation for quantum computers, which have become a reality with the development of Noisy Intermediate scale Quantum (NISQ) devices [3]. The applications of these technologies are vast and ever-growing, ranging from quantum-communication protocols, or the design of drugs, fertilizers, and environment-friendly materials, to model complex phenomena in fields like artificial intelligence, finance, and logistics [3]. The research in this field involves many leading groups in scientific institutions and numerous companies, making it one of the most intersectoral and interdisciplinary endeavour to date in science and technology. While many challenges are still ahead in terms of size, circuit depth or qubit robustness against decoherence, the emergence of such problems has also fostered the collaboration across fields and led to the exploration of new paradigms, such as topological protection, and even seeking inspiration from nature to preserve quantum coherence [6, 8]. Additionally, existing platforms are being used to simulate the real-time dynamics of lattice-gauge theories relevant to fundamental interactions in higher dimensions, which might otherwise be rather limited, and at times impossible,, with conventional quantum simulation methods [54–56]. A pioneering proof-of-concept coding of the Schwinger Hamiltonian into an ion system has been successful, allowing researchers to observe particle-antiparticle pairs popping out of a vacuum from a vacuum and measure their entanglement [57]. This approach represents an example of a new theory-experiment relationship, where Theory can design experimental implementations from the available resources, also in combination with classical and/or digital optimizations. Then, experiments perform the physical simulations that provide theoretical insight.

In addition, quantum technologies are potentially relevant in the investigation of macroscopic quantum-coherence emergence in biological systems, known as quantum biology [7, 58]. While photosynthesis is a paradigmatic example in which macroscopic quantum effects have been demonstrated, they may not necessarily be critical to the function [8, 59]. However, other cases, such as how our sense of smell or birds' mapping abilities work, are currently under experimental and theoretical investigation [6], notwithstanding the microscopic effects in the brain [60].

Lastly, we notice that quantum mechanics is the microscopic theory accounting for all kinds of phenomena, from the cosmology of the universe down to the quarks as elementary constituents of matter, but the microscopic world cannot be seen. While this fact can be met as a conceptual limitation, it also presents a unique opportunity when we consider current quantum technologies and simulators in particular. Quantum many-body physics addresses phenomena on a size scale and complexity level – ever-growing with current technologies– that stand close to this classical emergence boundary, from the simplest material compounds to the simplest biological molecules. In fact, quantum technologies are now capable of producing systems with progressively increasing size that show macroscopic quantum coherence. In this scenario, quantum many-body physics and its contemporary evolution in terms of quantum complex systems can indicate to us where to place the classical-to-quantum crossover, i.e., the border between size scales where we must use quantum mechanics instead of classical physics. Thus, the same models built up to describe the experimental data, which are essential to educate knowledge and intuition, could work as bridges between exciting microscopic theories and our (narrow) macroscopic view. This statement has significant cultural, epistemic, and educational dimensions. These considerations can impact the educational aspects, and combined with the progress in creating suitable digital and interactive tools, can make the storytelling and education of quantum many-body physics accessible, indeed a powerful potential in physics education. This is especially relevant in that the current scientific blossoming in quantum technologies presents a new challenge for society, while indeed citizens are already immersed in the second quantum revolution but lack the tools and educational contexts to develop an awareness of how quantum technologies will transform their lives [61]. From a Responsible-Research and Innovation (RRI) perspective [62], an additional tool-set is therefore needed, which in turn requires an approach based on Physics-Education-Research (PER) as well as Physics-Outreach-Research (POR) [61]. This challenge concerns also classical physics, but it acquires a special significance in the quantum domain, due to even more limited literacies that would be needed to grasp an understanding of just the essential concepts.

Following on all the above reflections, this perspective review is thus aimed at scientists active or interested in interdisciplinary research, providing the reader with an overview of the current status of these wide fields of research, while focusing on cold and ultracold-atomic platforms. As highlighted in the guidelines and competence framework from major public funding agencies [63, 64], interdisciplinary approach is a must in modern science, and this review is motivated from placing this concept in a practical context. Following the Quantum Flagship [65] classification, QT are the foundations for four pillars: quantum simulations, quantum computing, quantum metrology and sensing, and quantum communications. With a focus on hardware concepts, frameworks and devices, this review mainly connects to the first three pillars, where examples from cold and ultracold-atom platforms are numerous and specific. Contemporarily, the fields of quantum communications [66] and quantum internet [67] are rapidly expanding, in fact already pursuing the integration of quantum and classical communication systems to create shared infrastructures and ways to enable hybrid devices. The impact of theoretical design and experimental demonstrations in this field is potentially enormous in sectors such as healthcare, space exploration, banking, underwater communication, industry, and transportation. Closer to this review, quantum networks and information processing have an indisputable influence on the realistic design of toolsets for efficient and effective architectures involving large numbers of qubits. Importantly, given the steady and rapid evolution of the field, the overall picture distinctions and boundaries between the pillars' topics can move over time. While we refer the interested reader to [66, 68, 69] for comprehensive reviews about quantum communications, in the rest of this review we will only mention selected quantum communications developments that can extend, in the near term, the applicability of specific hardware.

While the content of this review aims to be accessible from different disciplinary expertises. To this purpose, a multilevel approach is adopted: each subject is presented in a pedagogically-oriented manner and accounting for the main experimental demonstrations. With such a broad setting and whenever possible, reference will be made to existing reviews that are specific to— and therefore can be used to zoom in—more limited parts. Besides an interdisciplinary perspective on the subject, with this review we wish indeed to provide readers with a resource made of links and connections among most diverse pieces of the puzzle, deepening those that cannot be found in existing literature and giving a compact summary of the rest.

The review is organized as follows. In Section II we provide the quantum-technology toolbox useful to discuss the physics applications of interest for the rest of the journey. We first dig into the concept of quantum simulators in Section II A, briefly accounting for the main QT platforms and then focusing on those based on quantum gases. The advent of quantum technologies has boosted the development of methodological theoretical and simulational frameworks essential to describe and predict the behaviour of QT; thus, in Section II B we describe in detail the tools of open quantum systems to model driven-dissipative quantum phenomena, including also the discussion of tensor networks as approximate methods especially relevant to model current experiments and linked to topics connecting to following sections. Then, in Section II C we review the main ideas concerning quantum computing, where we encompass fault-tolerant and NISQ devices, and the relevant question of quantum optimization and control that is transversal to quantum simulation as well. As a final theoretical and experimental tool we introduce, in Section II D quantum metrology and sensing, describing the well-known tools of atomic clocks and atom interferometry, and discussing the more recent relevant paradigms of squeezing, many-body entanglement, and their quantification.

With the platforms described and the theoretical toolbox at hand, the newly accessible physics is discussed. Section III reviews the toolbox uses to explore selected evergreen paradigms of condensed-matter physics (Section III A), including the well-known but yet surprising BCS-BEC crossover and different forms of commensurate-incommensurate phenomena, as well as freshly sprouting frameworks (Section III B) such as dissipation engineering, many-body localization or dynamical quantum phase transitions. Section IV is then dedicated to toolbox implementations in the field of fundamental interactions, focusing on the possibility of investigating via analogue systems the quark-gluon phase diagram of quantum chromodynamics (QCD), and via quantum simulators lattice-gauge field Hamiltonians. From the quark size to the universe, Section V discusses selected different routes, along which open problems in cosmology and astrophysics are envisioned to be explored by means of quantum technologies. One, in Section V A, is via the use of atomic clocks and atom interferometry for general relativity test, like the measurement of the gravitational constant $G$, the variation of fundamental constants or equivalence principle tests, or for detection of gravitational waves or the quest for ultra-light dark matter. One second, in Section V B, is about analogue simulations of gravity and black-holes physics with superfluid systems. Still concerning fundamental physics, Section VI sketches currently circulating ideas to test foundational concepts of quantum mechanics which are still not understood, e.g., how the collapse of the wavefunction work, with implications on the measurement problem and the classical-to-quantum crossover. We follow on the applications by picking one outside traditional perimeter of quantum physics, that is discussed in Section VII: here, we briefly account for the many routes in which quantum science and technology crosses chemistry and biology, including the quest for the persistence of macroscopic quantum-coherence effects in biological systems such as plants and the brain, and the prospects of quantum computing - already in the NISQ era - to address the engineering of molecules, bearing e.g., pharmaceutical and ecological interest. The journey is then completed in Section VIII, by opening a window into the Responsible Research and Innovation (RRI) aspects involved in quantum science and technology education and outreach. Remaining considerations are elaborated in Section IX, before proceeding to the concluding remarks in Section X.

With the wide-open landscape of subjects listed so far, it is comprehensible that this review places the fastly growing field of atomic technologies in the perspective of the variety of its applications, with a special attention to the most unconventional among them. The in-depth narrative of the individual topics is necessarily limited, especially whenever more recent and specific reviews are available from literature, which we thereby refer to. Along these lines, the choice of topics illustrated in the Section II among a vast range of possibilities, has been operated by reverse-engineering the theoretical and experimental concepts and tools touched upon when dealing with the chosen applications.

## II. THEORETICAL AND EXPERIMENTAL CONCEPTS AND TOOLS

In this section, we provide a compact overview of tools that are functional to our journey on atomic quantum technologies and their applications to quantum simulators, quantum computing, and quantum metrology and sensing. We begin in Section II A, by briefly summarizing the main existing experimental platforms based on atomic and molecular (AMO) devices in order to highlight the wide range of hardware-development and possibilities to non-expert readers. We will briefly discuss some of those based on solid-state technologies that are relevant to specific parts of this review. We provide a compact account of the essential physical traits characterizing the use of cold and ultracold-atomic platforms as quantum simulators, their potential, current challenges or limitations, and scalability. We then overview in Section II B the theoretical paradigms currently at the forefront of the description of QT, providing a link between quantum simulators, quantum computing, and quantum metrology and sensing. In fact, quantum simulations can on the one side be applied to investigate on classical computers or QT platforms the behavior of quantum matter for condensed matter, fundamental physics, and quantum metrology applications, and on the other side they can benefit from the existing classical and quantum protocols in quantum computing for accessible applications in the present NISQ era. Finally, this section will be completed with an overview of concepts and tools for quantum computing in Section II C, and in Section II D for quantum metrology and sensing.

### A. Quantum Simulators Platforms

Quantum simulators are controllable and tunable quantum systems used to simulate the behavior of other complex quantum systems that are observed in nature [9]. In more recent years, there has been significant progress in the development of quantum simulators, with the emergence of new technologies and experimental techniques, hand in hand with theoretical advancements.

#### *1. Platforms Overview*

Before describing in detail AMO platforms for quantum simulation, for the sake of completeness we briefly recall in this section those quantum technology platforms based on solid-state devices, that are currently the subject of extensive research at different levels of technological readiness and constituting NISQ-friendly hardware realizations of simulation, computing or sensing protocols.

The most commonly used solid-state platforms are based on superconducting circuits [70, 71]. These superconducting qubits hinge on the use of Josephson junctions working as non-linear, non-dissipative inductors. The latter, looped with a capacitor, provide anharmonic oscillator circuits with an uneven energy-level structure that can be externally controlled, therefore allowing for unambiguous addressing and manipulation of the quantum states forming the qubit. The most common type of superconducting qubit is the transmon qubit, originally proposed in [71], where two (or more) superconducting islands are joined with one (or more) Josephson junction(s). In so doing, the ratio of the oscillator-to-kinetic energy can be tuned to increasing values, providing exponential protection from low-frequency noise while preserving useful anharmonicity. Quantum information can be stored in the number of superconducting pairs (charge qubit), in the direction of their current (flux qubit), or in the phase of the oscillatory states (phase qubit). Superconducting qubits are highly tunable and can be manipulated with microwave pulses, making them ideal for use in quantum circuits. Superconducting qubits are currently most developed for quantum computers architectures. They have been used to realize a wide range of quantum computing operations, and they are a promising platform for building quantum computers in the future [72].

Regarding the currently known disadvantages of the use superconducting qubits for building quantum computers, we highlight the reduced connectivity compared to atomic realizations. Low connectivity may limit the depth of circuits that can be implemented on the device, also related to the typical coherence times at present in superconducting quantum device prototypes [73]. It is also important to consider that solid-state devices, and not just superconducting qubits, are prone to certain disparity between individual qubits, as these are fabricated, specially when compared to atoms or ions which are identical by nature.

In solid-state devices, qubits can also be realized using spins. Different platforms can be employed based on the material that constitutes the crystal lattice in which spins are trapped. Spin qubits in diamond [74] are a type of qubit that rely on the spin of electrons trapped in nitrogen-vacancy (NV) centers within the diamond crystal lattice. These NV centers are defects in the diamond structure that result from the replacement of two carbon atoms with a nitrogen atom and a vacant lattice site. The spin of the electron trapped in the NV center can be manipulated with electromagnetic fields, and the resulting spin states can be used as qubits. One advantage of spin qubits in diamond is that they can operate at room temperature, unlike other types of qubits [75]. Additionally, diamond is a hard and inert material, which makes it a promising platform for developing robust and scalable quantum devices. Spin qubits in diamond have been used to implement error correction algorithms [76], and so they are a valuable candidate for the development of fault-tolerant quantum technologies.

Spin qubits in silicon are another type of qubit that rely on the spin of electrons in silicon structures and, given that silicon is a widely used material in the semiconductor industry, it offers a potential pathway for integrating quantum devices with existing electronic technology [75, 77], which could enable the development of scalable and commercially viable quantum devices [77]. Spin qubits in silicon are typically formed by introducing a phosphorus atom into a silicon crystal lattice and then using the spin of the electron associated with the phosphorus atom as a qubit. Like diamond spin qubits, silicon spin qubits can operate at room temperature and have shown long coherence times [78]. Despite some challenges in achieving high-fidelity quantum operations, they remain as a potential avenue for the development of practical quantum technologies [75].

It is also worth mentioning polariton qubits, that are based on the quantum optics phenomenon of polariton condensation also denoted as *quantum fluid of light* [79]. Quantum fluids of light occur when photons and matter interact strongly, creating the polariton as a hybrid particle. Polaritons can then form a Bose-Einstein condensate: in fact, photons and matter become indistinguishable, forming a fluid-like substance that behaves like a superfluid. This unique behaviour has potential applications in quantum information processing, as well as in the study of fundamental physics [80]. Additionally, their study can foster the understanding of light-matter interactions and of the behaviour of quantum systems in extreme conditions [80]. Polariton condensates have also shown potential as qubits in quantum computing, envisioned to be based e.g., on superposition states with different orbital momenta [81]. The coherent and stable nature of these fluids makes them attractive candidates for storing and manipulating quantum information. Polariton qubits may have several advantages over other qubit types, given that they can be operate at room temperature and are compatible with traditional semiconductor fabrication techniques. Additionally, the strong light-matter interactions that give rise to polariton condensates offer the possibility of achieving high-fidelity quantum operations [82]. While there are still challenges to be addressed, such as the short coherence times, some of the properties discussed make them interesting qubit platforms [83].

Quantum well and quantum dot lasers have emerged as important applications of quantum technology in the field of photonics [84]. Quantum well lasers use ultra-thin layers of semiconductor material to confine the motion of electrons in one dimension, creating a quantized energy-level structure that can be used to produce light [85]. In quantum dots, electrons are confined in all three dimensions within nanometers sizes [86], with a degree of control of relevant system properties like their lifetime, and the precise engineering of their light-emitting properties at specific wavelengths to make excellent single-photon sources with applications in quantum communications [87], sensing [88] and quantum computing, in particular based on electron spins [89]. In particular, quantum dot systems possess the advantage that the electron coupling with external reservoirs can be precisely tuned electrically and that interparticle and spin interactions, along with their interplay, are strong enough to be observed [90]. Additionally, electron states in quantum dots are amenable to be effectively described by Hubbard Hamiltonians in specific regimes, more easily accessible in this hardware than in other atomic-based platforms, where thermal energy is much smaller than tunneling energy, and the latter much smaller than onsite repulsion energy [90]. Existing challenges in these platforms include the need for improved control over their size, shape, and composition to optimize their emission properties or the susceptibility of quantum dots to environmental perturbations [91], which can limit their coherence and quantum properties.

Another interesting platform for quantum physics applications is provided by the macromolecules known as molecular nanomagnets (MNMs) [92–94]. These molecules typically host large manageable low-energy spectra, where the corresponding eigenstates and eigenvectors can be suitably engineered. These levels can be used for storing and processing quantum information. In particular, MNMs have been used as qudits, enlarging the available logic space for computation, compared to qubits, building blocks of the more diffused solid-state or atom-based architectures. Moreover, qudits on MNMs recently proved to allow quantum error correction and fault-tolerant computation, even in the space of a single molecule. MNMs also allow a high degree of control to synthesize supramolecular structures (also transferable on solid-state devices), where notably the qudits, even if possibly interacting, can maintain separately their properties and coherence. For all these reasons, the role of MNMs for quantum information, simulation and sensing is gathering increasing relevance. Beyond these targets, several other applications have been envisaged, as for electronics (single-molecule transistors or superconducting devices) or spintronics (say via the novel chiral-induced spin selectivity).

*Hardware Developments towards Quantum Communication* Before proceeding to describe in more detail the main cold and ultracold-atom based platforms, we close this introduction by highlighting that despite the widespread successful near-term applications of NISQ devices based both in solid-state or in atom technologies, impactful applications still require large number of qubits that those available in the current hardware. In this direction, important research is devoted to explore different routes for reliable and long-distance connections within NISQ devices, a large research field falling under the umbrella of quantum networks and quantum information processing, and more generally quantum communication [66, 95] typically enabled by optical means [96]. A key challenge in implementing quantum networks is to distribute entangled flying qubits, i.e., quantum channels typically realized with photons, into nodes that are spatially separated. This operation is performed by means of suited quantum transducers that can write the entanglement properties into stationary physical qubits, in fact functioning as quantum memories, also using teleportation protocols as a concept introduced in this field with the seminal work of Bennett et al. [97]. Global and significant efforts are being dedicated over the last decades to advance the theoretical design and experimental realization of these crucial tools, with explorations that touch upon all the main quantum platforms, no matter whether atoms, ions, or solid-state based. A complete discussion of this field falls beyond the scope of this review, that focuses more strongly on quantum simulation hardware and corresponding applications, rather than on quantum information and algorithmic tools. While for comprehensive reviews, we refer the reader to [66, 95, 96], in relevant sections below we mention selected quantum communications developments that can extend, in the near term, the applicability of specific hardware.

*Ultracold Atoms* After having summarised predominant solid-state technologies, we can shift our attention to AMO platforms, the focus of this review. The development of ultra-cold atomic technologies has exploded since the very first realizations of the Bose-Einstein condensation of ultracold Bose gases in 1995 in the group of Carl Wieman and Eric Cornell at JILA [98] and of Wolfgang Ketterle at MIT [99], and of superfluidity in Fermi gases in 2003 in the groups of Debbie Jin at JILA [22] and of Wolfgang Ketterle at MIT [100]. While referring to seminal reviews for details [101–103], we here recall that in those early days, cooling of dilute atom gases down to nano-Kelvin temperatures in the quantum degenerate regime while remaining in the dilute gas regime, was only possible by means of combined techniques of laser cooling and magnetic trapping [104] followed by evaporative cooling, notwithstanding the required knowledge of low-energy scattering properties of atomic species to ensure efficient thermalisation via elastic collisions. Though in dilute conditions, the emergence of interaction effects has been since the very beginning a clear distinction from the original prediction of Bose and Einstein [105]. In fact, the steady and rapid progress in quantum-gases platforms capabilities results also from the emergence of a new scientific community where atomic and molecular physics, quantum optics, condensed matter, and quantum information scientists merged into, fostering cooperation, contamination, and cross-disciplinarity as values for the way of conducting research.

After almost three decades, the toolbox at hand for accurate control and manipulation of these systems under extreme quantum conditions has enormously flourished with tools to engineer system properties for quantum states realisations. Among these: the trapping geometries allowing for reduced dimensions (D) down to effectively 2D, 1D, and 0D systems also with the aid of optical lattices (see Section II A 3), and the introduction of uniform, boxed, potentials [106] allowing for textbook examples of superfluid behavior [107], as well as refined detection techniques like the quantum gas microscope [108]; the strength and range of the interactions, with the tuning of the short-range van der Waals interactions via the low-energy scattering length (see Sections II A 2), the use of ultracold ions with Coulomb-like potentials (Section II A 4), Rydberg atoms with $\sim 1/r^6$ (Section II A 5) and dipolar atoms with $\sim 1/r^3$ interactions (Section II A 6), and cavity-photon mediated tunable interactions depending on the number of coupled cavity modes (Section II A 7); the addition of disorder (see Section III B 1), or the coupling to environmental noise and dissipation mechanisms (see Section II B 1) and the possibility of investigating dynamical phase transitions (see Section III B 4). With reference to Figure 1, the main platforms and tools are briefly highlighted in the rest of this section, to provide a flavour of the capabilities at hand, before diving into the theoretical tools in Sections II B–II D, and finally into the applications.

### *2. Atoms with Tunable Short-Range Interaction Strength*

Atomic gases platforms are nowadays equipped with the powerful tool of the Fano-Feshbach resonance mechanism, allowing to manipulate the short-range part of the inter-atomic interactions via the tuning of the low-energy scattering length via changing an external magnetic field.

Before diving into the mechanism illustration, we recall the relevant fundamentals. The low-energy limit of the s-wave scattering amplitude is $f_s(k) \simeq [a^{-1} + ik]^{-1}$ in terms of the scattering length $a$ for $ka \ll 1$. The scattering length $a$ has an important physical meaning, as depicted in Figure 1a. It represents the degree by which the long-range (free particle-like) behaviour of the wave function is affected by the interactions. The corresponding interatomic potential can be accurately described by a pseudopotential $V(\vec{r}) = 4\pi\hbar^2 a/m\delta(\vec{r})$, with $m$ the atom mass and $\delta(\vec{r})$ the Dirac-$\delta$ function. The sign of $a$ provides useful information on the effects of the interaction. When $a$ is positive, the two-particle wavefunction possesses a zero before approaching its asymptotic form: thus, the short-range part of the wavefunction behaves as a bound state, while the long-range part as a free particle. Thus, during the scattering process the particles are bound for a finite amount time before decaying back into two free ones. When $a$ is negative, the wavefunction does not vanish at any distance from the origin, and no resonant intermediate states appear. When turning to the many-atoms system, positive $a$ values can allow for the formation of metastable pairs of atoms with wavefunction localized in real space, while negative $a$ values determine an attractive interaction $V(r)$.

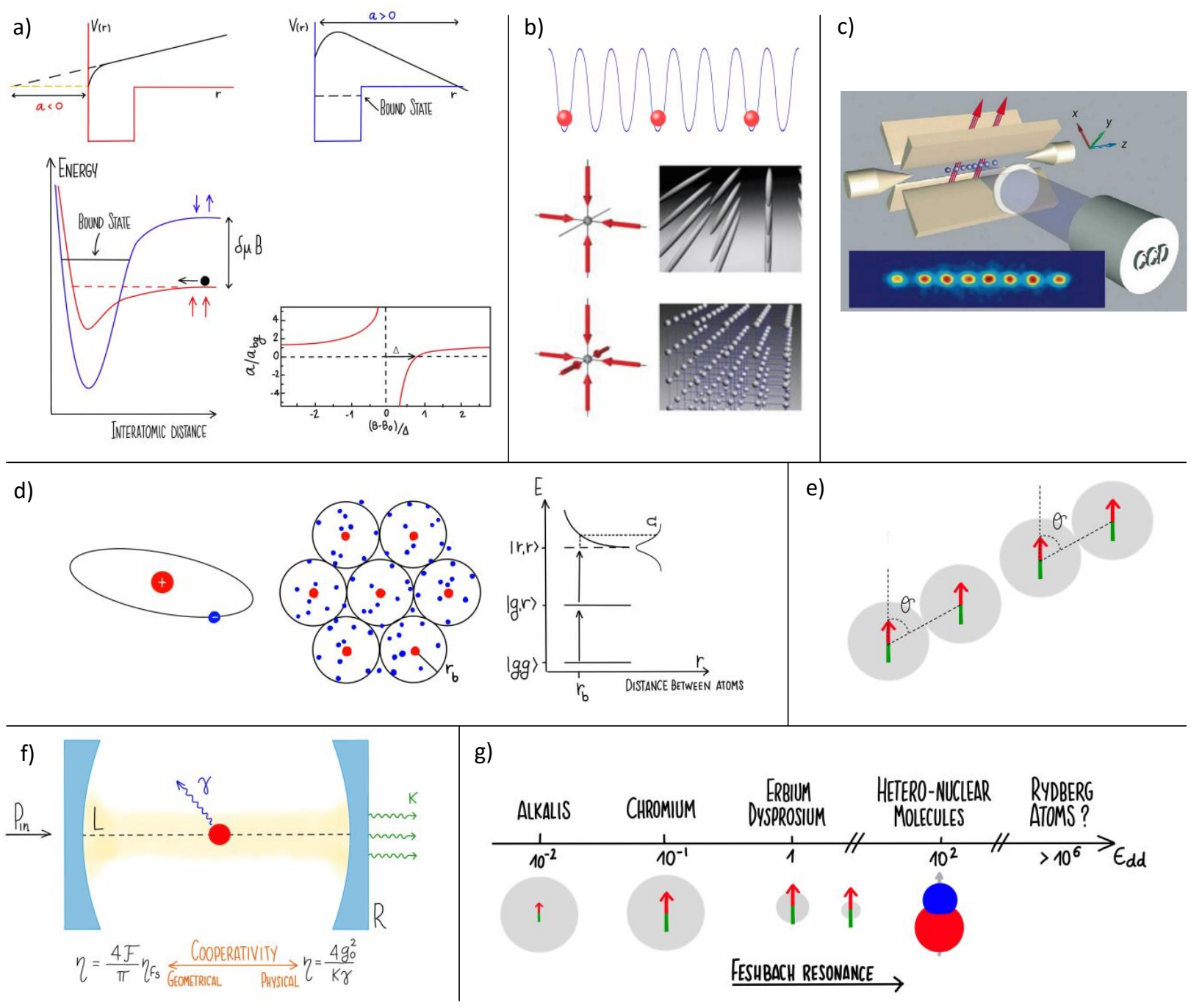

Figure 1: Schematic representation of the different atom−technology platforms considered in the Section II A. (**a**) Atoms with tunable interactions. Top: physical meaning of the scattering length $a$ illustrated through the simplest case of an attractive square−well potential. $a < 0$ (left): the wave function has no zeros in the physical $r > 0$ plane. $a < 0$ (right): a bound state appears. Bottom left: Fano−Feshbach magnetic resonance mechanism. A static magnetic field tunes the energy difference between the threshold of the open channel and the bound state in the closed channel. Bottom right: resulting scattering length vs. magnetic field. (**b**) Optical lattices. From top to bottom: 1D, 2D, and 3D optical lattices leading to 2D, 1D, and 0D−like confinement, respectively. Center and bottom images are from [109]. (**c**) Trapped ions. Linear Paul trap. Ions are trapped using static electric control fields combined with time-dependent radio−frequency oscillating electric fields to stabilize the confinement. In tight radial confinement, laser−cooled ions form a linear string (inset image for eight ions) with a spacing determined by the trade−off between external confining fields and ion−ion Coulomb repulsion. Image from [2]. (**d**) Rydberg atoms. Left: Rydbergs are atoms excited to very high energy states. Right: Rydberg blockade: the shift due to their strong mutual interaction displaces a doubly excited state out of resonance. Center: therefore, it is unlikely to have two Rydbergs closer than the corresponding blockade radius $r_b$. (**e**) Dipolar atoms. Atoms can be represented as dipoles, the interaction being thus necessarily anisotropic: in this 1D configuration, e.g., a rotation of the angle $\theta$ between the dipole direction and the direction of atoms line up, can turn the interaction from maximally repulsive (head to head) to maximally attractive (head−to−tail). (**f**) Atoms in optical cavities: the free−space cooperativity $\eta_{fs}$ related to the atom-photon strength $g_0^2$ is amplified into $\eta = 4\mathscr{F}\eta_{fs}/\pi$ by the number of round trips of the photon in the cavity, and measured by the free-spectral range $\mathscr{F}$, a geometric quantity. The system is open and out-of-equilibrium, due to unavoidable photon leakage and atomic spontaneous emission at rates $\kappa$ and $\gamma$, respectively. In fact, it can be shown that $\eta = 4g_0^2/(\kappa\gamma)$, so $g_0^2$ can be tuned by geometrical means. (**g**) Summary of energy scale for the interaction strengths from dilute (also with Fano-Feshbach resonances), to dipolar, to hetero-nuclear, to Rydberg atoms.

Further terms in the expansion of the energy scattering can be considered when energy dependence of thee scattering properties cannot be neglected. To lowest order, $f_s(k) \simeq [a^{-1} - r_0 k^2/2 + ik]^{-1}$. This introduces a new relevant parameter, the so-called effective range $r_0$. When positive, $r_0$ is a measure of the effective range of the interatomic potential. When negative instead, $r_0$ is unrelated to the interaction-potential range, it rather expresses the possibility of the two atoms colliding at low enough energy to remain for a finite time in a bound state and then decay back into two free particles: this is the process of resonant-scattering.

Reverting back to the tuning mechanism for the scattering length, the original seminal derivation of the theory has been independently worked out by Fano [110] and Feshbach [111]. In a nutshell, the Fano-Feshbach mechanism is a straightforward phenomenon occurring when a discrete bound state (also called closed channel) and a free-scattering state of two atoms (also called open channel) become resonant, thereby affecting the scattering amplitude. The resonance condition can be achieved by exploiting the different magnetic moments and responses to an external magnetic field of the scattering state in the open channel and the bound state in the closed channel. Typically, magnetic means are employed, using alkali atoms or rare-earth elements. In experimental atomic systems with s-wave ground states and zero electronic angular momentum, the collision properties are determined by the electronic Born-Oppenheimer potentials between two $s$-state atoms. These potentials depend on the total electronic spin, denoted as $S = S_1 + S_2$, with alkali atoms having quantum numbers of 0 and 1. The resulting spin-singlet state $V_S$ becomes the closed channel, while the spin triplet state $V_T$ becomes the open channel. By introducing a static magnetic field $B$, the energy difference between the two continuum levels can be adjusted due to the different magnetic moments of the two channels. This tuning of the magnetic field allows for control over the energy difference between the continuum threshold of the open channel and the bound-state energy of the closed channel [112]:

$$a(B) = a_{bg}(1 - \frac{\Delta}{B - B_0}),$$

where $\Delta$ is the width of the Fano-Feshbach resonance, occuring at magnetic field $B_0$, and $a_{bg}$ is the scattering length well away from resonance, in fact named background scattering length.

The scattering length can also be cast in the form $a = a_{bg} - mg^2/(4\pi\hbar^2\nu)$ [113] with $m$ the atom mass, in terms of the strength $g$ at wavevector $k = 0$ of the coupling between the open and the closed channels, the detuning $\nu$ from resonance. When the effective range $r_0$ plays a role, it can be cast in the form $r_0 = -8\pi\hbar^4/\ (\mathrm{mg}^2)$ [113]: this makes apparent that weaker couplings determine larger effective ranges and longer lifetimes of the resonant state. When considering the two-body problem, Fano-Feshbach resonances are classified as narrow or broad depending on whether $|a_{bg}/r_0| \ll 1$ or $\gg 1$, respectively. When considering the many-body atoms system, however, the relevant length scale becomes the inverse Fermi momentum $k_F^{-1}$. Thus, while the width of the magnetic resonances is a fixed trait of the two-particle interaction potential, in the many-body system this can be tuned - at least slightly - changing the atomic density.

Besides than with magnetic fields, the Fano-Feshbach resonance mechanism can be also implemented by optical means [114]. In contrast with the magnetic ones, optical Fano-Feshbach resonances allow for a more versatile external control of the resonance width, besides its location. The idea is that a photon with (typically optical) frequency $\nu$ excites the two atoms system into a roto-vibrational level. Modeling for simplicity the atom as a two-level system, with $|g\rangle$ and $|e\rangle$ separated by energy $h\nu_c$, the resonance position can thus be changed after tuning the laser frequency electromagnetic spectrum. These resonances are called optical Fano-Feshbach resonances. While more versatile, optical Fano-Feshbach resonances suffer from the fact that the excited roto-vibrational state is inevitably subjected to irreversible decay even at zero collision energy with a characteristic time $\hbar/\gamma$, with $\gamma$ the spontaneous emission rate. Therefore, the scattering length is complex, with an imaginary part describing collisional loss. In addition, both the resonance width and shift depend on the laser intensity $I$. Notwithstanding these difficulties, the use of optical resonances has also been developed in optical cavities, that are amenable to accurate control and engineering of atom-photon interactions [115].

### 3. *Atoms in Optical Lattices*

Optical lattices have emerged as a powerful tool in the field of AMO quantum technologies due to their ability to trap and manipulate individual particles with high precision [116, 117] and low temperatures in periodic potentials, while not suffering from non-accurately controllable complications of solid-state technologies, that can typically arise from phonon-like interactions and fabrication defects. These potentials are created by the interference patterns of laser beams, which can confine particles in stable positions, thus amenable to load atoms in specific spatial patterns. In turn, this can be realised in the presence of disorder [118], variable geometries and different dimensions, either spatial (3D to 0D) or even synthetic [119, 120]. It is thus not surprising how atoms in optical lattices allow for the study of a wide range of phenomena [109] (see Figure 1b). Breakthrough examples of physics accessible with optical lattices include the seminal observation of superfluid (SF)-to-Mott insulator phase transition [25], emergence of dissipation from superfluid behavior [121], dissipation-induced coherence [122], Josephson physics [123–126]. An important property of these systems is that the related dynamical scales correspond to frequencies $\sim$ Hz-kHz allowing for the experimental observation of the dynamics of the system in real time; compatible with typical lifetimes ($\gtrsim$1 s) of the prepared atomic states. Optical potentials have been used to trap and cool not only neutral atoms but also molecules [127, 128].

They constitute essential building blocks for various quantum technologies [125] including quantum computers [129], sensors [130], and simulators [117, 131, 132]. Moreover, technological development allowed for single-site controllability, probing and imaging with the appearance of quantum gas microscopes, both in bosonic and fermionic systems [108, 133].

In fact, another important versatility of optical lattices is that technological advancement has allowed to trap and cool both bosonic and fermionic species [29, 134–136], as well as spin mixtures or Bose-Fermi mixtures, which have emerged as a promising platform to create controllable species-dependent potentials and tunable interactions for atoms or molecules [137], with applications to, e.g., the study of topological phases and exotic superfluidity [138, 139]. Optical lattices present several advantages in comparison with other QTs. In particular, the system sizes that are achievable, the discrete nature of the system, the flexible geometry and the single-site resolution make them a perfect platform for simulation and Hamiltonian-based protocols. However, they are less-adapted to gate-based approaches, as for e.g., trapped ions. This is partially due to the ratio between their coherence time, limited mostly by background radiation pressure, and gate time—as the system frequencies are $f \sim$ Hz-kHz small compared to trapped ions with $f \sim$ MHz—as well as their lower experimental repetition rate compared to other platforms. We note that despite these challenges, progress is steadily being achieved. If used for quantum computing, atoms in optical lattices offer the advantage of providing qubits made of individual neutral atoms and consequently identical by construction. However, lack a modular construction of the hardware, like it is possible for solid-state technologies, makes the scaling more challenging despite the steady progress. We refer to ref. [140] for a complete discussion on the commonly utilized theoretical methods for analyzing the cooperative responses of atomic arrays, and a deep-dive into recent advancements and potential future uses of planar arrays as adaptable quantum interfaces connecting light and matter. Moreover, we refer to [141] for a theoretical and experimental review of optical dipole traps for neutral atoms.

#### *4. Trapped Ions*

While optical lattices represent an ideal platform for simulation and sensing, and the progress towards quantum computing (QC) is still ongoing (with notable progress thanks to Rydberg-atom technologies, see Section II A 5 below), trapped ions were originally better technologically suited for gate-based quantum technologies [142, 143] and they are still nowadays one of the leading cold-atomic platform for digital approaches due to continuous development [144, 145]. This is due to their ability to prepare high-fidelity and long-lived qubits [146–149] with short gate times (with characteristic frequencies $f \sim$ MHz) and high repetition rates, while remaining a suitable hardware for Hamiltonian-based approaches. The development of linear Paul traps, using a combination of radio-frequency and static electric fields (see Figure 1c), allowed for the confinement of ions and consistent manipulation of the quantum states of the ionic chains with high precision and increasing number of elements, overpassing 100 qubits. Once the ions are trapped, by using laser pulses or microwave fields it is possible to perform single-qubit operations, entangling gates, and other quantum operations with high fidelity [150–152]. In fact, even N-body interaction terms have been recently encoded [153].

One of the significant advantages of trapped ions over other quantum technologies is their long coherence times [144] due to their low coupling to the environment. Trapped ions have coherence times on the order of seconds, or event minutes [154], significantly longer than other qubit technologies.

As for optical lattices, scalability represents one of the challenges, as these platforms lack the modular nature of solid-state QC, leading to certain operations becoming slower as the system is scaled up or dissipative effects becoming more relevant. Substantial efforts have been made towards scalable approaches [155] and progress is continuously made towards larger ion traps while maintaining high coherence times and fidelity. Constant assessment on the technological challenges for the construction and operation of a trapped ion system requirements has also been regularly made [144, 145]. All these combined efforts have allowed for QC applications where trapped ion qubits have been shown to perform some of the seminal QC protocols, such as Shor's algorithm for factoring large numbers, and Grover's algorithm for searching unsorted databases, with high fidelity [156].

Importantly, an essential aspect of trapped ions is the fact that they are also suitable for Hamiltonian-based or analog physics, as one can program the coupling to the trapped modes, engineer the interaction between the ions, and thus access a wide variety of long-range coupling configurations [157, 158]. This analog-digital duality has made trapped ions one of the most flexible platform also in quantum simulation [159–161], e.g., of quantum chemistry problems [162] or also in sensing [163]. This flexible nature and applications together with novel techniques for the probing of the quantum state [164] in efficient and scalable ways, have made ion traps one of the leading quantum technologies.

### 5. Rydberg Atoms

Another cold-atomic platform that has experienced substantial development in recent times is based on the use of Rydberg atoms and their unique properties [165]. The Rydberg atoms platform consists of ensembles of individual highly excited atoms (see Figure 1c), i.e., characterized by large principal quantum numbers $n$ [166, 167], confined in optical lattices or arrays of optical tweezers [168] using the gradient force of a focused laser beam to trap, transport and manipulate single particles, which are separated by a few micrometers. In fact, the Rydbergs benefit from the same level of control through optical addressing techniques that we discussed in Sections II A 1 and II A 3, enabling great controllability over the individual atoms within the system.

By driving the atoms to highly excited Rydberg states, interactions scaling as $\propto 1/r^6$ – beyond the $\mu$m scale—can be achieved. When in the Rydberg state with quantum number $n$, the atoms possess two crucial properties. Firstly, their lifetime scales as $n^3$, significantly longer than the lifetimes associated with low-lying transitions (typically in the range of $100\mu$s for $n \sim 50$). Secondly, they exhibit substantial dipole moments between states $n$ and $n-1$ with opposite parity, scaling as $n^2$. Consequently, these properties give rise to significant interaction strengths $V$ corresponding to frequencies $V = /h \geq 1$MHz for $n \sim 50$, at distances approximately 5 $\mu$m [169]. The presence of interactions modifies the excitation dynamics with a mechanism that is known as Rydberg blockade, observed for the first time in 2009 [170] (see Figure 1c): in essence, only one of any two atoms that are spatially close by can be excited, due to the shift introduced by their strong mutual interaction, which displaces the doubly excited state out of resonance [169].

These properties make Rydberg atoms ideal for quantum computing [171], simulation [172], and sensing [173] applications. Rydberg technology allows, as was the case of trapped ions, for gated-based protocols [174] or analog-based approaches [169]. Additionally, their long lifetimes allow for long gate sequences and simulation protocols [165, 175]. Another advantage inherent to the use of Rydberg systems is their strong dipole moments, which facilitate long-range interactions between them, increasing their flexibility and applicability in creating multi-body gates [176].

Unlike trapped ions, Rydberg-based systems are easier to scale up, as is the case of optical lattices. On the other hand, challenges remain that are associated with the increasing size, such as the need to isolate the Rydberg atoms from external perturbations originated by collisions with other atoms or photons: in fact, these dissipative effects can cause decoherence, which can reduce the timescale for which we can perform operations with high fidelity. In more recent years, the development of optical tweezers has rendered Rydberg atom technologies even more flexible [177, 178], as it allows precise arbitrary positioning of the Rydberg atoms, enabling the creation of specific quantum states, simulating particularly relevant physical geometries, and exploiting the long-range nature of interactions to create specific gates and protocols [179]. In very recent years, Rydberg atoms were used to successfully implement high-fidelity two qubit operation in parallel [180], in scalable, highly connected 2D systems [181]. For an exhaustive description of Rydberg atoms we refer to the more recent review [169].

### 6. Atoms with Dipolar Interactions

Similarly to Rydberg atoms, another relevant platform characterized by such long-range interactions is dipolar gases. As for Rydbergs, these dipolar gases can be confined as well in optical lattices, traps and cavities, benefiting from the technological control achieved in those platforms. Since they have been envisioned [182], dipolar gases have been associated to a number of intriguing quantum states, such as Wigner crystals [183, 184], ferrofluids [185], systems with roton-maxon excitations [186], checkerboard supersolids [187], Haldane insulators [188], the emergence of quantum scars [189], and Fulde-Ferrell-Larkin-Ovchinnikov phases [190]. These states emerge due to the interplay between quantum fluctuations and frustration effects due to interactions.

In the absence of dipoles, atoms in the ground state interact through short-range van der Waals interactions, which decay as $1/r^6$ and are typically isotropic due to the spherically symmetric electronic cloud of most atoms in the ground state [191], contrasting the long-range dipolar interaction. Different platforms are available to investigate the effects of dipole-dipole interactions (DDIs) in ultracold gas contexts. For instance, electric dipole moments can be induced using heteronuclear molecules [128, 192–194] or Rydberg atoms [56, 195, 196] in an electric field, or by employing light-induced dipoles [197]. Moreover, elementary particles can possess permanent magnetic dipoles even in the absence of an external field. Consequently, the impact of magnetic DDIs on quantum gases can be studied under full rotational symmetry, even at extremely small magnetic fields.

In dipolar gases experiments, various species of atoms have been used depending on the specific research goals and experimental setups. Some commonly studied species include Dysprosium (Dy), Erbium (Er), Chromium (Cr) and Dysprosium-Erbium mixture (Dy-Er) [191].

Generally the DDI interaction scaling as $1/r^3$ [191] is also anisotropic and can be either attractive or repulsive depending on the relative orientation of the dipoles. Specifically, its elastic component varies as $1 - 3\cos(2\theta)$, where $\theta$ represents the angle between the relative position of the particles and their polarization direction. In one-dimensional (1D) geometries, unique characteristics come into play [198, 199]. An illustrative example is the exactly solvable Lieb-Liniger gas [200], where the many-body excitation spectrum becomes identical to that of a free Fermi gas in the limit of infinite contact interaction strength $g_{1D} \to \infty$, referred to as the Tonks-Girardeau gas [201]. In fact, tuning of the quantum liquid density causes a crossover from a strongly interacting Tonks regime to a quasi-crystal, while the low-energy system remains characterized by Luttinger liquid behavior [202–205].
Theoretical descriptions based on the local density approximation [198], time-dependent Hartree method [206], and diffusion Quantum Monte Carlo simulations [207, 208] have been developed to capture the behaviour of the system across weakly and strongly interacting regimes. More recently, independent tuning of the van-der-Waals contact potential and of the dipolar interaction strength by means of dipoles rotation, have made accessible richer regimes, investigated both experimentally [189] and theoretically [209].

Notably, the development of stable degenerate quantum gases composed of atoms with strong dipolar forces as the primary interactions [210] has revealed the coexistence of crystalline order and superfluidity, known as supersolidity [211], that have recently realized experimentally [189, 210, 212–215]. Other studies [216] revealed the possibility of self-bound droplets and droplet assemblies.

For a more complete description of dipolar gases, we recommend the theoretical review [217] for a theoretical review, and the more recent experimental [191].

### *7. Atoms in Optical Cavities*

Optical-cavity quantum electrodynamics (QED) explores the interaction between light and matter in confined electromagnetic modes, with the purpose of achieving strong coupling between quantum emitters and the electromagnetic field and enabling the study of light-matter interactions at the quantum level. Matter in QED cavities is an out-of-equilibrium, driven-dissipative system, whose description requires open quantum system methods (see Section II B 1). As such, it can serve as an ideal platform for interdisciplinary research on noise and dissipation engineering. For a comprehensive understanding of atom processes in optical resonators we refer to the pedagogical work in [218], while an extensive review [137, 219] outlines the capabilities of optical cavities.

As sketched in Figure 1f, the concept is that the free-space cooperativity $\eta_{fs}$, i.e., the probability for the an atom to scatter a photon in the solid angle, is amplified into $\eta = 4\mathscr{F}\eta_{fs}/\pi$ by the number of round trips of the photon in the cavity, as measured by the free-spectral range $\mathscr{F}$. The latter is a geometry-related quantity involving the cavity fabrication characteristics. The single-atom cooperativity $\eta$ thus accounts for enhanced atom-photon interaction strength with respect to $g_0^2$, that is achievable by geometric means. Photon leakage at rate $\kappa$ and atomic spontaneous emission at rate $\gamma$, besides possible external pumping make the system open, driven-dissipative. In fact, it can be shown that $\eta = 4g_0^2/(\kappa\gamma)$, so that the $\eta$ design can result in tuning the strength of the coupling $g_0^2$ with respect to dissipations $\kappa\gamma$ [218].

In the strong-coupling regime, coherent evolution dominates despite dissipative processes, leading to collective Rabi oscillations [220] of the trapped atoms. Integrating cold atoms into optical cavities allows for the investigation of mechanical effects resulting from atomic motion interacting with the cavity field. Friction exerted by the emitted radiation acts as an effective cooling mechanism [221], achieving sub-recoil cooling limits [222]. This technique is more efficient in atomic ensembles, where the coupling strength scales with the square root of the particle number. Meanwhile, photon leakage not only dissipates energy but also carries information about atomic dynamics, enabling applications in feedback control and non-destructive measurements of atomic states or quantum spin squeezing [223]. Cavity photons can simultaneously serve as external optical lattices and probes, as demonstrated in previous works [224, 225].

Atomic ensembles in high-finesse optical resonators exhibit intriguing phenomena, such as long-range atom-atom interactions mediated by photon exchange [226]. These interactions collectively modify the field inside the cavity, affecting each atom in a position-dependent manner. Thus, the coupling between light and atoms strongly depends on their positions, emphasizing the role of interference effects.

Thermal atoms driven by a transverse classical pump and interacting with a vacuum cavity mode create an interesting phenomenon known as superradiance [227], in which self-organization is observed with a macroscopically populated cavity mode. The self-organization of thermal atoms was first experimentally demonstrated in [228], with self-organization revealed symmetry-breaking of translational symmetry. Self-organization of a BEC in a cavity was also observed [229].

Theoretical works have predicted a phase transition achievable by coupling a transversely laser-driven Bose-Einstein Condensate (BEC) to a single-mode cavity, with non-trivial ordering arising due to photon-mediated interactions [230], and cavity-mediated fermionic superfluidity in reduced dimensions is envisioned [231].

Besides the possibility of tuning the atom-photon interaction strength, the development of multimode cavities has led to tunable range of the interactions. Unlike single-mode cavities indeed, multi-mode cavities support transverse electromagnetic modes resonating at the same frequency: the interactions mediated by these modes possess finite-range interactions in the transverse direction [232]. The adaptation of the field to particle distribution in highly degenerate multimode cavities allows to explore novel conceptual systems and the study of crystalline and liquid-crystalline ordering [35, 233]. The comprehensive phase diagram of this complex system was investigated [234], along with additional insights provided in [235, 236]. Multimode cavities also establish connections to neural network and spin models, enabling investigations into dislocations, crystal boundaries, and phonon spectra. Initial investigations into this relationship were presented by [237, 238], while extensions involving fermionic atoms [239, 240] and local couplings using multimode cavity QED have been explored [35, 241].

Cavities can be categorized into standing-wave and travelling-wave (ring) cavities. Standing-wave cavities consist of two mirrors that create a stationary pattern through light reflection, forming a standing wave. Ring cavities, on the other hand, confine light in a circular path through total internal reflection. Unlike standing-wave cavities, ring cavities do not create a standing wave by bouncing light back on itself. However, by introducing two counter-propagating beams into the cavity, a ring cavity can support a standing wave. This introduces an additional degree of freedom where the resulting standing wave mode can freely rotate by adjusting the phase of one of the input modes. In this configuration, momentum exchange occurs between the field modes, enabling atoms to exert a back action on the optical field. This back action is sensed by all trapped atoms in the cavity, leading to coupled motions and the emergence of rich cooperative behaviour such as Recoil Induced Resonances (RIR) [242]. Experimental investigations of cold atoms in ring cavities have observed superradiant behaviour [243, 244], characterized by Superradiant Rayleigh Scattering (SRyS) [245] and Collective Atomic Recoil Lasing (CARL) [242]. Recent experiments have further explored the properties of the supersolid phase within ring cavities [246, 247].
Many applications of trapped atoms in cavities, such as quantum simulation or quantum computing, exploit space-programmable interactions. In particular the connectivity network, or graph, is usually dictated by geometry. Tunable non-local interactions are crucial for some applications like the simulating information scrambling in black holes and mappings of hard optimization problems onto frustrated classical magnets [248]. Some results [115, 248] show the experimental realization of programmable non-local interactions in an array of atomic ensembles within an optical cavity, offering also a test case for experimentally observing the emergent geometry of a quantum many-body system.

For similar reasons, also the potential importance of quantum electrodynamics in waveguides for quantum information applications has been widely recognized, see e.g., [249]. In essence, quantum emitters such as atoms or molecules, can interact even strongly with radiation propagating along a waveguide. In this way, high-entangled states from strong coupling or collective phenomena, as superradiance, arising from correlations between photons and bound states of them, can be observed, similarly to the case of cavity QED. In the same context, striking experiments have been performed quite recently with cold atoms, semiconductor quantum dots, quantum solid-state defects and superconducting qubits. For instance, in recently developed cold atomic systems coupled to nanoscopic photonic waveguides [250–252], these nanoscale devices can mediate long-range atom-atom interaction in a similar way to standard optical cavities leading to potential applications not only in quantum simulation [253], but also in quantum communications [254–256].

We close this platforms' overview by remarking that current atom technologies allow for a remarkable tunability of strength and range of the interactions (see Figure 1g), from short van-der Waals equipped with Fano-Feshbach resonances to Rydberg-like $\sim 1/r^6$, dipolar $\sim 1/r^3$, Coulomb $\sim 1/r$, and multimode QED-cavity mediated mechanisms intrinsically tunable from infinitely long to short range with increasing the number of coupled cavity modes.

### 8. *Miniaturization and Atom Chips Technology*

One of the problems that cold atom technologies have experienced over the years is the limitations of miniaturization, currently requiring moderate setup sizes and optical tables. This poses certain challenges, particularly relevant in the field of quantum sensing where the ability to produce portable, small and robust experimental setups can be crucial.

Instruments employing atomic ensembles at room temperature have achieved successful miniaturization through the utilization of micro-electro-mechanical systems (MEMS) alkali vapor cells [257]. Cold-atom generation has been demonstrated by employing a single laser beam reflected from different cell geometry structures [258, 259], while research in this direction is still going on, with the goal of investigating alternative cell geometries to reduce overall size [260]. That being said, it is important to note that, while there are promising opportunities for compact instrumentation [261], via micro-fabrication techniques involving optical elements [262–264] or via the development of grating chips [265, 266] for on-chip magnetic confinement of laser-cooled atoms, systems typically require traps that need liter-range volume loading from vacuum apparatuses. The miniaturization of ultra-high vacuum (UHV) vacuum packages [267] remains limited and many cases constrained by conventional bulk machining methods used for chamber materials, compromising partially the scalability potential provided by micro-fabrication techniques so far.

The substantial progress made in this miniaturization effort is described in depth in the reviews [266, 268, 269].

## B. Theoretical and Simulational Paradigms for Atomic Systems

### 1. Driven-Dissipative Quantum Systems

As we discussed in detail in Section II A, current developments of quantum technologies have allowed for a high-degree of control over the experimental platforms. In particular, we are currently in a unique position in which we can derive simple microscopic models based on well-understood approximations to describe closed quantum systems, and even more interestingly characterise accurately the dominant couplings of quantum technological platforms to their environment. This change of paradigm has enriched our description of the devices and enabled us to access new physical regimes that could not be explored in the context of closed quantum systems. In this section, we discuss how the use of open quantum systems, starting from their theoretical characterisation, plays an essential role in current—and future—quantum technologies.

*Open Systems: Reservoir Engineering* Traditionally in QT, one of the main approaches consisted in the cooling of the system to its ground state [104, 270] in a consistent manner. In this way, the problem revolved around modifying the coupling terms and coherent drives in our physical system, in order to map the problem of interest into the experimental device at hand that was then subsequently cooled. In the field of quantum simulation, this approach was typically denoted as *Hamiltonian engineering.* This technique was widely applied in state preparation for consistent particle loading in experimental devices [271] or entangled state preparation [272], with numerous examples to this day, typically to study condensed-matter analogs in so-called *crystals of light* with cold atoms [117, 132]. An obvious limitation to this protocol is that the number of problems that can be mapped is directly dependent on the classes of Hamiltonians that the device can represent faithfully. However, in this section we will describe how from the very early development it was understood that this idea could be extended to general dissipative couplings, with particular success in the fields of quantum simulation and quantum metrology.

The fact that these systems are also coupled to their environment represents certain challenges to our ability to describe them as the bath typically consisting of a large number of degrees of freedom; but also, provides novel ways of controlling and probing these systems, enlarging the class of problems we can map into the QT platforms and the phases of matter we can access. More concretely to represent OQSs, we are required to understand the set of necessary approximations that render a numerically effective and controllable description of both system and environment. Once this description is provided, typically in the form of stochastic or master equations as described in the following sections, exploiting dissipative coupling offers a new toolbox for the engineering of quantum many-body states, which is often described as *reservoir engineering* [273]. The idea of exploiting both coherent and dissipative coupling was originally imported from the field of Quantum Optics into cold atomic platforms were it quickly became the standard. Some of the most relevant examples of this technology transfer, now widely applied, include optical pumping [274, 275] and laser cooling [104]. In addition, current temperatures achieved in cold atomic experiments, e.g., fermionic quantum gas microscopes [276–278]—systems particularly challenging to cool in the presence of vanishing scattering lengths, would not have been possible without the use of dissipative engineering in the form of sympathetic cooling [279], where an atomic species—typically harder to cool to physically relevant temperatures—is immersed into another atomic species that can be cooled to lower temperatures, e.g., a bosonic species in a BEC state [280]. Some other relevant techniques for dissipative cooling and state preparation in AMO platforms include combining this with dark-state driving [281, 282] or more complex cooling schemes, achieving temperatures even within the lowest Bloch band [283]. For a wide set of early examples, see the dedicated review [273].

Up to this point, the applications that we have considered used the environment as a cooling reservoir where energy could be transferred from the system in a controlled manner. However, dissipative coupling can also be used for a wide range of applications including the creation of entangled states [284, 285] or topologically protected states [286] which could be used for quantum enhanced metrology as discussed in Section II D or quantum computing, see Section II C.

Furthermore, beyond these consistent cooling and state-preparation protocols presented here, the inclusion of dissipative dynamics can also give raise to fundamentally new phenomena. In particular, dissipation can alter the behaviour of known phases of matter drastically or it can be used to infer or probe some inherent properties of the system as particle statistics [287]. It can even generate new phases, with some early examples in [284, 288, 289]. The emergence of these new phases of matter has lead to the study of non-equilibrium critical behaviour and new universality classes in driven-dissipative scenarios [290, 291].

An interesting case that has drawn substantial attention recently is the case of measurement-induced phase transitions [292–294], a subclass of dynamical phase transitions that appear due to the interplay between projective measurements and coherent evolution leading to drastically different entanglement properties in the different phases of monitored systems, with the difference been encoded only in non-linear quantities of the system's state, making its probing rather elusive [295]. This particular example is an illustrative instance of a fundamental aspect of the OQSs framework, as they allow to also consider their relation with foundational quantum concepts, such as information scrambling [296], the role of measurement [297] or the emergence of classical behaviour [298, 299].

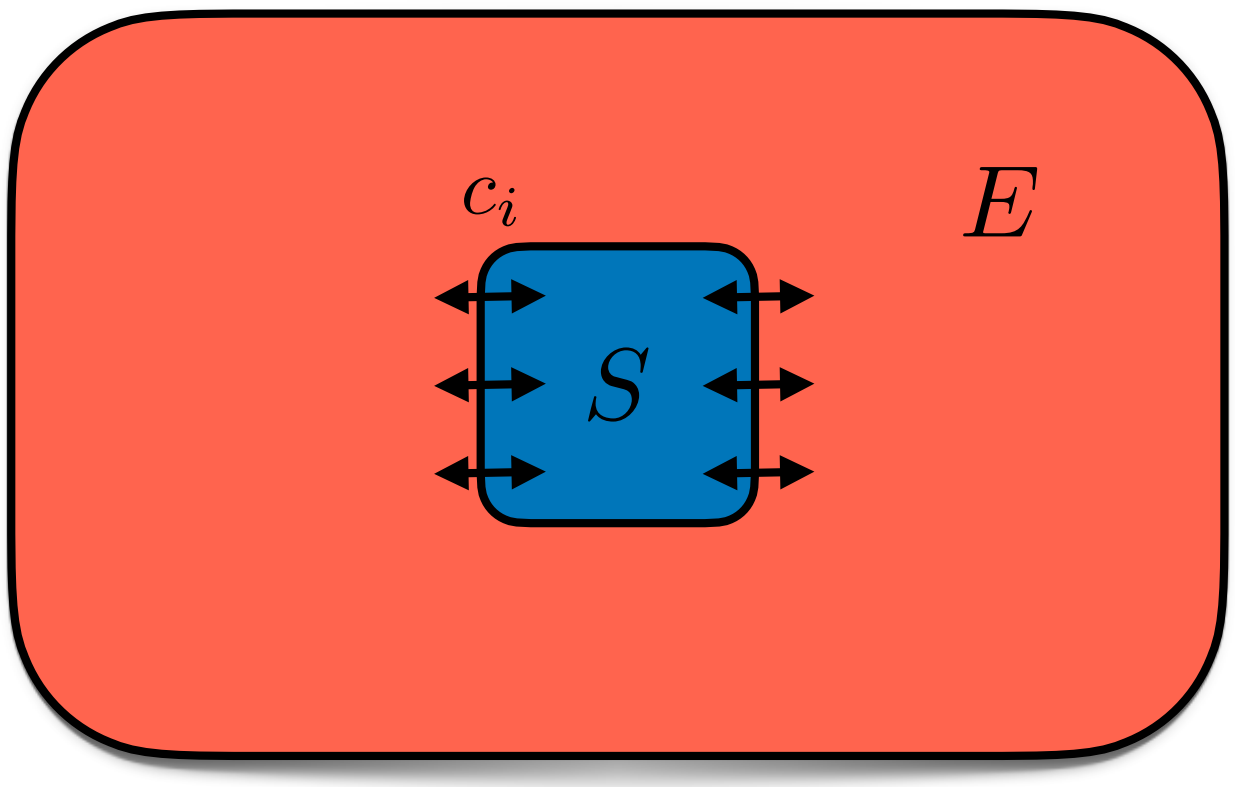

Figure 2: Diagram of an open quantum system. The physical system of interest ($S$) is immersed in a typically larger environment ($E$) also denoted as the bath. The $S-E$ environment coupling can be described by a discrete set of dissipative channels $c_i$ under the suitable approximations.

Notably, the necessity to characterize these non-equilibrium dissipative phases also stimulated developments of existing theoretical techniques and approaches developed for equilibrium, as those based on Green functions. This formalism is particularly useful in the description of transport problems in correlated media beyond 1D and amenable for long-range interacting systems, thus constituting a useful framework in the physical description of, e.g., biological matter. As we do not cover in detail these methods in our review, we refer the reader to the existing recent literature on the topic [300].

*Describing Dissipation in Quantum Systems: Gorini-Kossakowski-Sudarshan- Lindblad Master Equation* As we have just described, despite experimental efforts that guarantee the always-improving isolation conditions of quantum systems, an open quantum system approach is necessary to guarantee both a more complete description at relevant bath-system timescales and also as a controllable novel form of interaction with quantum systems, accessing new physical regimes.

As a result, we require a formalism to describe the dynamics of systems weakly coupled to their environment, as shown in Figure 2,— while we later on discuss ways in which this condition can be relaxed. Here, we discuss the main approximations that allow us to write an evolution equation for an open system with Hamiltonian $\hat{H}_{\text{tot}} = \hat{H}_{\text{sys}} + \hat{H}_{\text{env}} + \hat{H}_{\text{int}}$ based on a set of energy and timescale hierarchies that we detail:

- Born Approximation: this approximation assumes that system and environment are weakly coupled, implying that the dynamics induced by this coupling $H_{\text{int}}$ are small compared to the system or environment ones given by $H_{\text{sys}}, H_{\text{env}}$.

- Markov Approximation: Firstly, it implies the S-E coupling to be frequency-independent over short timescales, that we can define with a rate $\Gamma$. Secondly, it requires that the information transmitted from the system into the environment, decays exponentially fast in a way that this information cannot return to the system $S$ itself. This, together with the Born approximation, implies that the environment can be considered static throughout the evolution.

  This approximation is valid in a wide range of scenarios; formalising the condition that the bath correlations $\langle b^{\dagger}(t)b(0)\rangle \to 0$ decay in timescales $\tau_{\text{env}}$. These $\tau_{\text{env}}$, on the other hand, are given by the inverse of the spectral width of the bath [301]. E.g., if we assume a bath in thermal equilibrium, the decay of correlations will be much faster than the rate of change of the system $t \sim \Gamma^{-1} \gg \tau_{\text{env}}$ [302, 303].

These approximations, together with the rotating-wave or secular approximation [36, 302] neglecting the fast rotating terms of $\hat{H}_{\text{int}}$ to guarantee that the resulting density matrix after the evolution is physical (trace-preserving and positive semidefinite), allow us to derive the desired description. It is relevant to mention that in the context of quantum optics and AMO systems all of these approximations can be summarised as simply the presence of a dominant frequency in the total system that we can relate to the characteristic energy of the physical system $\omega_{\text{sys}}$, which is much larger than typical coupling strengths with the radiation fields $\Omega_0$ or the related detunings $\Delta$.

Moreover, we have specified that typically $\tau_{\text{env}} \lesssim \omega_{\text{sys}}^{-1}$ [36, 301, 302]. Thus, we can justify all these approximations by writing:

$$\Gamma, \Omega_0, \Delta \ll \omega_{\text{sys}} \,. \tag{1}$$

The standard equation that fulfills the required approximations for the evolution of an OQS, is denoted the Gorini-Kossakowski-Sudarshan-Lindblad master equation describing the evolution of the reduced system density operator $\rho$:

$$\dot{\rho} = \mathscr{L}\rho = -i\left[\hat{H}_{\mathrm{sys}},\, \rho\right] + \sum_i \frac{\Gamma}{2}\left(2\hat{c}_i\rho\hat{c}_i^\dagger - \left\{\hat{c}_i^\dagger\hat{c}_i,\, \rho\right\}\right)\,, \tag{2}$$

with the jump operators $\hat{c}_i$ characterising the $i$-th dissipation channel between system and environment. A crucial point is that these operators appear after tracing out the environment degrees of freedom and, as a result, the description of the OQS is done in terms only of the system-only Hilbert space reducing dramatically the theoretical and numerical description.

While the type of channels is extremely large and system-dependent—from e.g., background gas collisions or inelastic light-scattering for a cold-atomic system, to parasitic fields or dissipative coupling to defects in solid state systems - the effects that those channel produce on the system can be related to two important physical phenomena: dissipation and decoherence. The first one, dissipation, describes the loss of energy or particles from the system into the environment. Decoherence instead is characterised by the transfer of information from the system into the environment due to its indirect probing of the system. At the level of the physical system, this produces the dynamical destruction of coherence between the states competing with the superposition arising from coherent dynamics. Both dissipation and decoherence are fundamental concepts in the description of OQS and their implications.

Finally, while the master equation allows for a system-only description of the OQS, the exponential nature of the Hilbert Space size produces relevant limitations to the sizes we can compute numerically as the density matrix $\rho$ has dimension $dim(\mathscr{H}) \times dim(\mathscr{H})$. One possible avenue to overcome this limitation consist of the use of stochastic unraveling techniques that allow us to describe these systems as an average of random trajectories. These techniques are the topic of the next section as, apart from their numerical utility, they are deeply connected with specific experimental setups and ways in which we can interact and probe quantum systems linking to quantum measurement theory and quantum information [304].

*Stochastic Unraveling Description of Open Quantum System* In this section, we detail some of the existing theoretical and numerical methods that allow for the efficient description of open quantum systems within the Markovian approximation, based on some form of stochastic averaging.

The first modelisation of the bath in this direction comes from the field of quantum optics where a quantum stochastic calculus formulation allowed to describe the effect of the environment as a set of noise terms leading to the evolution of the system governed by Quantum Langevin-type equations or master equations [305–307].

Importantly, these equations where then quickly formulated in term of stochastic Schröedinger equations [308] meaning that the effect of the noise in the evolution equation could be described in terms of probabilistic random individual trajectories that can be efficiently sampled numerically. Moreover, each of the individual trajectories would correspond to the measurement record of a single run of the corresponding experiment. Thus, these techniques do not only are interesting to provide a sample of the statistically averaged reduced density operator of the system but also are intimately connected to the information that the measurement apparatus or bath would extract from a given quantum system. While this idea is inherently in the realm of quantum optics, the considerations are in fact rather general and this approach has been approach widely in cold atomic and solid-state systems. Examples can be found in these reviews [36, 309, 310].

This powerful formalism remains rather mathematical, nevertheless, substantial work was performed in the early years of the development of the theory to connect it to the physical properties of the bath and the specific experimental scenarios. In particular, we would like to highlight two relevant scenarios that have develop to constitute their own standalone frameworks for OQS: photodetections characterised via quantum jumps, Figure 3a, and homodyne detection described via quantum state diffusion, Figure 3b.

**Quantum jumps: Monte Carlo wave-function method** This approach was independently introduced in different contexts of quantum optics from laser theory to the description of radiating atoms [311–314], with all revolving around the process of photodetection. The idea is rather simple: when a photon is detected on the system's apparatus this implies a sudden change on the state of the system, or jump, e.g., a two-level atom decays from the excited state emitting a photon into the background field, see Figure 3a. Consequently, it is possible to associate the state of the system in a single realisation with the stochastic measurement of the detector. This was observed experimentally already in the early years of the theory by monitoring the change of an ancilla in optical systems [315–317], and later on also in solid-state ones [318, 319]. More recently, the degree of experimental control has allowed not only to observe such quantum jump but also reverse it as it is occurring [320].

This technique unravels the master equation evolution into a set of individual trajectories where the system undergoes time evolution and experiences a set of jumps at random times and locations that are sampled according to well-defined probability rules. The evolution of the master equation is then recover from a set of pure state evolutions—numerically requiring $dim(\mathscr{H})$ instead of $dim(\mathscr{H}) \times dim(\mathscr{H})$ at the cost of the sampling over trajectories. Despite its origin in quantum optics, this technique can be applied to any Markovian open system, in particular, it is widely uses in cold atomic systems. For a detailed review on the method and extensive example applications, see [36, 309]. The essential point is that this formalism, and the ones presented below, can be extended to situations in which there is no measurement apparatus but rather the measurement is performed by certain environmental degrees of freedom that remain not monitored, not requiring then experimental probing of the system.

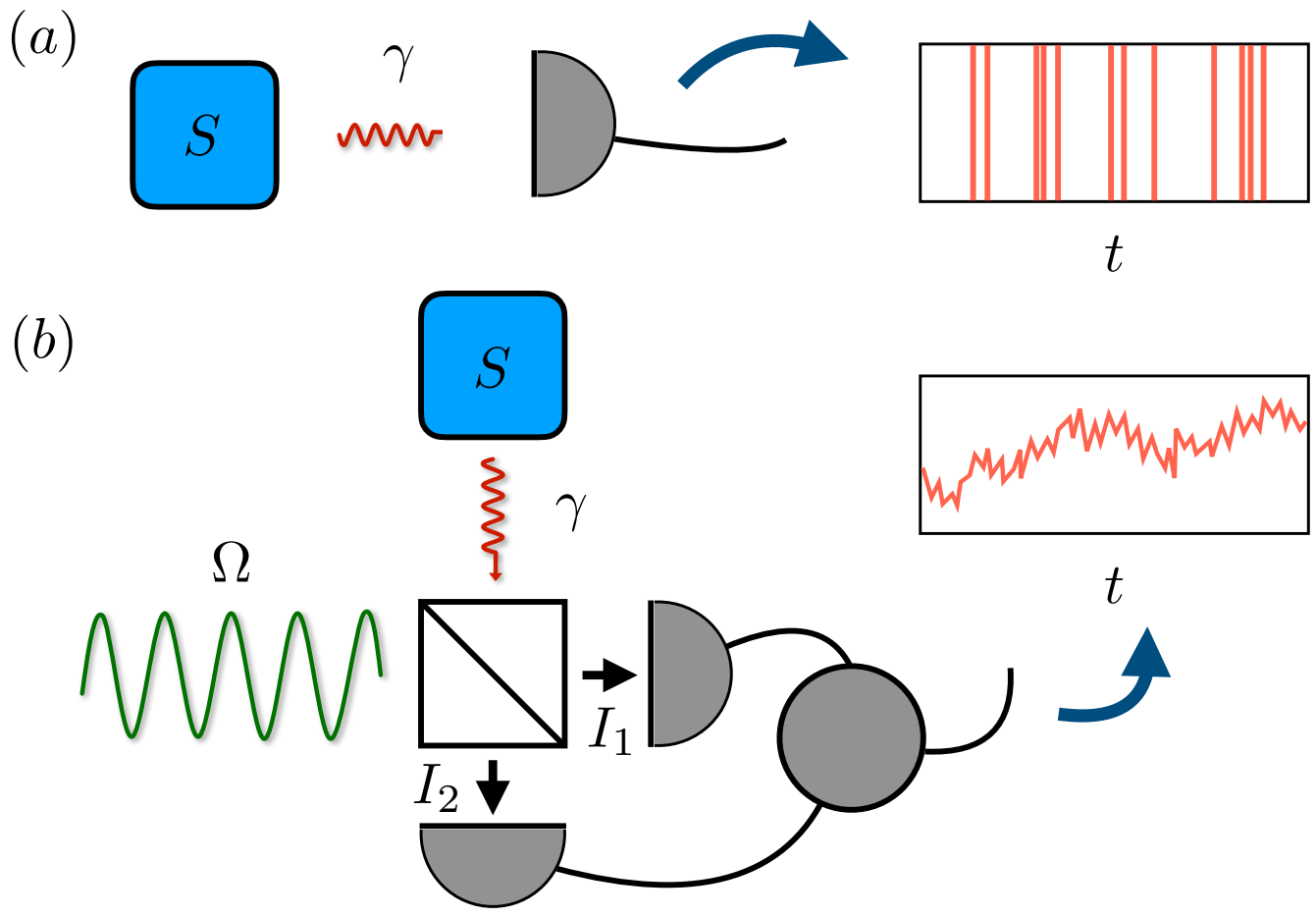

Figure 3: Diagrams for the stochastic unraveling of an open system: (**a**) A physical system $S$ emits photons into the environment with a characteristic rate $\gamma$; when these are recorded via our photodetector, we gain the information that the system has undergone a quantum jump; (**b**) The output photons of our system $S$ is instead coupled to a strong oscillatory signal $\Omega$ through a homodyne detection scheme of intensities $I_1, I_2$ at the output of a beam splitter. The system is then weakly and continuously monitored leading to a noisy diffusive behavior instead of projective collapse.

**Quantum state diffusion** The second physical scenario that we consider is the one of continuously and weakly probing the system by means of homodyne detection, i.e., by coupling the output from the physical system to a strong oscillator field [304, 306, 321], see Figure 3b. In this case, the photons incoming from the system are not directly measured and, as a result, the system will only be weakly perturbed and undergo a diffusive evolution that can be model in terms of a noise term. These diffusion equations, see e.g. [322, 323], both in terms of density operators and pure states, allowed for a similar treatment to classical control theory of these systems leading into not only a framework where it is possible to describe the measurement process but also quantum control and feedback [310, 324, 325]. This diffusive evolution behaviour has also been observed experimentally [326, 327], and its effect on the system as characterised theoretically [328, 329] has been modelled and mitigated in a variety of scenarios (see here the first experimental example [330]), or completely reverted [331].

The quantum state diffusion allows then to describe a system under weak continuous monitoring whose master equation—to be seen as the statistical average of all possible measurement outcomes at all possible times – is reduced to a set of noisy random realisations; an approach that is complementary to the quantum jumps, where the advantage to use either of them depends typically on the characteristics of the model being studied. In addition, there is another relevant specific subclass of bath description in the weak measurement limit, that we discuss in the following section.

**Collisional models** The unraveling techniques presented above are not the only possible ways in which the exponentially large number of degrees of freedom of the environment can be treated or discretized. In the following, we consider a specific non-perturbative method that has gained interest in recent years, partially due to its applicability in discrete physical systems such as gate-based QC [332, 333] or even more generally in applications outside physics as described in Section VII. This is the case of quantum collisional models (CMs) [334, 335].

Quantum collisional models were first introduced by [336], and provide a bath description that is both discrete in time and space. More specifically, in this framework the system is coupled to a small subset of bath degrees of freedom at a time, turning the complex bath-system coupling Hamiltonian into a reduced *quasi-pairwise* discretization of their interaction that allows for a more efficient description of the coupling. It is immediate to understand how this approach could be beneficial in certain scenarios describing spatially structured systems, e.g., lattices, linear traps or quantum circuits where the system itself possess a certain space structure. While the approximation of considering that the system interacts with a reduced portion of the bath at a time instead of all the normal modes at once, seems rather drastic, applications have been found in a wide range of systems. While referring to the relevant reviews [334, 335] for a detailed list of applications, significant examples are e.g., in quantum thermalisation [333, 337], quantum transport [338, 339] or in state preparation [340] among others.

The last relevant aspect of CMs that we would like to discuss in this review is that the time discrete nature of the interaction allows for a natural framework to incorporate structured baths beyond the Markovian approximation and also beyond the weak-coupling regime. In this sense, CMs constituted an ideal framework for a non-perturbative modelisation of non-Markovian systems, see [341, 342].

In the following, we will discuss some general aspects of the existing approaches to describe with OQS beyond the standard Markovian limit.

*Beyond the Born-Markov Approximation* As it is not the main topic of this work, in the following we will highlight a set of relevant scenarios for the platforms and frameworks that we include in the review, while for detailed studies we refer to [343–346].

While both the Born and the Markov approximations are applicable in a wide range of systems—particularly in the context of cold atomic systems, due to the typical timescale separation as we have discussed—there are certain scenarios in which these are no longer reasonable assumptions [347].

For example, it is easy to imagine scenarios in which the system is no longer weakly coupled to the environment, e.g., if we introduce the system into a cavity it will strongly couple to the radiation field. Even if this constitutes a challenge for the description, the cavity modes can mediate long-range interactions with complex space dependence [35, 115, 348] that we can use for programmable quantum simulation and computing, such as modelling information scrambling, that is a fundamental problem in e.g., cosmology and quantum information [349, 350]. We hope this motivates the reader to understand how as the degree of control develops, it is important to consider descriptions beyond the standard framework to model new possible applications.

In addition to the relaxation of the weak-coupling condition, there are other relevant scenarios where we cannot assume that the Markov approximation is preserved, as the bath does not remain unchanged by the coupling to the system. This can be either because their size is comparable or because its correlations do not decay fast in comparison with the system timescales. This is a common problem in solid-state systems, but also present in cold atoms under certain conditions, e.g., by considering coupling to untrapped states from the optical trap [351, 352] or by the presence of a spin or fermionic bath [353].

Furthermore, in our derivation of the GSKL master equation, we required a third approximation namely the rotating-wave or secular approximation [302] that guaranteed that our dynamical map was preserving the physical properties of our state as time evolves. This can also be avoided by deriving the Bloch-Redfield master equation [354], a perturbative master equation that can represent non-trivial bath spectra and that can be solved numerically for moderate system sizes.

In general, there has been substantial work in solving the general problem of OQSs in the absence of these approximations: through perturbative master equation approaches [355, 356], through the use of stochastic methods such as quantum jumps [357] or quantum state diffusion approaches [358], through collisional models [334, 342], or through a hierarchy of pure states approaches [359]. We refer again to the relevant reviews for a detailed list.

Before we move on, there are two relevant aspects of non-Markovian systems that should be considered. The first is that biological systems, object of Section VII, are often subject to non-Markovian dynamics [360] as it has been reported in experiments on chemical compounds involved in photosynthesis, where reltively long-coherence times and bath relaxation rates comparable to the system dynamical timescales have been observed [361]. Thus, it is important to consider a general, non-necessarily Markovian description for biological systems as OQSs. For a detailed discussion see [345]. Finally, there is a relevant aspect on non-Markovianity related to fundamental questions on quantum information. There has been extensive work in defining measures of non-Markovianity in terms of some quantification of the information exchange between system and environment, e.g., via Quantum Fisher information flow [362] or distinguishability measures [363].

All in all, in this Section II B 1 we have discussed a set of theoretical tools that allow for the description of driven-dissipative quantum systems in a feasible manner, discussing scenarios in which the different approximations suffice, and also means to extend them. However, in any of these frameworks we often require the manipulation of large matrices due to the exponential growth of the Hilbert space with the system size. As a result, many of these techniques cannot be applied directly to more than 10–20 sites/particles. Motivated by this limitation, in the following section we discuss one of the leading approximate methods to tackle the simulation of large quantum systems, that is tensor networks, with special focus on their 1D version, denoted as matrix product states.

*2. Quantum System Description via Tensor Networks*

As we have just discussed, the major bottleneck in the theoretical and numerical description of current quantum technology platforms is the quickly increasing dimension of the Hilbert space. This requires finding ways in which we can manipulate and perform operations with large objects, but this method quickly becomes unfeasible due to the exponential growth. As an example, let us consider a set of $N$ 2-level atoms (two hyperfine states of a neutral atomic species), here the configuration space is given by $2^N$. If we consider the computational storage of the quantum state associated to our system this task becomes quickly impossible. In particular, with $N = 4$ we are only required to store $2^4 = 16$ complex coefficients to define our quantum state. However, as we increase the size, e.g., $N = 300$ atoms, then our space would be have a dimension of $2^{300}$ and the associated Hamiltonian matrix would have a size of approximately $10^{82}$ B if using double precision, a number larger than current estimates of the number of barions in the observable universe ($\sim 10^{80}$). This example should quickly make the reader aware of how exact methods, even exploiting specific properties of the system such as certain conserved quantities or constraints, quickly become untreatable in current—and near-future—technologies.

Consequently, we are required to consider approximate methods to describe quantum technology platforms. In this section, we briefly introduce the framework of tensor networks an approximate method that has been been developed in the last 30 years for the computation of both equilibrium and dynamical properties of quantum systems. The origin of tensor networks is connected to the development of the density matrix renormalization group (DMRG) [364, 365] that was designed for the iterative computation of equilibrium properties in one-dimensional quantum systems. A few years later, it was shown that the ground states of the systems addressed by DMRG could be efficiently represented [366, 367] by means of matrix product states (MPS), which are the essential components of tensor networks in 1D. The ability to map the DMRG problems into the language of MPS enabled the rapid development of the field with techniques that could describe also time-dependent processes (t-DMRG) [368–370].

The basic principle behind matrix product states is the fact that, as observed in nature, low-dimensional and low-energy quantum systems present a moderate amount of entanglement. This is well-characterised in literature by the existence of bounds and so-called area laws [371] limiting the bipartite entanglement in physical systems to the size of the boundary between subsystems in contrast to the entanglement growing with the system's volume [372] as it would for a random state in the Hilbert space. Thus, MPS constitute a low-entanglement *ansatz* representation of the states in our quantum system providing an efficient truncation of the Hilbert space to the relevant physical region, providing accurate representations in one-dimensional gapped systems. It is important to mention that this description was also extended to the representation of operators via so-called matrix product operators (MPOs) [373–375], allowing to deal with non-local and long-range systems. Furthermore, extensions to larger dimensions have already been developed [376] while optimizations in these cases are rather system-dependent, and moderate success has been found in modelling quantum many-body systems in 2D and 3D.

In this section we will highlight some of the relevant tools developed in the field of MPS that are applicable to the topics discussed in this review, focusing on the one-dimensional case but also highlighting the link with OQS techniques, particularly with the stochastic unraveling methods discussed in Section II B 1, and potential links to machine learning and QC. For a detailed description of the formalism we direct the reader to some of the excellent extensive reviews in literature [377–379].

*Matrix Product States (MPS) and Operators (MPO)* The simplest approach to understand how to represent a quantum state via MPS is by considering two decompositions. The first one, related to quantum information, is the Schmidt decomposition [380] that represents a pure state in terms of two orthonormal bases of two partitions $A$ and $B$ $\{|i\rangle_A\}, \{|j\rangle_B\}$:

$$|\phi\rangle = \sum_{i,j} \lambda_{ab} |i\rangle_A |j\rangle_B \,. \tag{3}$$

The second one is the singular value decomposition (SVD), that allows to express any generic matrix $M$ with $dim(M) = d_1 \times d_2$ in a product of matrices of the following form:

$$M = USV^\dagger, \tag{4}$$

with $U$ composed of orthonormal columns satisfying $U^\dagger U = I$ and $dim(U) = d_1 \times \min(d_1, d_2)$; $S$ a diagonal matrix of $dim(S) = \min(d_1, d_2) \times \min(d_1, d_2)$ formed by the singular values $\lambda_1 < \lambda_2 < ... < \lambda_r$, and we denote as Schmidt rank $r$ the number of non-zero entries of $S$ and, $V^\dagger$ composed of orthonormal rows as $VV^\dagger = I$ and $dim(V) = \min(d_1, d_2) \times d_2$.

Ee then consider a generic quantum state representing a discrete system of $M$ quantum systems with individual

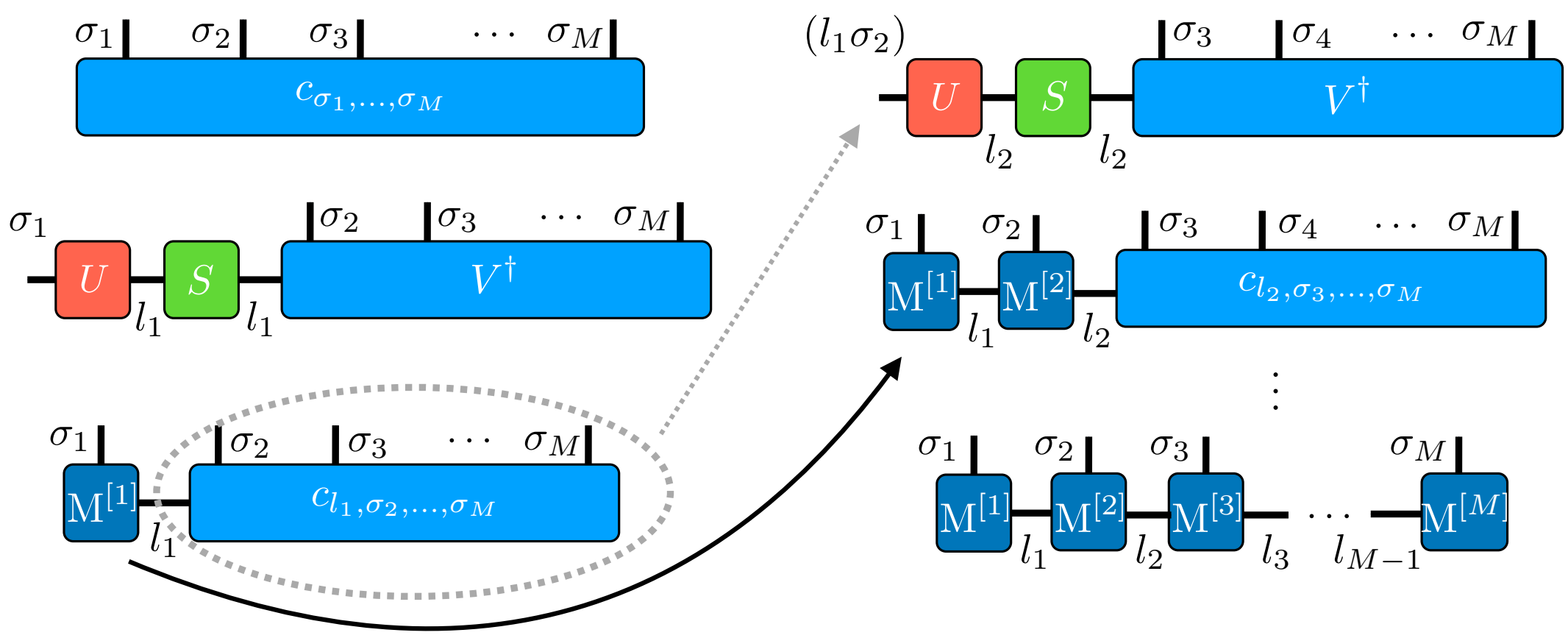

Figure 4: Construction of an MPS representation of a generic state: A quantum state representing a $M$-unit system, e.g., $M$ particles or lattice sites, requires a set of complex coefficients given by the product of the local dimensions $\sigma_i$, with $\dim(\sigma_i) = d$, generating a $M$-legged tensor of dimension $d^M$. By performing iterative local SVDs for each dimension $\sigma_i$ we can write the state as a set of product of local matrices $\mathrm{M}^{[i]}$ for each combination of $\{\sigma_1, ..., \sigma_M\}$. These matrices can be compressed in a controlled way, according to their bipartite entanglement with the rest of the system. Figure taken from [381].

local basis $|\sigma_i\rangle$ of dimension $d$, written as:

$$|\phi\rangle = \sum_{\sigma_1,\ldots,\sigma_M} c_{\sigma_1,\ldots,\sigma_M} |\sigma_1, ..., \sigma_M\rangle \,, \tag{5}$$

where we require a set of $d^M$ complex values $c_{\sigma_1,\ldots,\sigma_M}$ to represent our state. By iteratively performing SVD to the local Hilbert spaces, as described schematically in Figure 4, we can transform the $M$-indexed coefficients into a product of local matrices of the form:

$$|\phi\rangle = \sum_{\sigma_1,\ldots,\sigma_M} \sum_{l_1}^{r_1} \sum_{l_2}^{r_2} \ldots \sum_{l_M}^{r_M} M^{\sigma_1}_{l_1} M^{\sigma_2}_{l_1,l_2} \ldots M^{\sigma_{M-1}}_{l_{M-2},l_{M-1}} M^{\sigma_M}_{l_{M-1},l_M} \,. \tag{6}$$

Crucially, this transformation can be done without loosing any information if we keep the maximum matrix dimensions. However, the key aspect of MPS compression is that every time we perform SVD, the Schmidt rank associated to the matrix is given by the amount of entanglement between the two bipartitions connected by the corresponding link. In many relevant physical cases this Schmidt rank is rather small, trivially being $r = 1$ for the case of a product state, allowing for the reduction of the matrix dimension. In fact, the best approximation to each local matrix with a maximum dimension $D_{max}$ is that associated to the largest $D_{max}$ singular values and singular vectors.

Thus, we can obtain the matrix product form and compress the state in a controlled way, as in every iteration we can monitor the amount of error in the representation as the sum of the singular values not being stored $\epsilon_i \equiv \sum_{l_i=D_{max}+1}^{r_i} (\lambda_i)^2$. Moreover, we can relate this compression with the idea of MPS as a low-entanglement *ansatz* if we consider the definition of Von Neumann entropy [380], for a given bipartition with singular values $\{\lambda_i\}$:

$$S_{vN} = -\sum_{l_i}^{r_i} (\lambda_i)^2 \ln (\lambda_i)^2 \;\rightarrow\; \max (S_{vN}) \leq \ln (D_{max}) \,. \tag{7}$$

As we mentioned before, this matrix product structure can be generalised for operators in the form of MPOs [373, 374], where the local tensors now will have an additional index $\sigma_i'$. These formalisms allow for the straightforward product of states and operators and the computation of overlaps and expectation values, see e.g., [377].

*Algorithms with MPSs and MPOs: Equilibrium, Dynamics and Open Systems* The first applications of MPS methods, where related as we mentioned to the calculation of equilibrium properties. Particularly for the calculation of ground states, these revolve around the idea of adapting DMRG protocols [364, 365] to MPS language through local updates of tensor network to iteratively minimize the energy variationally. The development of MPOs allowed to also successfully apply to long-range interacting systems [375]. Moreover, further advances have rendered the method well-suited for the calculation of not only ground states [377] and low-excited states but even thermal states [373] in dissipative settings. In addition, MPS methods can also be used to understand certain spectral properties and the structure of excitations in the lower part of the energy spectrum [382, 383]. Other protocols have developed frameworks to compute highly-excited states [384] in conditions of low entanglement or have also extended the formalism to continuous systems [385]. This wide range of applicability in equilibrium phenomena together with the ability to approximate large systems made MPS a suitable approach to study a wide range of one dimensional critical phenomena where the failure of the low-entanglement *ansatz* could characterise criticality [386, 387].

In parallel to this development, MPS has also been widely used in the study of dynamics. Originally, based on local tensor updates similar to Suzuki-Trotter decompositions, this is the case of time evolution block decimation (TEBD) [368–370]. Later on, variational tangent-space methods where also successfully employed in computing time evolution in the MPS/MPO language through the time-dependent variational principle (TDVP) [388, 389].

Quite crucially, as any protocol based on MPS when a tensor update is performed, either for DMRG or for time evolution, the error in the re-compression into MPS form can be monitored and keep to a suitable value. This error is typically the limiting factor in the maximum system sizes and simulation times that can be achieved.

Natural extensions to the study of open quantum systems where also developed in the context of tensor networks. Particularly, early examples of the use of MPOs where dedicated to the representation of density operators [373]. In general, most protocols that we have discussed so far are readily applicable to vectorized density matrices, where the local Hilbert space of each individual tensor is doubled and adapted via a Choi's isomorphism [390]. Importantly, all the stochastic methods described in Section II B 1, are also compatible with time evolution algorithms through MPS since they are typically based on the use of pure state evolution. In the context of Markovian open systems, extensions to the DMRG protocol for the computation of steady states were developed [390]. More recently, extensions to the study of dynamics in non-Markovian systems were developed [391, 392]. For further details we refer to this topical review [393].

*Beyond One-Dimension* While the applications of matrix product states are quite general and new examples are constantly developed, it is important to consider that MPS constitutes a subclass of the bigger tensor network family. This is particularly relevant when we consider the applications of tensor networks in contexts beyond quantum simulation and its links with QC, as we discuss in the following section.

The first generalisation of MPS to two-dimensional systems is the case of projected entangled pair states (PEPS) [376]. In parallel to these, tree-tensor networks (TTN) and, particularly, multiscale entanglement renormalization ansatz or MERAs was developed [394], these methods were also suitable for the study of multidimensional systems. These more general structures of tensor networks are an active area of research with new protocols being currently developed, e.g., for thermal states in 2D [395]. One of the most relevant differences with the case of MPS is that the optimization of the computational routines are often heavily system-dependent and, as a result, algorithms need to be tailored to the problem at hand. Despite this, there are some general strategies for the efficient development of tensor networks, with new algorithms being developed [396, 397]. We refer to this recent review for further details [398].

*Tensor Networks Beyond the Simulation of Quantum Matter* One of the first field of applications of tensor networks (TN) outside the description of quantum many-body systems was the simulation of lattice gauge theories. This link was originally established for quantum simulators [54, 399] more generally, but was quickly rephrased into TN language [400–402]. Since then, TN have been widely applied in a large amount of fields: from the description of holographic theories in cosmology [403] to their links with conformal field theories in critical systems [404, 405], to the study of dynamically constraint models [406] or to the understanding of quantum thermalisation [407].

In this section, we would like to focus our attention on a set of applications of TN that are specially pertinent for this review and will be the focus on Section II C. That is their utility in the development of current QC algorithms, partially due to their synergy with machine learning approaches.

Perhaps the simplest way to understand the role of TN in QC is the idea that TN offer a natural framework for information compression, in the same way that they provided a compressed state description through a controllable *ansatz*. Then, one could apply TN to the efficient simulation of QC circuits [408]. This has allowing to devise protocols for efficient tomography [409], for error mitigation [410] or for data classification [411]. Furthermore, the TN framework can also accelerate existing variational quantum computing approaches [412] and provide compact descriptions of random circuit networks [413].

In parallel to these developments, the use of TN in QC has been enhanced by its connection with the field of machine learning. While initially machine learning approaches were applied to the study of quantum matter as an alternative approach, solving the many-body problem [414, 415] or helping in the characterization of phases of matter [416]. It was quickly understood that the underlying similar mathematical framework between both disciplines would be able to create a synergistic development of both fields.

Since then, the TN formalism has been applied to the network structures native to ML, formally understanding their equivalence [417], allowing for the compression and simplification of ML algorithms by exploiting the TN structure [418], characterising their expressivity [419] or being able to adapt tensor network algorithms such as DMRG to their language [420]; also intertwining with Bayesian based-protocols [421] or generative modeling [422]. In the same way, ML approaches can be used to accelerate operations in TN [423].

All in all, the use of TN has enabled to access larger systems, model experimental setups where exact methods were no longer feasible and access new physical regimes in the context of quantum simulation, from the benchmarking of the devices [424], to mapping interesting problems, e.g., in quantum chemistry [425] or in non-linear systems [426]. These long-standing benefits of the use of TN in quantum simulation, have only recently transferred to the QC framework [3, 427, 428] and its likely that many of the potential use cases and optimizations are yet to be revealed.

In this section, we have summarised some of the most relevant aspects for the description of open quantum systems in numerically treatable approaches. We have also discussed in detail one of the leading techniques for the simulation of quantum many-body and quantum computing given by tensor networks. Finally, we have connected these to more general ideas of complex network theory and linked it with biological and more generally applied physics systems. However, it is important to highlight that other widely used methods for the simulation of quantum systems exist but will not be covered in this review. In particular, we highlight the following methods that were applied to certain studies in the following sections and direct the reader to the extended reviews included: quantum Monte Carlo approaches [429], Keldysh formalism [430], transfer matrix methods [431], applied in quantum optics not to be confused with the formulation of path integrals in TN language [432], and truncated Wigner approach [433, 434].

### C. Quantum Computing

Existing quantum technologies bear a long-term promise in the implementation in quantum computers, with all their potential applications. Chemistry [435], pharmaceutical [436], finance [437], and machine learning [438] are some of the many example fields.
From a purely theoretical point of view, quantum computing algorithms that have been already demonstrated to solve certain types of problems faster than their existing classical counterparts [439]. In the last 40 years, many quantum algorithms [440] that confirm this advantage have been designed.

These theoretical proposals gave impetus to an intense experimental activity aimed at realising a quantum computer. As we showed in the previous section, there is a broad range of completely different quantum technologies that can be employed as quantum computing elements like qubits, or quantum gates. Despite different hardware realizations, all experimental proposals for a quantum computer must satisfy the so-called five Di Vincenzo criteria [441]: a set of essential conditions that serve as a benchmark for assessing the feasibility and practicality of quantum computing systems. The criteria encompass five fundamental requirements:

- *Architecture scalability* . The qubit is a well-defined physical system that can be isolated from the external environment. Moreover, increasing the number of qubits does not modify the device's functioning principles.
- *Initialization ability*. It is possible to initialise qubits in a custom desired state.
- *Coherence*. Qubits have long coherence times.
- *Universality*. It is possible to perform quantum gate operations that form a universal set of gates. In this way, every quantum operation can be realized as a composition of gates from the universal set.
- *Addressability*. The device is capable of addressing and selecting the qubit that is intended to be measured, isolating it from the remaining ones and performing the measurement without information leaking or cross-talks.

A device that satisfies all five Di Vincenzo criteria is a quantum computer that is capable of running quantum algorithms of any sort, without any limitation in the instance size, of the computation time. On the contrary the current stage in the technological development of quantum computers is often referred to in literature as NISQ-era (Noisy Intermediate Scale Quantum Era) [442], a name that is explanatory of the characteristics of the available technology today and what we predict in the near future. In fact, such devices are affected by several types of noise and errors, that come from imperfect experimental realizations or limited sampling.

Before providing an overview of theoretical strategies to deal with noise and errors, we want to comment briefly on the first criteria since it often plays a pivotal role in many quantum computing applications, like chemistry, pharmaceutical, and finance [435, 436, 438], while we refer to review work [66, 67, 443] for a complete dive in quantum computing, quantum communication and quantum internet.

Architecture scalability, the ability to scale up the number of qubits and interconnect different quantum platforms, often necessitates the interconnection of many Quantum Processor Units (QPUs), that may also differ in terms of experimental realization [444].

While the QPUs communication open up the possibility of constructing a quantum network, and so realizing the so-called Quantum Internet [445], it is also a significant challenge, requiring among other resources the realisation of a quantum memory. A quantum memory is used for storing and processing qubits for photon transmission with high fidelity, enabling synchronization within quantum computers, extending the range of quantum communication via repeaters, and distributing entanglement between distant nodes [445]. Various implementations, including cold neutral atoms and doped crystals, support storage and retrieval functionalities, enabling efficient entanglement transfer. However, achieving high fidelity transfer remains a significant challenge for network scalability despite advancements in polarization qubits [445].

Although we refer to the recent review work [443] for a deep dive in quantum computing topics, in Section II A we summarised the different experimental platforms, indicating relevant differences, and main advantages and challenges. Though not at all specific to atom technologies, for completeness in this section we shift the focus onto the theoretical progress that has been made in quantum computing regardless of the specific hardware realisation, to deal with errors and noise. In the course of the years, this has come about with two different strategies: error-correction codes and a hybrid quantum-optimization approach. Finally, we introduce in Section II C 3 the concept of quantum control, which is essential in both strategies above.

#### 1. *Fault-Tolerant, Error-Corrected Quantum Computation*

The practical realization of a quantum computer is subject to the presence of imperfections, errors and noise. Thus, it is natural to develop protocols that reduce or completely mitigate their impact. This has led to the development of error-correcting codes. An initial strategy that deals with the presence of noise is encoding quantum information redundantly using additional redundant qubits, as discovered in [446]. In the extra-qubits strategy-based algorithms, the quantum information is spread over multiple physical qubits to generate a logical qubit [446]. Most transformative algorithms, such as [439], require error-corrected qubits. As a result, many protocols, including stabilizer and topological codes, have been developed to protect logical qubits [447]. Nevertheless, the process itself of detecting and correcting errors is also prone to noise, which means that error correction alone cannot guarantee the long-term storage or processing of quantum information and alternative approaches are required.

Fault-tolerant quantum computing is a theoretical framework for building quantum computers that can continue to operate correctly even in the presence of errors caused by noisy quantum hardware or imperfect control operations. The idea is to design quantum algorithms and hardware in such a way that errors can be detected and corrected on the fly, without disrupting the computation. The Quantum Fault-Tolerant threshold theorem allows for the execution of large quantum computations by suppressing the quantum error rate below a certain threshold. Proof of this result can be found in [448–450]. The challenge of lowering noise levels remains significant, but some progress has been made in both algorithmic and hardware development. Hardware improvement on both improvements in the quality of the qubit properties and also in providing hardware components and architectures that are less prone to errors. Results regarding hardware improvements of solid-state platforms such as electron spin qubits in silicon [451], and spin qubits in diamond [452] have been presented; as well as optical platforms' ones [453]. In the quantum error correction algorithmic literature, some authors [454, 455] connect different quantum error correction algorithms and protocols with the achievement fault-tolerance computation.

Moreover, there exist additional strategies towards fault-tolerant quantum computing. Here we introduce two of them. The first one is making use of topological qubits to overcome the current capability of traditional qubits to resist noise and errors [456], while the second one is based on using quantum annealing [457], which utilizes a different type of hardware and algorithms than gate-based quantum computing. This approach connects with the quantum-simulation related ideas discussed in II B 1.

Although fault-tolerance operations are starting to become achievable only in very recent years [458], we are now entering the transition phase towards a fault-tolerant quantum computing era, following on the initial steps have been implemented for different hardware platforms, including solid state devices, such as electron spin qubits in silicon [451], spin qubits in diamond [452] and optical devices [453]. Moreover, as described in Section II A 1, when commenting on nanomagnet technologies, fault-tolerant schemes for fault-tolerant quantum information have been also proposed exploiting qudits, as those synthesizable on magnetic nanomagnets [92–94].

#### 2. *Quantum Optimization*

While fault-tolerant architectures are one of the main routes to design QC hardware that can cope with imperfect implementations, there are other approaches to reduce the noise impact, like the creation of on-purpose designed quantum algorithms.

One of the most common ways to achieve this goal is to turn the design into a quantum optimization algorithm. In fact, optimization problems are ubiquitous in both classical and quantum frameworks, including machine learn-ing [438], chemistry [435, 436], finance [437], and more. All these different scenarios rely on the same basic optimization structure: finding the best from a set of potential solutions, subject to a set of constraints. Optimisation aims to find a solution that maximizes or minimizes a particular criterion, such as profit, cost, or energy consumption. Thus, the essential task is focused in mapping the problem into the optimal cost functions given the system's constraints, transforming the problem into finding the extreme of a given cost function.

In the quantum context, the optimization problem is encoded in the quantum computer itself [459]. The most prominent example is the Variational Quantum Eigensolver (VQE) [460], which we describe in more detail in the section below. Even if we present the basic scheme of VQE as a textbook model of variational hybrid optimization algorithms, many results [461, 462] show how to improve VQE and hybrid variational algorithms performances for specific applications sourcing from more advanced techniques. In the VQE problem, similar to *Hamiltonian engineering* as in Section II B 1, the objective is to find the lowest energy state of a given quantum system. This can be used to find the ground state properties of a given quantum model, and also to map a relevant, otherwise hard to solve, problem into the quantum one. VQE is part of the family of hybrid quantum-classical variational optimization algorithms [463]: the hybrid core idea is to use a quantum computer only to tackle specific tasks of an algorithm, leaving other tasks to a classical computer. In this way, the quantum computer would be used to exploit a (possible) quantum advantage while a classical computer can manage tasks like storing data, arithmetic, and in general, any processing operation that is simpler to implement in classical hardware, for which typically classical efficient algorithms exists. Section II C 2 is dedicated to quantum variational algorithms, while referring to [464] for a complete review on this topic.

For the sake of completeness, we want to mention that it is also possible to tackle the optimization process only using a quantum device, in what is known as a quantum annealer [465], a specific type of analogue quantum computer. In fact, the computational model of quantum annealers is completely different from other QC algorithms presented in this review. Thus, while referring to more recent reviews such as e.g., [466], for a complete theoretical and experimental perspective, we here only recall quantum annealers are devices that implement the adiabatic driving from a well-known ground state of an initial Hamiltonian $H_0$ to the unknown ground state of the problem Hamiltonian $H_p$. In so doing, the solution of the given optimization problem is encoded, based on the adiabatic theorem of quantum mechanics.

*Quantum Variational Algorithms* Quantum variational algorithms, originally proposed in [460, 463] are a family of hybrid quantum-classical algorithms that use classical optimization techniques to find the ground state of a given Hamiltonian. In the NISQ era, quantum variational algorithms are especially important because they can be implemented on the available quantum hardware [464], which is currently limited in terms of the number of qubits, the circuit depths (i.e. the longest path in a quantum circuit, in terms of the number of executed gates), the presence of noise or the imperfections in gates that can be implemented. Thus, shifting the focus to a resource-based approach where the optimization can be performed considering the hardware-constraints on top of those of physical origin. This way the quantum circuit can be parametrized in terms of the set of available quantum operations. Every circuit then can be described in terms of a set of variational parameters, describing the gates applied, their position and times or their angles when applicable.

As we introduced in the previous Section II C 2, a prominent example of quantum variational algorithms in the NISQ era is the Variational Quantum Eigensolver (VQE), which was first proposed by [460]. The VQE algorithm consists of three main components: a parametric quantum circuit that prepares a trial wavefunction, a measurement of the energy of the trial wavefunction, and a classical optimization loop that updates the parameters of the trial wavefunction to minimize the energy. The typical hybrid algorithm scheme is represented in Figure 5. The quantum computer executes the parametric state preparation that is then measured, while a classical feedback loop extracts the outcome from the quantum measurements, manipulates and stores the information, performs statistical processing, computes the relevant quantities and feeds back the quantum computer with a new set of variational parameters, found using the classical optimization subroutine.

Figure 5 can also be viewed as an outline of the basic scheme for a computational optimization algorithm, in fact, other quantum variational algorithms that fit this scheme and are commonly used nowadays. The latter include the Quantum Approximate Optimization Algorithm (QAOA) [467], which is used for combinatorial optimization problems, and the Quantum K-Means Algorithm (QK-means) [468], used for clustering. These quantum variational algorithms, including the VQE, are well-suited for the NISQ era for several reasons, including their ability to cope with noise and other imperfections in current quantum hardware due to their parametric nature [469–471].

As we notice above, Parametric Quantum Circuits (PQCs) are typically [440] describe the state of the circuit in terms of a collection of parameters $\{\vartheta\}$ used to prepare the trial wavefunctions. If we consider starting from an initial reference state $|\psi_0\rangle$, while the solution to the problem is encoded in the target state

$$|\psi(\theta^*)\rangle = U(\theta^*)|\psi_0\rangle, \tag{8}$$

characterised with the optimal parameters $\theta^*$. Even at this general level, we can firstly see that the *expressibility* of the PQC ansatz, encoded within $U(\vartheta)$, is important to guarantee that the target state is actually reachable using a particular variational ansatz, meaning it exists a set of parameters $\vartheta$ for which $|\psi(\vartheta)\rangle = |\psi^*\rangle$. On the other hand, also *trainability* is important, where this term quantifies how easy it is to actually find optimal parameters $\vartheta$ and, in so doing, identifying the target state, via information extraction from the final quantum state.

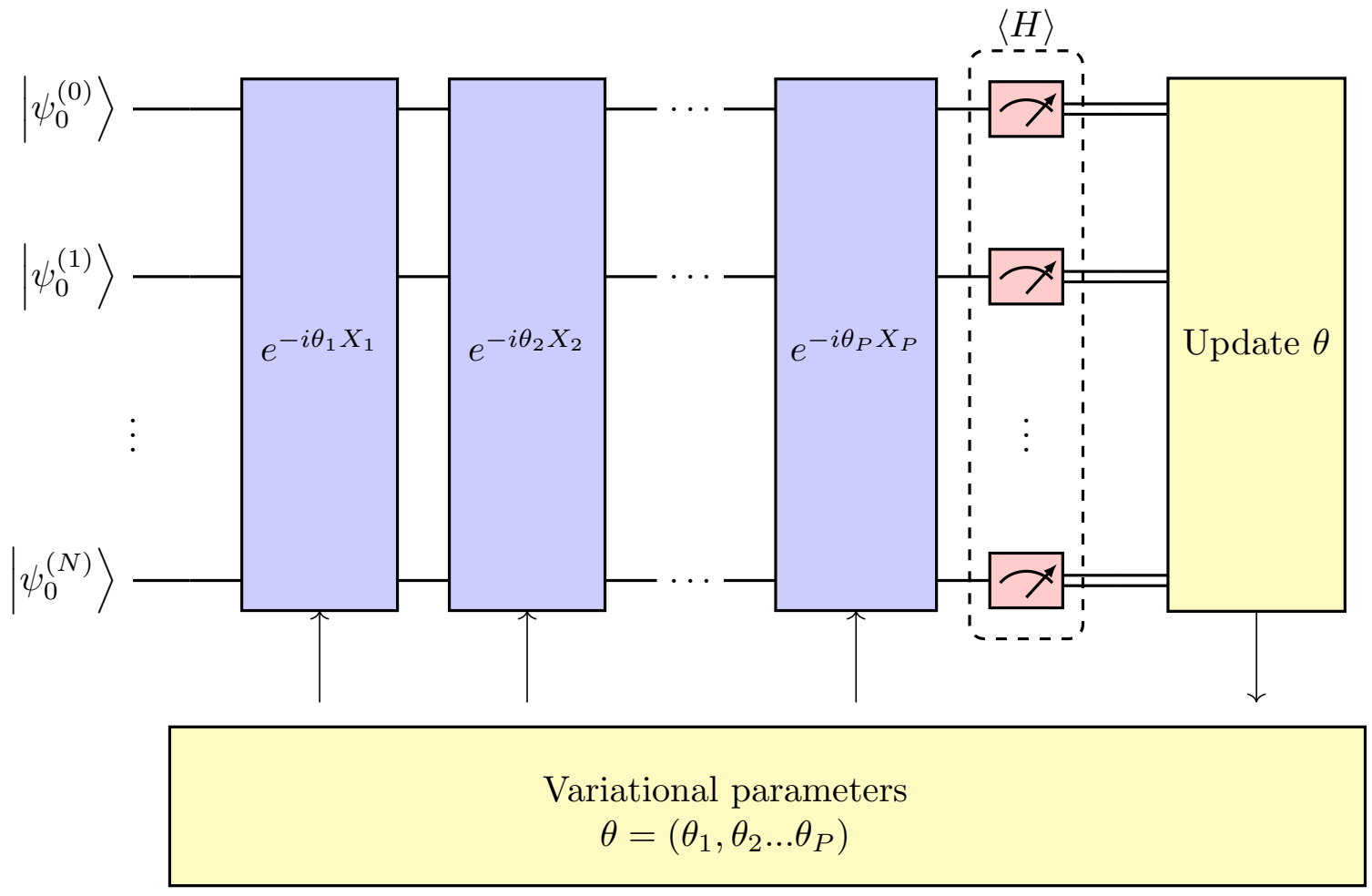

Figure 5: Basic scheme of variational hybrid quantum algorithms. A $N$-qubit quantum register is initialized in the variational state $|\psi_0^{(0)}, \psi_0^{(1)}, ...\psi_0^{(N)}\rangle$ at the step $i = 0$. The state evolves under the application of several parametric quantum gates (purple boxes) until a quantum measurement is performed (red boxes). The mean value of the relevant observables, e.g., the Hamiltonian of the quantum system, is then computed using a classical processor (yellow boxes) that feeds back to the quantum processor a set of new variational parameters, and the entire procedure restarts. New parameters should be predicted using a classical iterative minimization sub-routine so that when convergence is reached at $i = I$, the solution to the original problem is found.

In most cases there is a trade-off between expressibility and trainability of a quantum circuit [472] due to both experimental and theoretical limitations. Experimentally, we have already discussed the presence of noise and errors in quantum circuits, while theoretically there is a large open challenge in hybrid variational computing, the appearance Barren plateaus. Barren plateaus are large regions in the parameter landscape, in which the cost function has vanishing gradients, which prevents efficient optimization. It was recently shown [473] that the expectation value of the gradient of the cost function corresponding to randomly initialized parametric quantum circuits vanishes exponentially with the number of qubits. In [474], a solution based on the connection between barren plateaus and the locality of the cost function is proposed. In particular, it has been proven that if the cost function is local, in the sense that the associated quantum observable contains only 1-qubit operators, the effect of Barren plateaus is not so severe. In such cases, the Barren plateau exists, but the variance of the cost function only decays polynomially with the number of qubits, if the quantum circuit length scales at most as $\log(N)$, with $N$ number of qubits. Other strategies are proposed to avoid Barren plateaus, such as classical shadows [475], classical optimization strategies [476, 477], or measuring low-depth analytic gradients [478].

### *3. Quantum Control*

Quantum control theory development stems from the implicit goal of controlling quantum phenomena [479]. Controlling, manipulating, and steering quantum systems towards desired states and objectives is both the ultimate need of all quantum technologies and the final goal of quantum control as a theory; so the rapid evolution of the first led to a similar improvement in the second over the last three decades [480]. A classical control theory exists as a field of engineering and applied mathematics. Quantum control, instead, differentiates due to the unique characteristics of microscopic quantum systems, such as entanglement and coherence, or quantum measurement, which have no classical counterpart. Such effects require establishing a new theoretical foundation and systematic methods for manipulating and controlling quantum systems , in what was denoted as quantum control theory.

This field has already achieved significant successes in physical chemistry [481], atomic and molecular physics [482], , quantum optics [324], and fundamental aspects of quantum mechanics [483]. Recent advancements highlight the essential role of quantum control theory for the future application of quantum technologies [484].

The problem of defining *controllability* for quantum systems has been studied extensively. Various notions of con-trollability, have been proposed [485]. Moreover, controllability criteria have been defined for finite-dimensional closed quantum systems while, regarding infinite-dimensional quantum systems or open quantum systems, only limited results have been obtained [485] as we briefly discuss later.

The open-loop strategies for quantum control are ones in which the control actions are predetermined and executed without real-time feedback or adjustment based on the system's current state or measurement outcomes. When developing a one-loop strategy, one can employ either a coherent or incoherent quantum control strategy. Coherent control strategy involves manipulating the states of a quantum system by applying semiclassical potentials while preserving quantum coherence, with many successful applications in physical chemistry, spin systems, and also permit to enhance multidimensional nuclear magnetic resonance (NMR) experiments' sensitivity in the presence of relaxation [486]. Incoherent control instead allows for the destruction of coherence during the control process and has been introduced to enhance quantum control capabilities for uncontrollable quantum systems [487]. Despite these achievements, one-loop strategies are not suitable for quantum systems affected by noise resulting from coupling with uncontrollable environments. Closed-loop learning control and quantum feedback control are the two main different strategies that have been proposed to overcome this obstacle: the first achieved great success by iteratively operating with new samples to control quantum phenomena in chemical reactions [481]; while the latter has improved system performance in various tasks, including controlling squeezed states and entangled states, state reduction, and quantum error correction [488].

Finally, we want to highlight that the measurement process (typically the weakly continuous one), can be part of control protocols: we dedicate Section III B 3 to this particular framework, often denoted as quantum measurement and feedback control [304].

For an exhaustive review of the topic of quantum control theory and for a complete description of its application the authors recommend previous topical reviews, e.g., [486].

### D. Quantum Metrology and Sensing

We now turn to the last major theoretical and experimental toolbox useful to the purposes of this review: quantum metrology and sensing. Quantum metrology and sensing commonly denotes the use of a controllable quantum state to measure or detect properties of another system, classical or quantum. For instance, to use the sensibility of a quantum state to an external field to measure the field itself, with high accuracy. In the last two decades, quantum metrology and sensing have increasingly turned towards strongly correlated systems as a promising paradigm, for which we refer to the more recent review [2]. Within this framework, we focus here on quantum gases, since they offer a powerful platform for developing quantum measurement protocols and devices, both benefiting from technological developments in the variety of QTs described in Section II A, and also connecting to the engineering of quantum states of matter via the advancements in atomic clocks [489] and atom interferometry [490–495].

In this section, we first remind the reader briefly of the principles of operation of atomic clocks and atom interferometers [494] as transversal tools in atom-based quantum technologies. Then, with this toolbox in mind, we review fundamental theoretical aspects related to quantum metrology and sensing that will be essential to some of the applications described in Section V.

#### 1. Atomic Clocks and Atom Interferometry

*Atomic Clocks* Precise clocks operate based on the same fundamental concept of resonance, thus involving systems that possess a natural oscillation frequency and an external driving that can be tuned across this resonance. In fact, frequency standards have been historically based on observations of cyclic motion of celestial bodies, the design of macroscopic mechanical resonators like pendulums or miniaturised electronic resonators, all significantly limited by the dependence of the operational parameters and stability from external agents. The development of atomic clocks, leading to an outstanding progress in more recent times, represents a remarkable breakthrough to overcome these limitations.

While referring to [489] for a very comprehensive review, we here summarize the main concepts and facts. To start with, these novel frequency standards provide the current base unit of time, that is derived from the electronic ground-state hyperfine transition frequency in caesium, while accurate optical frequency standards are becoming a secondary standard. High-performance atomic clocks have crucial impacts and potential implications, besides providing the standard definition of time and frequency. In fact, they provide the worldwide coordination of atomic time; they are a critical component of global navigation satellite systems; they help in deep space navigation and communication networks and are part of inertial sensor technology like absolute gravimeters, gyroscopes, and gravity aided navigation; they have important applications in geodesy, i.e., in the form of a chronometric leveling method to connect height systems between countries, based on a frequency comparison between two remote optical clocks via optical fibers, microwave or optical satellite links [489].

Atoms are extraordinary natural oscillators. Identical atoms of the same species have the same natural oscillation frequencies given by the modes of the electron wave functions, and can therefore serve as ideal frequency standards. Thus, the concept underlying the use of atomic transitions as a frequency reference is to produce an oscillatory signal in resonance with the atoms' (or molecules') natural oscillations, and then accurately counting the oscillation cycles to measure the time intervals. In particular, the atom is prepared in one (of two) of its quantum states, associated with one of its natural oscillations. Then, a local oscillator is used to generate radiation around the natural oscillation frequency, and a mechanism is designed to detect the change of the atom's state and in particular the resonance condition reveling synchronicity between the local and the atoms' natural oscillations. This synchronization ensures the accuracy and stability of the clock [489].

While the atomic hydrogen maser is one important and still widely used standard realization since its introduction in the 1960s, more common methods to achieve synchronization are based on atoms' absorption and use lasers. The degree of synchronization is limited by noise inherent to the measurement protocol and by dissipative couplings to the environment, especially of electromagnetic origin. Frequency standards are in fact characterized by statistical and systematical errors. The former arise from measurement fluctuations and are characterized in terms of fractional frequency errors: they are estimated basically by performing averages over different probing time duration of the fractional frequency deviation between the local oscillator frequency—assumed to be perfect—and the clock atoms' frequency as related to the reference oscillator frequency [489]. One commonly used indicator is the Allan variance [496], quantifying how the precision on the measured frequency improves with longer averaging times. Systematic errors are more challenging to estimate, since their origin may be unknown or not completely understood. On the other hand, they are crucial for stability, since its statistical improvement with increasing averaging time is strongly affected by systematics. One effective way of dealing with systematic errors is to compare the performance of different versions of the same clock [489]. Designing a good clock requires three essential conditions related to the concept of resonance: stability, high-frequency operation, and narrow-line resonances.

Stabilization of the local-oscillator frequency to an atomic transition is accomplished by extracting a sensitive discriminator signal $dS/df$, with $S$ the signal obtained from the atomic sample and $f$ the frequency shift of the applied radiation, and using it to provide a feedback mechanism. This requires maximizing the value of the atomic-transition frequency $f_0$ and of the rate of change $dS/df$ of $S$ vs. $f$, while minimizing the fluctuations $\delta S$. The governing parameters involved are the signal-to-noise ratio, the quality factor $Q$, and the measurement duration $T_m \propto 1/\Delta f$. Atoms are prepared in one of the clock states and the clock transition resonance is excited near the frequency that maximizes the value of $(dS/df)/\delta S$.

The desired high frequencies and narrow linewidths of the clock transitions have for many years been guaranteed by exploiting microwave—in particular hyperfine—transitions. While even higher frequencies might in principles be exploited as those involved in Mössbauer spectroscopy, Mössbauer clocks suffer from a number of limitations: significant systematics, primarily associated with pressure effects in the host material, dispersive line shapes, technical limitations to observe coherences and realize sufficient collimation of the local oscillator. The optical region of the spectrum comprises suitable narrow-linewidth transitions available in many atoms. However, only recently improved techniques have been made available for locking lasers to stable reference cavities, thereby meeting two requirements for optical atomic clocks: narrow spectra lasers and a convenient method to count cycles of the stabilized laser local oscillators. This has enabled the development of optical atomic clocks with high stability and precision.

In optical atomic clocks, systematic shifts can be classified as those due to environmental perturbations and observational shifts [489]. Environmental shifts are caused by external factors such as present electric and magnetic fields, while observational shifts are related to instrumental and observational effects specific to the clock's experimental setup. Of course, also relativistic effects determine fundamental and universal observational shifts, which are precisely the subject of current investigations, see Section V.

*Atom Interferometry* The fundamental concept of particle-wave duality suggests that matter can exhibit interference phenomena similar to light. This has been experimentally demonstrated for various systems, including electrons and neutrons [497] and, more recently, even large organic molecules [498]. On the other hand, atoms offer precise and accurate controllability due to their internal structure and the possibility of manipulating their quantum states with light. In contrast with e.g., electrons, atoms also lack electric charge, minimizing unwanted external electromagnetic interactions. For this reason, atomic interferometry has been strongly developed, in conjunction with precision measurements [494].

As conceptualised in Figure 6, atom interferometers are basically similar to optical Mach-Zehnder interferometers, in that they use light-pulse atom-optical elements instead of mirrors and beam-splitters. By timing the laser pulses appropriately, one can manipulate the populations and create superpositions of states, similar to the behavior of light in an optical interferometer. The principle of operation of mirrors and beam-splitters is that laser light is made to interact with atoms resonantly to specific atomic transitions, thus coherently driving the populations of the ground and excited states into Rabi oscillations. Thus, a $\pi/2$ pulse, i.e., corresponding to a quarter oscillation period, acts as a 50:50 beam-splitter by creating an equal probability of ground and excited states. Similarly, a $\pi$ pulse acts as a mirror [495].

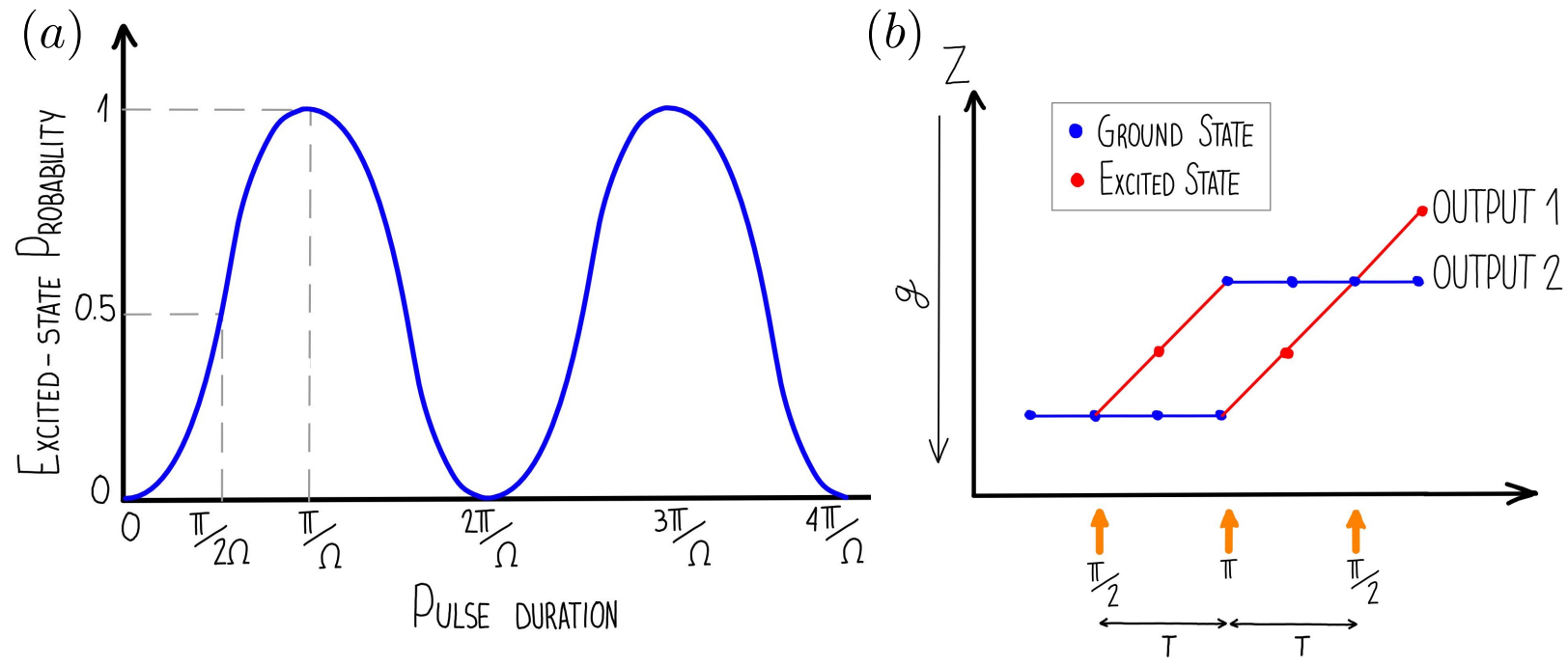

Figure 6: Conceptual operation scheme of an atom interferometer. (**a**) Concept for the case of beam splitters and mirrors. The probability of detecting an excited two-level atom fluctuates as the interaction time is manipulated by laser pulses. If the energy of the laser photons matches the energy difference between a ground and an excited state, the population undergoes Rabi oscillations characterised by a frequency $\Omega$. A pulse of duration $\frac{\pi}{2\Omega}$ creates a balanced 50:50 superposition state, whereas a pulse of duration $\frac{\pi}{\Omega}$ results into a population inversion. (**b**) Concept for an atom-interferometry Mach-Zender protocol. A sequence of three laser pulses is applied to a cluster of atoms in their ground state. First, the initial pulse splits the atomic beam by generating an equally-weighted superposition state as described in (**a**), inducing both momentum transfer and spatial separation. After a duration of time $T$, a $\pi$ pulse mirrors the atomic beam by inverting the populations of the ground and excited states as in (**a**), subsequently making their trajectories converge. Finally, one more $\frac{\pi}{2}$ beam-splitting pulse mixes the populations, leading to interference patterns in the output port populations. The interferometric phase, which is influenced by the external fields such as the gravitational one $\vec{g}$, can be assessed by quantifying the number of atoms in each port. The sensitivity of the system relies on the size of the enclosed space-time area encompassed by the atom trajectories.

Therefore, these laser pulses can in principle create superpositions of internal as well as momentum states, given that atoms absorbing the photons get a finite momentum kick and makes them travel along different paths simultaneously. During these paths, a different phase can be accumulated due to the distinct interactions with the given external fields, gravity being one such example. This phase difference can then be detected after recombination of the atomic beams. In the simplest configuration, this operation can be accomplished by a sequence $\pi/2 - \pi - \pi/2$ of three light pulses applied, effectively corresponding to the matter-wave equivalent of a Mach-Zehnder (light) interferometer.

Since the first laboratory demonstration of atom interferometers [499], a number of variants have been developed with diverse applications, which can be reviewed in e.g., [494]. Selected applications are discussed in Section V A.

### *2. Entanglement as a Resource for Quantum Metrology and Sensing*

Atom interferometry poses several challenges due to various sources of uncertainty, which can be classified as either device-related or statistically driven [500]. Experimental techniques have made considerable progress in reducing the former, reaching a level comparable to or even lower than statistical errors [43, 49, 499–502]. To achieve even greater precision, research efforts are now focusing on addressing the issue of statistical uncertainty, especially in the context of quantum phase estimation [503, 504]. The conceptual map in Figure 7, adapted from [2], provides a glimpse of the current status of metrological gain sensitivity in atomic platforms, including ions, Bose-Einstein condensates, and cold thermal (i.e., not condensed) atoms.

One potential solution to mitigate statistical uncertainty lies in leveraging entanglement, specifically through the use of quantum squeezing, to start with [223, 505, 506]. Quantum squeezing allows for the reduction of uncertainty in a specific observable below the fundamental Heisenberg limit, at the cost of increased uncertainty in a conjugate observable, as introduced already for light [507]. Experimental setups involving atomic spin squeezing have been successfully proposed and implemented using either collision-driven interactions or light-mediated interactions in optical cavities [137, 223, 508–512]. While most of the early research has been focused on squeezing schemes for internal states of the atoms, more recently proposals for momentum-states squeezing have been pushed forward [513–515], the latter inspiring the more recent realization [514], and preluding to further progress with the proposal of more advantageous simultaneous squeezing in internal and external degrees of freedom [516].

The presence of entanglement provides an advantage for sensing, as it can enhance considerably the sensitivity of the considered protocol for estimation [2, 173, 491, 503, 517, 518]. More in detail, sensitivity denotes here the variance (uncertainty due to classical and quantum errors) $\Delta Q$ in the estimation of the desired quantity $Q$, calculated on a varying (up to infinite) number of repeated measures on a chosen interrogation time. This amounts to the minimum detectable value for $Q$ and for unitary interrogation time. Beyond on the specific details of the sensing protocol and on the decoherence timescales [173], sensitivity depends critically on the spatial multi-partite entanglement between the different subsets of the sensor.

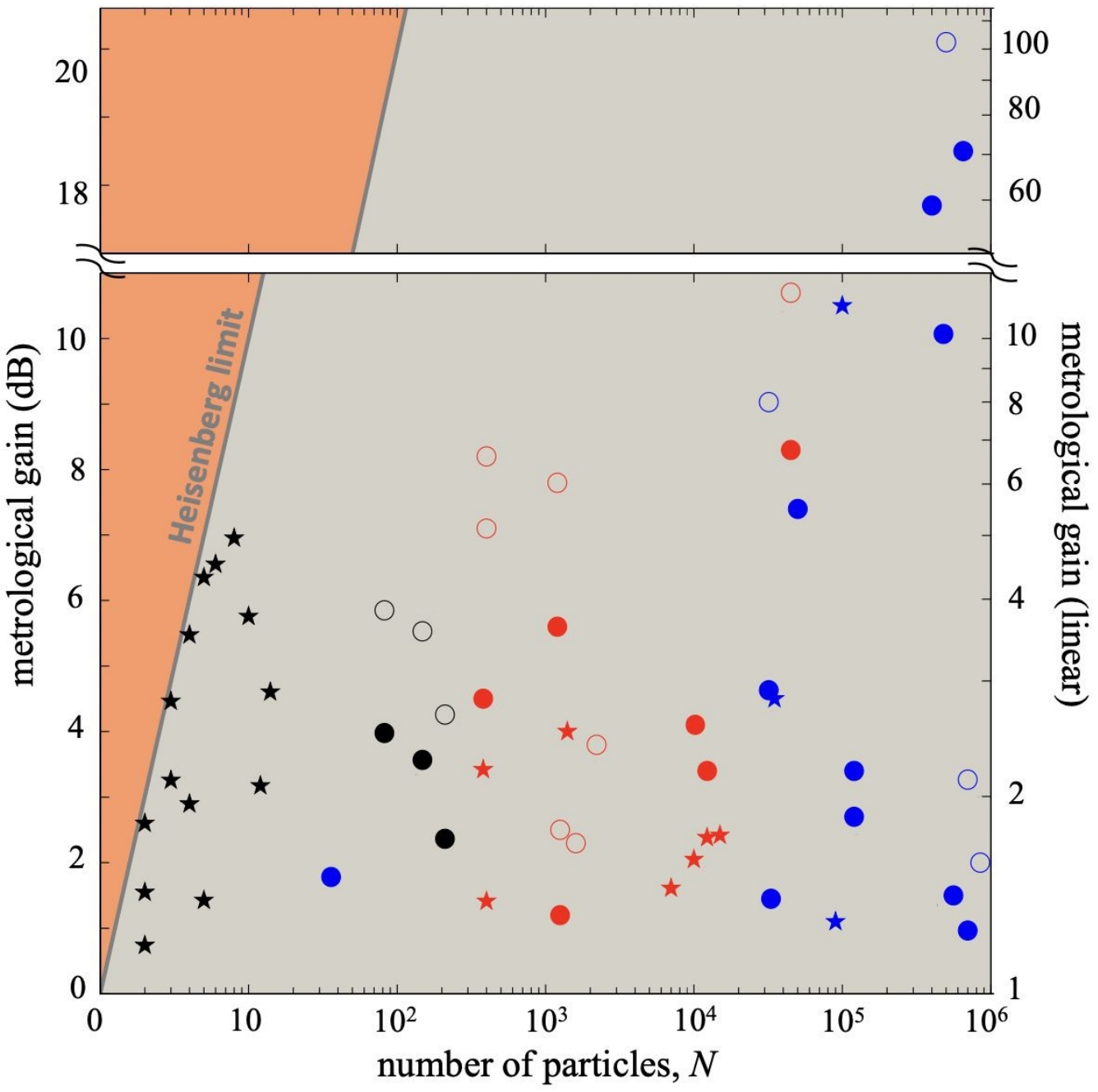

Figure 7: Conceptual map of the achieved gain $\Delta\theta_{\rm SQL} = 1/\sqrt{N}$ in phase sensitivity over the standard quantum limit, vs. the total number of atoms $N$ or, when fluctuations are present, its average. Figure adapted from [2], to provide an idea of the current technological landscape. The logarithmic scale [left, in dB, $10\log_{10}(\Delta\theta_{\rm SQL}/\Delta\theta)^2$] and the linear scale [right, $(\Delta\theta_{\rm SQL}/\Delta\theta)^2$] are employed to display the gain. As a reference, the solid thick line represents the Heisenberg limit with $\Delta\theta_{\rm HL} = 1/N$. Different colors refer to different experimental platforms as follows. Black: trapped ions. Red: Bose-Einstein condensates. Blue: cold thermal ensembles. Different symbols refer to different ways of estimating sensitivity, as follows. Stars: gains directly measured in phase sensitivity, acquired through full phase estimation experiments. Circles: expected gains based on the characterization of the quantum state, as computed e.g., from $\Delta\theta = \xi_R/\sqrt{N}$ using the spin-squeezing parameter $\xi_R$, or as $\Delta\theta = 1/\sqrt{F_{\rm Q}}$ using the quantum Fisher information $F_{\rm Q}$. Filled (open) circles: results obtained without (with) the subtraction of technical and/or imaging noise. We refer to [2] for the correspondence of each symbol to a different experiment.

Multi-partite entanglement (ME) is generally defined as follows. For a $d$-dimensional discrete system with $N$ components (e.g., sites), $c$-partite entanglement, with $1 \leq c \leq N$, implies that a partition $\{|\psi_i\rangle\}$ exists, where the maximum number of components in a single $|\psi_i\rangle$ is $c$. The tensor-product state $|\psi\rangle = \prod_{\otimes_i} |\psi_i\rangle$ is then said to be $c$-producible, or to have entanglement depth $c$. In addition, a system is said to host $c$-partite entanglement if it is $c$-producible but not $(c+1)$-producible. Instead, the number $h$, with $\frac{N}{c} \leq h \leq N - c + 1$, of disentangled subsets is the degree of separability [517–519]. The usual separability observed for product states corresponds to $h = N$ and $c = 1$. Note that the subsystems are not necessarily physically adjacent sites or parts of the system, e.g., in a lattice, but can be distant in general. When $c = N$, $|\psi\rangle$ is said to host *genuine* ME. Note also that $c$ can even diverge with $N$, $c \sim N^l$, $0 \leq l \leq 1$.

For mixed states, $c$-producibility in $h$ subsets holds if $\rho$ can be decomposed (generally not uniquely) as

$$\rho = \sum_{\tilde{\lambda}} p_{\tilde{\lambda}} \, |\tilde{\lambda}\rangle\langle\tilde{\lambda}| \,, \tag{9}$$

where $p_{\tilde{\lambda}} > 0$ without any lack of generality, and $|\tilde{\lambda}\rangle$ are c-separable states in $h$ subsets, not necessarily with the same space-partition. If $c = N$, then Equation (9) is still valid, trivially with a single partition and in every decomposition. In general, the $c$-producible decomposition $|\tilde{\lambda}\rangle$ is not orthogonal, thus $\rho$ is not diagonal. Moreover, the producibility of $|\tilde{\lambda}\rangle$ is generally lost in other decompositions.

Multi-partite entanglement can be lower-bounded by a quantity, called quantum Fisher information (QFI), $F_Q$ [2], expressed for pure states as sums of two-point connected correlations of local operators. The same quantity is also effective for mixed states, although the precise expression is in general more involved, given that it does not imply standard one- and two-point correlations only, unless specific symmetries are present [520]. More in detail, if $c$-entanglement, but not $(c+1)$-entanglement, is present, then the inequality

$$F_Q[N] \leq 4\,k\,c\,N\,, \tag{10}$$

holds [504]. This implies that its violation signals at least $(c+1)$-partite entanglement. The ultimate limit $F_Q[N] = 4\,k\,N^2$, when the state $|\psi\rangle$ hosts genuine multi-partite entanglement ($c = N$), is called the Heisenberg limit. Other notable limits can be formulated, also involving the separability $h$, and extending therefore that in (10). Overall, the QFI reveals a class of entangled states, obeying a necessary condition for super-shot-noise sensitivity of a quantum sensor.

More in detail, the QFI bounds the sensitivity of a quantum sensor via the so-called Cramer-Rao bound:

$$\Delta Q \geq \frac{c}{\sqrt{F_Q}}\,, \tag{11}$$

where $c$ is a classical statistical factor, depending on the experimental details: the maximum reachable sensitivity scales as the inverse of the square root of the QFI.

The quantification of entanglement has sparked a recent debate in both the quantum information and many-body communities, driven by experimental observations in quantum gases, see e.g., [521–527]. Textbook examples of quantum devices effective for sensing are trapped ions and Rydberg atoms devices, atomic clocks, ensemble sensors, spin sensors, magnetometers [528] (as SQUID-based technologies), single electron transistors, optomechanical devices, or even elementary particles, as muons and neutrons [173]. Concerning the QFI, a weaker counterpart, the so-called classical Fisher information, has been measured in [529]. Alternatively, multi-partite entanglement has been estimated on multi-qubit devices, without the use of local operators as for the QFI, and using instead a long-range Ising Hamiltonian [530].

Focusing on the theoretical level, various sensing archetype schemes have been elaborated, mostly employing maximally-entangled states. Moreover, they are often based on interferometric schemes, as Rabi, Ramsey, or Mach-Zender like protocols. Indeed, exploiting quantum states, various estimation problems of different physical quantities can be mapped on the suitable phase estimations. Other fundamental applications concerns the study of the Einstein-Podolsky-Rosen paradox, and more in general for the verification of quantum mechanics against alternative theories, see e.g., [531–534].

Due to that central importance of maximally-entangled states, a large part of the efforts in the research on quantum metrology and sensing is devoted to the efficient synthesis of these states, as via bosonic Josephson junctions (mostly in their ground-states), suitable atomic collisions or atom-light interactions, mostly in cavities [2], and their characterization in terms of the quantum Fisher information content. In particular, Bose-Einstein dynamics experimentally proved particularly suitable for actual entanglement generation [535, 536].

Generating useful entanglement has also widely exploited quantum statistics [537] and long-range interactions [223, 538–540]. On the other hand, the impact of many-body finite-range interactions has also been considered [239], as they can as well induce the build-up of long-range correlations.

Relevant examples of highly-entangled states are realizable with sufficiently high fidelity, in ultra-cold atom or trapped-ion experiments. Some of them are GHZ states, NOON states and squeezed states. GHZ states are natural generalizations of cat states with $N \geq 2$ spins: $(|\uparrow \ldots \uparrow\rangle \pm |\downarrow \ldots \downarrow\rangle)/\sqrt{2}$. NOON states are instead superpositions of multi-well states with a unique populated well. The simplest example for them is obtained in a double quantum well with $N$ bosons, as combinations of two classical states where only a well is macroscopically populated: $(|N,0\rangle \pm |0,N\rangle)/\sqrt{2}$. GHZ and NOON states are useful for textbook protocols for quantum sensing, as Ramsey or Rabi schemes. However, these states are typically very fragile against decoherence, due to magnetic frustrations and particle losses, respectively. Also for this reason, a large attention has been devoted to the creation of more stable highly entangled states, for instance the so-called squeezed states. In general, squeezed states are characterized by the fact that the sum of the variances for conjugate variables is fixed to $\hbar/2$ [2, 173, 504]. The coherent states are a particular squeezed states, where the mentioned variances are equal, to $\hbar/4$.

GHZ and squeezed states have been created and manipulated in trapped-ions setups: one example for GHZ states is reported in Figure 8, adapted from [2]. There, the long-range Coulomb force is an ideal ingredient to prepare ions in highly-entangled states. We also mention that similar magnetic properties are exploited in molecular quantum nanomagnets—nanosized molecules with a certain number of ions interacting with each other via a strong Heisenberg (anti-)ferromagnetic coupling—, still to produce and manipulate entangled states for quantum simulation, information, communication and sensing, see e.g., [92–94] and references therein.

Various remarkable examples of sensing have been performed, using atoms and ultracold atoms. A major example being atomic clocks, as described in Section II D 1. There, the textbook approach exploits the extreme precision of resonant frequencies of long-living isolated atoms (the caesium atoms being the most-used ones). In the last years, enhancements of atomic clocks have been proposed, exploiting quantum properties [541], as well as optical magnometers and magnometers with Bose-Einstein condensates [504].

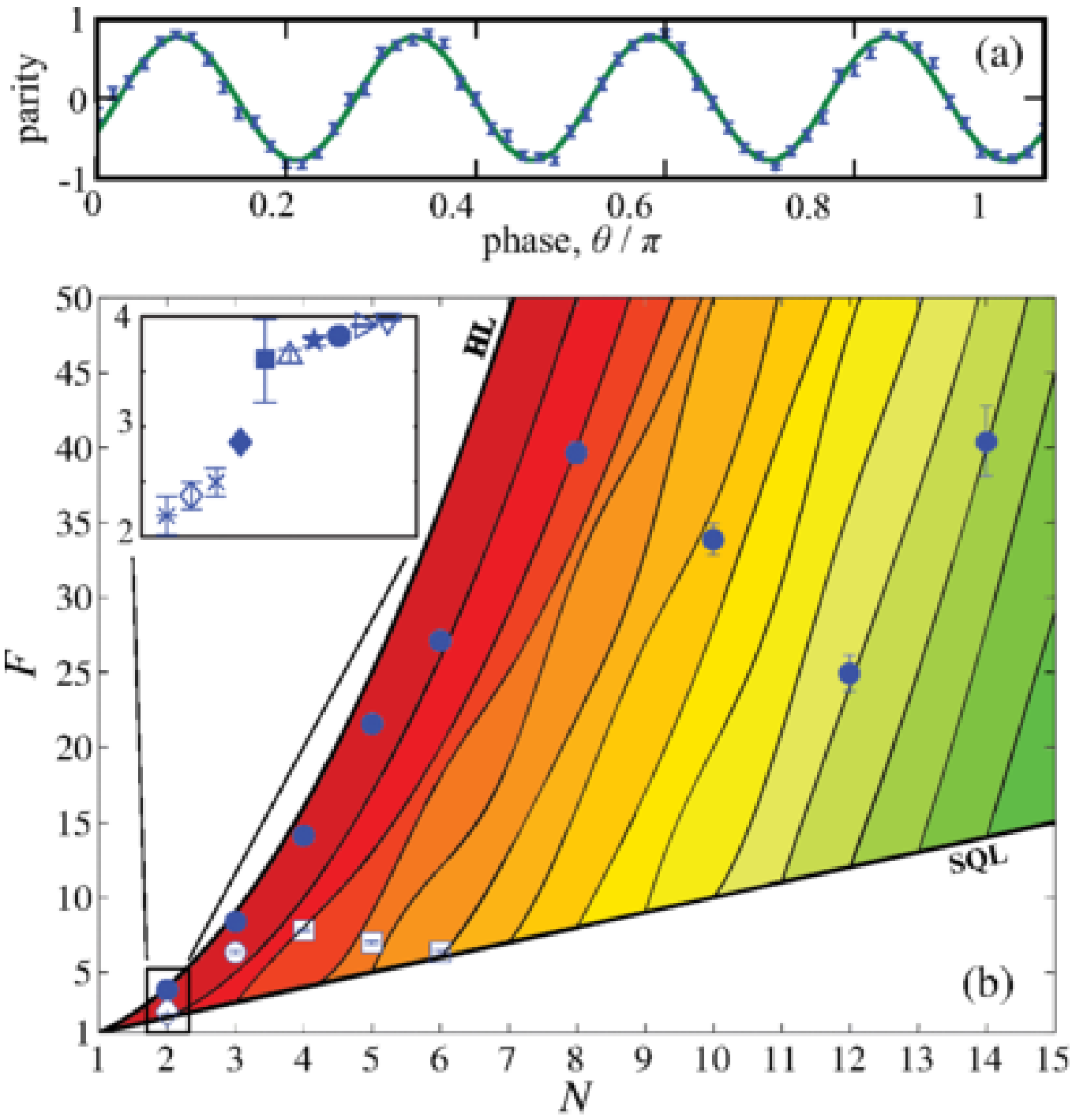

Figure 8: Showcase of phase sensitivity for ion Schrödinger cat states. Figure adapted from [2]. (**a**) Typical parity oscillations achieved using cat states, characterized by a distinctive period of $2\pi/N$ (here illustrated for $N = 8$). (**b**) Overview of experimental realisations. The Fisher information $F$ obtained from experimentally extracted visibilities $V$, is presented as a function of the number of qubits $N$, specifically expressed as $F = V^2N^2$. Upper thick line: Heisenberg limit (HL) $F = N^2$. Lower thick line: standard quantum limit (SQL) $F = N$. Thin lines: bounds for valuable $k-$particle entanglement, employed to perimetrise the shaded region corresponding to $(k + 1)-$particle entanglement. Notice that the darker red region indicates useful genuine $N-$particle entanglement, while the lighter red region signifies useful $(N - 1)-$particle entanglement, and so forth. Inset: zoomed$-$in view for the specific case with $N = 2$ ions.

Concerning more fundamental physics-oriented purposes, precise tests of gravity (equivalence principle or gravitational constant) have been performed using interferometric schemes on ultracold atoms setups, as discussed in Section V A.

## III. CONDENSED MATTER AND MANY-BODY PHYSICS

The engineering of quantum states of matter for simulating condensed-matter systems, or other applications, usually implies the concept of quantum phase transition. Quantum many-particle systems can undergo phase transitions even at zero temperature, usually driven by the competition between kinetic and interaction energies, or else between two length scales, across a critical point [542]. Therefore, their behaviours is influenced by a variety of parameters such as the dimensionality of the system, the strength and range of the interactions, or the amount of disorder. Remarkable results in ultra-cold atoms have given access, under accurately controlled conditions, also to paradigms falling outside this concept.

The enormous development of experimental techniques has led to the possibility of quantum simulate textbook condensed-matter physics problems. For example, the achieved possibility of combining interactions with strong synthetic magnetic fields has allowed to place a Landau gauge Bose-Einstein condensate in and near the lowest Landau level and perform in such an experiment quantum simulations of known and new effects related to quantum Hall physics, including e.g., spontaneous crystallization driven by condensation of magneto-rotons [543], as well as the realization of fractional quantum Hall states [544].

Experimental quantum simulators of hydrodynamic flow are currently developed with strongly correlated fermionic or bosonic atoms, that can be exploited to inform theories of fermion transport relevant to electron, neutrons, and quark systems. For example, universal behavior stemming from scale invariance has been probed in a strongly-interacting fermionic Lithium fluid under hydrodynamic and very low temperature conditions, finding that the sound diffusivity behaves similarly to what observed in liquid Helium 4, i.e., a strongly-interacting *boson* fluid, instead [545]. The high accuracy and control achieved in these experiments is quite remarkable, more recently including also new techniques allowing for spatial fluctuation thermometry of these quantum fluids [546]. Also thanks to the development of boxed confining potentials made with sheets of laser light, accurately enough homogeneous geometries are realized, that reproduce textbook studies of first and second sound in compressible Bose fluids [107], More recent advances in ultracold $K$-atom impurities embedded in a Fermi sea of Li atoms has led to the first controlled quantum simulation of the physics of Fermi polarons, as predicted by Landau theory of Fermi liquids, in an experiment with ultracold-atoms mixtures [547].

In this section, we make the specific choice of focusing on selected examples, classified among well established, evergreen, paradigms in Section III A and among freshly sprouting ones in Section III B. In the former case, we step into the BCS-BEC crossover, as an example of phase transition characterised by the continuous evolution of a single order parameter across asymptotic behaviors, and commensurate-incommensurate quantum phase transitions, due the competition between two types of potential energies, that favor different system configurations. In the latter case, we focus on quantum phase transitions based on completely different concepts, like the breaking of the thermalisation hypothesis or the case of dynamical quantum phase transitions.

The choice for these particular examples is based on their physics relevance and to their transversal applicability outside condensed-matter physics.

## A. Condensed Matter and Many-Body Physics: Evergreen Paradigms

### *1. BEC-BCS Crossover*

Here we address one relevant example of phase transition in which the characterising order parameter continuously evolves across limiting behaviors under the tuning of an internal parameter. In particular, this is the case of the crossover from a Bardeen-Cooper-Schrieffer (BCS) [21] type of weak-coupling superfluidity or superconductivity, to a Bose-Einstein condensation (BEC) of tightly bound, point-like, composite bosons made of two fermions, which is one of the most captivating and useful paradigms born in condensed matter physics as a tool to understand the puzzling behavior of high-$T_c$ superconductors, to be then applicable beyond material's science in the physics of fundamental interactions, nuclear matter, and cosmology [1, 548]. However, it is with the advent of ultracold atomic (Fermi) gases [22, 100, 549, 550], that the BCS-BEC crossover has been explored in length and depth under highly controlled conditions and as a joined effort of experimentalists and theorists. There are a number of useful reviews focused on the subject [102, 114, 548, 551–554], so that in this section we summarize the main ideas while providing a different perspective.

In essence, the main concept in this problem is that while the attraction strength between fermionic particles is progressively increased, all the fundamental system quantities smoothly interpolate between the two BCS and BEC limiting regimes. The ground-state firmly maintains the same kind of spontaneous symmetry-breaking, while one same order parameter evolves from the BCS-related gap function to the BEC condensate fraction. Thus, for example, at $T = 0$ the chemical potential evolves from the Fermi energy to half the binding energy of the composite boson, the critical temperature from the exponentially-reduced BCS expression for the fermions system to that of a non-interacting BEC of (half as many) composite bosons. A similar crossover physics may occur in the non-broken symmetry phases, while varying the interaction strength. Figure 9 illustrates the schematic phase diagram encompassing the BEC-BCS crossover and the various phases investigated so far.

The inception of the BCS-BEC crossover can be traced back to the formulation of the BCS theory itself, in fact during a time when alternative theories were emerging aimed to explain superconductivity in metals as a BEC point-like bosons, namely bound states of two electrons [556]. In their influential paper, Bardeen, Cooper, and Schrieffer sought to highlight the distinctions between their theory of strongly overlapping Cooper pairs and the BEC of point-like bosons. The theoretical groundwork for the BEC-BCS crossover theory was then laid by Eagles [557], focusing on its potential applications to excitons in semiconductors. Then, Leggett [558] and Nozières and Schmitt-Rink [555] advanced the idea by developing the formal theory at zero and at the critical temperature for superconductivity, respectively. Noticing that an increase of interaction strength in mean-field approximation for the fermionic system, would have led to unphysical divergent critical temperatures in the BEC regime, Nozières and Schmitt-Rink (NSR) recovered the correct BEC limit after including gaussian pairing fluctuations in a form of Random-Phase approximation. The work of NSR has then turned out to be especially relevant in light of the discovery of high-temperature superconductors (HTSC), when the universal log-log Uemura plot for the critical vs. the Fermi temperature of a number of conventional and HTSC superconductors [559]. Uemura et al. noticed how the HTSC were gathering along a universal line lying somewhere in between the two opposite regimes of BCS-like superconductivity and BEC of point-like bosons, the universality being later on explained by Pistolesi and Strinati when using a description in terms of the correlation length [560]. Since then, extensive research has been conducted in this domain, exploring the entire phase diagram, including the normal non-superfluid phase, and proposing the existence of novel regimes such as the pseudogap phase characterized by the presence of non-condensed composite bosons. In addition, it has emerged as a valuable concept in the study of the QCD phase diagram and the equation of state in neutron stars [5, 561, 562].

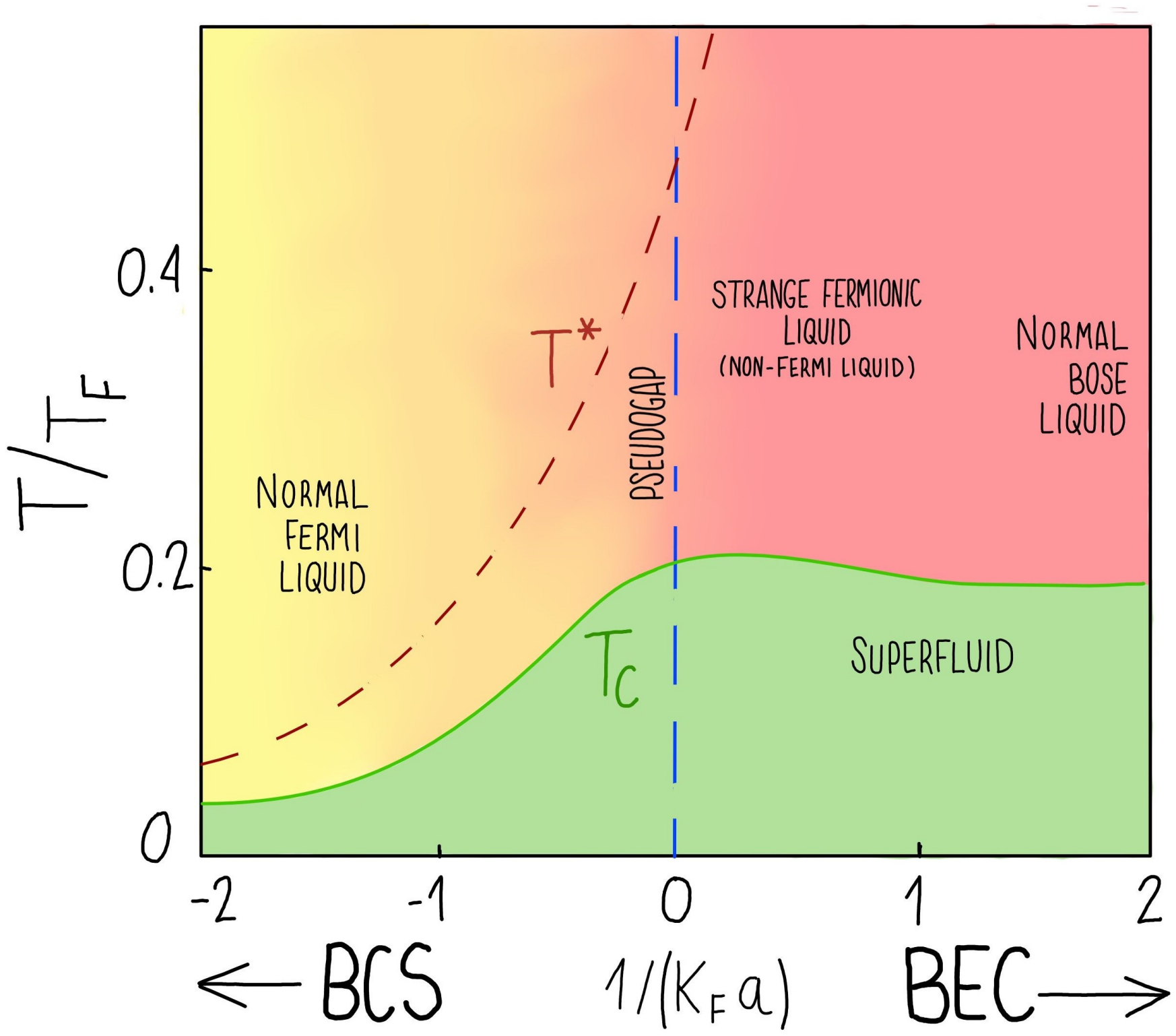

Figure 9: BEC-BCS crossover. Schematic phase diagram within the parameter space defined by temperature $T$ (in units of the Fermi temperature $T_F = E_F/k_B$, where $E_F$ represents the Fermi energy) and inverse scattering length $1/(k_F a)$ in units of $k_F^{-1}$. The green lower line separates the superfluid phase with broken symmetry from the normal phases (i.e., non−superfluid phases). The critical temperature $T_c$ exhibits an increasing trend with the inverse scattering length, starting from the BCS limit and approaching the BEC value $T_{BEC} \simeq 0.218 T_F$ of a gas composed of non−interacting bosons. Notably, $T_c$ passes through an optimal value around the resonance point $(k_F a)^{-1} = 0$. The predicted value of $T_{BEC}$ can only be obtained by considering pairing fluctuations, as first emphasized by Nozières and Schmitt−Rink [555]. Otherwise, without accounting for these fluctuations, $T_c$ would indefinitely increase with the strength of pairing. The red dashed line represents the dissociation temperature $T^*$ at which composite bosons, formed by two fermions, are disrupted by thermal fluctuations. This temperature can be viewed as indicative of the opening of a pseudogap in the single−particle spectral function. On the BCS side of the phase diagram, the critical and dissociation temperatures coincide, while in the BEC limit, they differ.

The interest in the BCS-BEC crossover has extraordinarily grown following the realization of Bose-Einstein Condensation [98, 99] and BCS superfluidity [22, 100] in ultracold quantum gases. The possibility of tuning the pairing strength between fermionic atoms allowed by the Fano-Feshbach (FF) resonance mechanism [563] described in Section II A 2, has led to the proposed resonance superfluidity [549, 550, 564, 565] as the route to realise high-$T_c$ superfluidity in Fermi gases. Resonance superfluidity was then experimentally realized first at JILA in the group of Debbie Jin [566, 567], followed by the Ketterle group at MIT [568], opening the door to the investigation of the BCS-BEC crossover under controlled conditions. In fact, as highlighted in Figure 10 from [549], it was immediately apparent how a universal behavior of superfluidity comprises the most diverse bosonic systems, like liquid $^4$He and the BECs of alkali atoms, and fermionic systems like $^3$He, the conventional superconductors, the HTSC, and the Fermi gases equipped with Fano-Feshbach resonances.

Since the first realization, a number of textbook experimental investigations have been performed, like studies about the evolution of vortex lattices [569] or the second sound [570] in the crossover. More recently, even more flexible platforms are being setup, that involve fermionic molecules with dipolar interactions. A two-dimensional

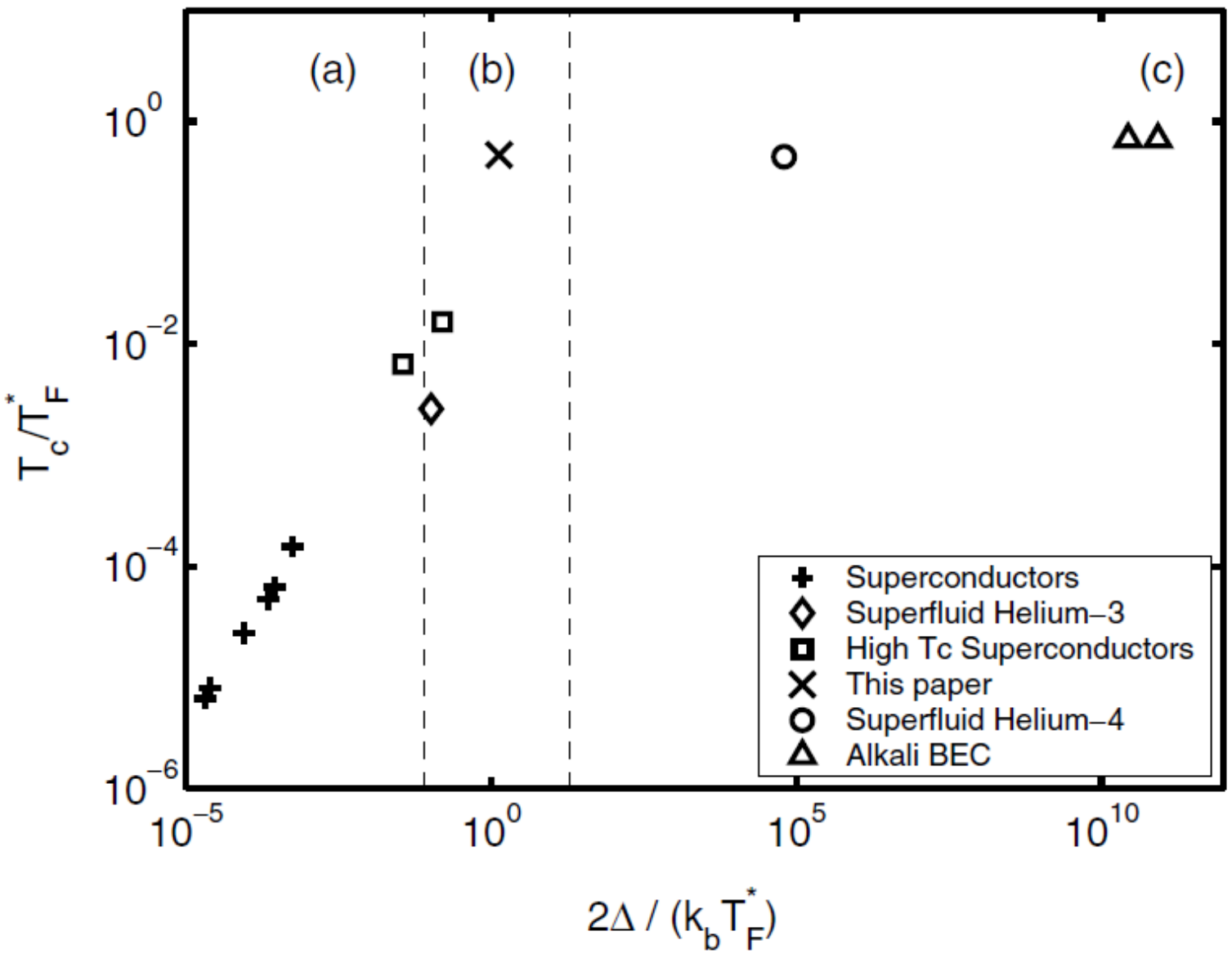

Figure 10: Universal behavior of superfluid systems. Log-log plot of the transition temperature $T_c$ as a function of the normalized gap energy $2\Delta$ relative to the effective Fermi temperature $T_F^*$. The different regions correspond to BCS systems (**a**), the crossover regime (**b**), strongly bound composite bosons exhibiting BEC-like behavior (**c**). In (**a**,**b**), $2\Delta$ signifies the the energy required to break apart a fermion pair and $T_F^*$ the effective Fermi temperature built from the system density. In (**c**), $2\Delta$ represents the minimum energy necessary to separate the composite boson into two fermions, that is the ionization energy leading to a charged atomic core and an electron, and $T_F^*$ is the corresponding ionic Fermi temperature. Image from [549].

Fermi gas of spin-polarized potassium-rubidium polar molecules is created with tunable dipolar interactions that are also elastic for efficient evaporative cooling to reach quantum degeneracy [128]. Understanding losses in reactive molecular gases such as KRb [571, 572] is critical for harnessing their anisotropic long-range interactions for quantum technology applications. Experimental control on spin exchange dynamics has been achieved [573, 574], paving the way to the investigation of many-body phases and nonequilibrium dynamics in long-range interacting systems with reduced dimensionality. Non-local fermion pairing has been directly observed in an attractive gas made of fermionic $K$ atoms, that in fact implements a Fermi-Hubbard model, and has allowed to investigate the BCS-BEC crossover including the formation of a pseudogap [575]. The Fermi-Hubbard model has been implemented also in a system of fermionic Lithium atoms in 2D optical lattices, and exploited to quantum simulate frustration- and doping-induced magnetism [576].

In fact, the emergence of Fermi gases [22, 568, 577, 578] has transformed the crossover physics from a phenomenological approach to a framework that can be explored through microscopic theories.

The conceptual map in Figure 11, adapted from [579], summarizes the theories developed so far in the domain of cold gases, in the relevant parameter space defined by $-(k_F a)^{-1}$ and $(k_F r_0)^{-1}$, the former driving the crossover between BEC ($-(k_F a)^{-1} < 0$) and BCS ($-(k_F a)^{-1} > 0$) limits and the latter distinguishing the resonance width from narrow ($(k_F r_0)^{-1} \ll 1$) to broad ($(k_F r_0)^{-1} \gg 1$) [113], as introduced in Section II A 2. The so-called one-channel models are constructed based on the BCS Hamiltonian, utilizing the single parameter $U = 4\pi\hbar^2 a/m$ in terms of the tunable scattering length $a$, which facilitates the description of broad resonances. Broad resonances have been extensively explored through self-consistent theories that incorporate pairing fluctuations [580–584] and zero and finite temperature Quantum Monte Carlo (QMC) simulations [585–588]. However, recent advancements in quantum gas experiments [215, 589, 590] have made intermediate resonances accessible, necessitating further theoretical examination. Intermediate width-to-narrow resonances are as well relevant for the equation of state of matter in neutron stars [5, 561] that is characterized by $k_F|r_0| \simeq 1$. Nevertheless, the theoretical treatment of narrow resonances faces several unresolved questions, primarily due to the need of introducing the finite width as a second parameter [113, 114, 591]. While QMC results are available [592] for a one-channel model that emulates the finite width using well-barrier potentials [593], these results are limited to the regime where $(k_F|r_0|)^{-1} \gg 1$.

In contrast, two-channel, boson-fermion (BF) models explicitly incorporate the resonant (boson) state composed of two fermions, representing the underlying FF mechanism: in fact, BF Hamiltonian models represent the many-body formulation of the original two-body FF mechanism, the resonant boson composed of two fermions being the bosonic mediator of the pairing attraction among the fermions themselves. Initially introduced in the context of high-temperature superconductivity [594, 595], the BF model has been proposed for ultracold atoms within a mean-field formulation [549, 550, 564], and subsequently developed using Random-Phase-approximation (RPA) methods [564, 596–598], as well as various forms of self-consistent RPA [599, 600]. To account for a wide range of FF resonance widths, the inclusion of particle-hole fluctuations has been accomplished using the powerful Functional Renormalization Group (FRG) approach, albeit in a parameterized manner [601–603] (see Figure 11).

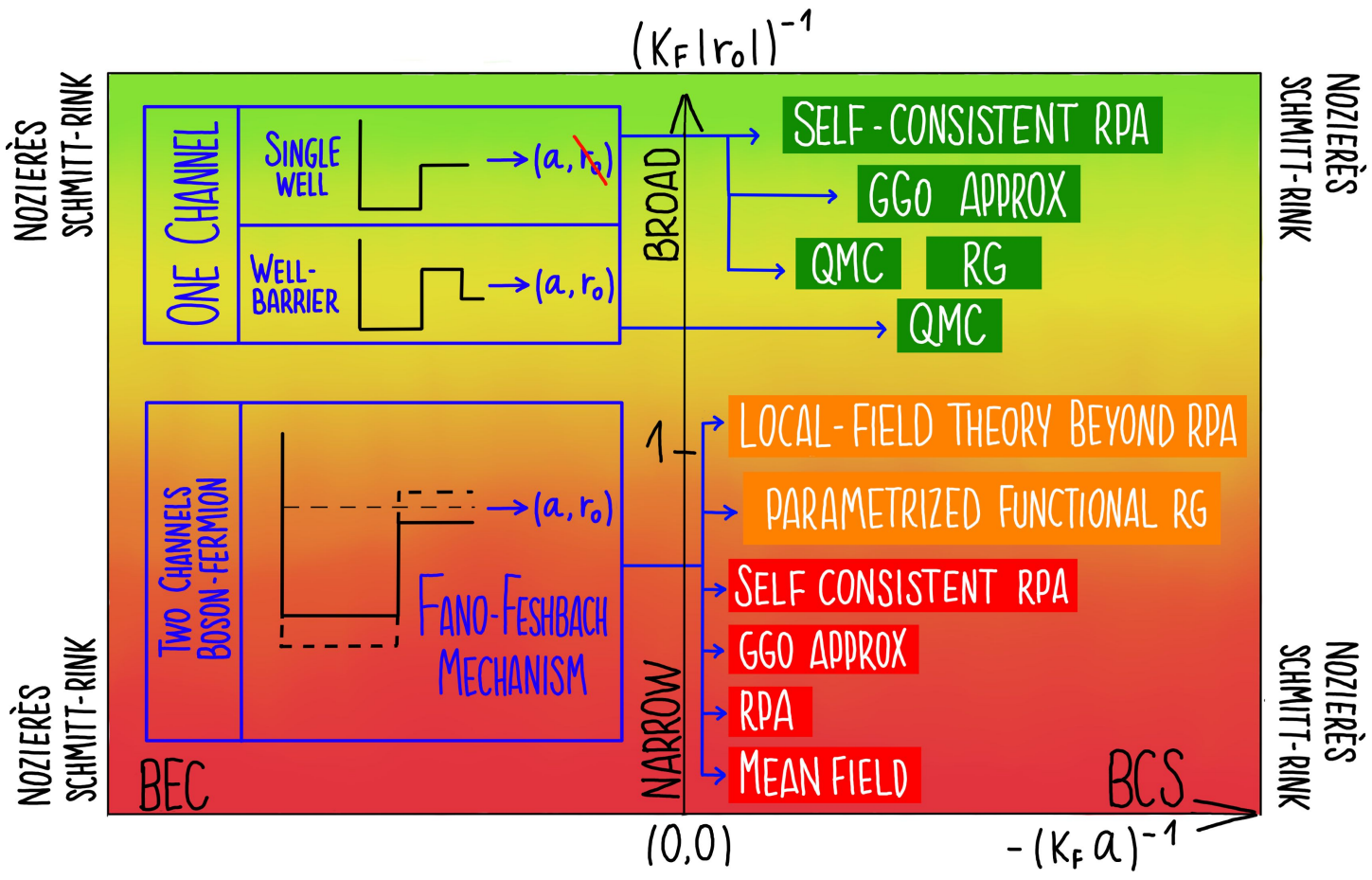

Figure 11: Conceptual map of BEC−BCS crossover theories within the parameter space defined by $-(k_F a)^{-1}$, which govern the transition between the BEC and BCS limits, and by $(k_F r_0)^{-1}$, characterizing the width of the resonance from narrow to broad, respectively. On the left side are general model−frameworks, including one or two−channel models. On the right side, the corresponding theoretical or Quantum Monte Carlo (QMC) methods employed to investigate the crossover are illustrated, covering the narrow (indicated by a red stripe), intermediate (represented by orange), and broad (depicted by green) regions. The boson-fermion local-field theory (BFLF) serves as a bridge for intermediate−to−large values of $(k_F r_0)^{-1}$, encompassing fluctuations through a comprehensive local−field theory of the boson−fermion Hamiltonian. For an account of the different theories and corresponding references, see the main text. The conceptual map is adapted from [579].

The intermediate regime that bridges the gap between narrow and broad FF resonances has been more recently addressed the boson-fermion local-local field (BFLF) theory [579], providing a unifying framework to treat fermionic atoms interacting via narrow to intermediate-width Fano-Feshbach resonances, with inclusion of exchange and correlation effects beyond mean field. The BFLF theory builds on the so-called local-field factor theories developed in the 70s [604, 605] to study the density and spin response in low-density metals, when the long-range Coulomb interactions are not negligible and lead to significant correlation effects, and providing a description quantitatively competitive with best quantum Montecarlo simulations [606]. In essence, the local-field factor concept enters the system's response function, where it incorporates the exchange and correlation effects beyond mean-field in a self-consistent manner, in fact corresponding to the introduction of irreducible vertex corrections [606]. Indeed, it is intimately connected to the pair-correlation function embodying the exchange and correlation hole and therefore to the structure factor that, in turn, is related to the response function via the fluctuation-dissipation theorem. This leads to a set of self-consistent equations to be numerically solved iteratively, to determine all the system's properties. In the BFLF, the method is built on top of the boson-fermion Hamiltonian, especially suited to describe narrow-to-intermediate FF resonances, and in a generalized manner to include density, spin, and superfluid pairing fluctuations in amplitude and phase, i.e., characterizing the Higgs and the Goldstone modes, respectively. The BFLF theory embodies the relevant symmetries and exact results such as the Hugenoltz and Pines theorem establishing the conditions for the gapless nature of the excitations, and recovers the known limits of BEC and BCS behaviors, extending to the whole crossover and to narrow-to-intermediate width FF resonances the predicted Gor'kov and Melik-Barkhudarov $T_c$ suppression due to the interactions.

The availability of a unifying theory of the BCS-BEC crossover in the whole phase diagram in Figure 11, may allow to investigate beyond mean-field whether universal behavior along the crossover holds under more general conditions, provided that the correlation length is used as a driving parameter [560]. The correlation length is in fact the microscopic measure of the size of a Cooper pair, as calculated from the variance of the pair correlation function, and crosses over from very large to very small values from the BCS to the BEC regimes.

Indications along these lines have been provided for different microscopic systems within different mean-field theories. In [607], the case of fermions interacting via a shape-resonance, in the form of a well-barrier potential, has been investigated within one-channel Hamiltonian in mean field, capable of describing broad-to-intermediate range resonances. As displayed in Figure 12a,b, when plotted vs. the correlation length, the superfluid chemical potential and the condensate fraction of a number of different microscopic realisations of the same Hamiltonian after tuning the potential parameters, collapse in only one universal curve. Similar behavior occurs for the case studied in [608] of fermions interacting via a separable potential characterized by a strength and a range like that introduced by NSR [555], in the additional presence of a spin-orbit coupling, and treated within the conserving $GG_0$ approximation [599]. As displayed in Figures 13a,b, universal behavior holds for the chemical potential at $T_c$ and the critical temperature, provided that the spin-orbit coupling is weak enough to keep the system below a topological phase transition. These results extend the findings from Pistolesi and Strinati [560] to different contexts.

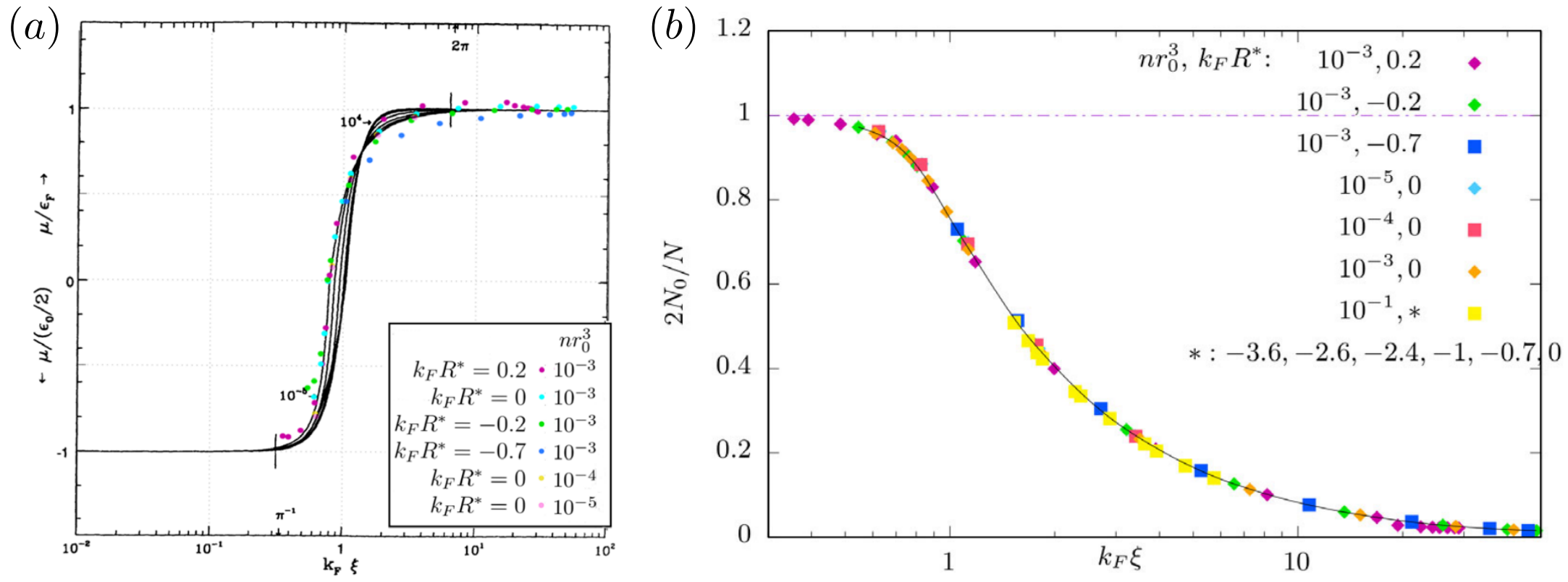

Figure 12: Universal behavior of fermionic superfluids in the BCS-BEC crossover. Fermions interacting via a shape-resonance in the form of a well-barrier potential, within one-channel Hamiltonian treated in mean field. (**a**) Chemical potential $\mu$ at temperature $T=0$ vs. correlation length $k_F\xi$ in units of the Fermi wavevector $k_F^{-1}$. $\mu$ is normalized to the Fermi energy $E_F$ when $\mu>0$ (BCS side) and to half the binding energy $E_b/2$ of the composite bosons when $\mu<0$ (BEC side). The model parameters are the scattering length $a$, effective range $R^*$ in $k_F^{-1}$ units, and the diluteness parameter $nr_0^3$ in terms of the density $n$ and the width of the well $r_0$. Each symbol refers to a different realisation of the set of model parameters, as in the legend. (**b**) Same as in (**a**), but for the condensate fraction $2N_0/N$ at $T=0$ in units of the number $N$ of fermions. Notice how both quantities show universal behavior, irrespective of the microscopic details. From [607].

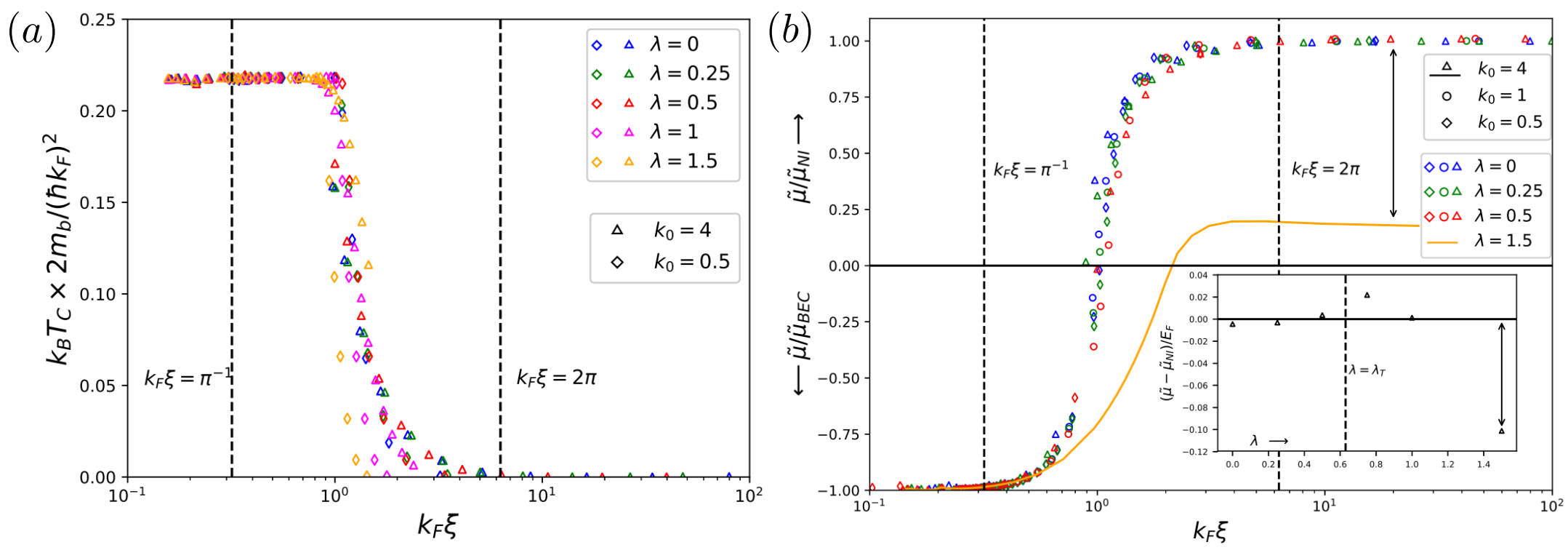

Figure 13: Universal behavior of fermionic superfluids in the BCS-BEC crossover. Fermions interacting via a model separable potential characterized by a strength $g$ and a range $k_0^{-1}$ like that introduced by NSR [555], in the additional presence of a spin-orbit coupling with strength $\lambda$, and treated within the conserving $GG_0$ approximation [599]. (**a**) Effective chemical potential $\tilde{\mu}$ at $T_c$ vs. the pair correlation length $k_F\xi$. $\tilde{\mu}$ is normalized as in Figure 12a. Each symbol refers to a different realisation of the set of model parameters $g$, $k_0$, and $\lambda$, as in the legend. In the inset, a specific region is examined, highlighting the difference between $\tilde{\mu}$ and its non-interacting value $\tilde{\mu}_{NI}$ as a function of $\lambda$. For larger $\lambda$ values, pair correlation lengths $k_F\xi>2\pi$ no longer correspond to a weakly interacting BCS regime, regardless of the interaction range (represented by $k_0=4$). (**b**) Log-plot of the critical temperature $T_c$ multiplied by the effective mass $m_b$ of the non-condensed resonant pairs, vs. $k_F\xi$. Different values of $\lambda$ are represented by symbols in the legend. Squares correspond to $k_0=0.5$, triangles to $k_0=4$. The coupling $g$ ranges from 1.25 to 30. In both (**a**,**b**), the vertical dashed lines at $k_F\xi=1/\pi$ and $k_F\xi=2\pi$ indicate the thresholds for the deep BEC and BCS regimes, respectively. Notice how both quantities show universal behavior, irrespective of the microscopic details, provided that $\lambda<\lambda_B$ be below the threshold for the transition to a topological state and the nature of the fluid changes. From [608] (Permission to use this content granted by Creative Commons Licence `https://creativecommons.org/licenses/by/4.0/`).

### *2. Commensurate—Incommensurate Transitions*

Let us now address a second paradigmatic problem that can be accessed with atomic technologies under controlled conditions, and lying outside the conventional case of the competition between kinetic and potential energies. Here, we consider quantum phase transitions due the existence of two types of potential energies for the system, that can lead to different configurations. This is the case when the phase transition is e.g., driven by the competition between two different length scales, yielding diverse complex phenomena. A non exhaustive list of this type of transitions, appearing in remarkably diverse either classical or quantum systems, include charge density waves [609], nano-contacts between solids [610], dislocations in crystals [611], adsorbed monolayers [612] in the form of noble gases on graphite substrates [4], bio-molecular transport [613], emergence of chaotic structures in metal-insulator transitions in Peierls systems [614] and spin glasses [615], Josephson junctions [616], quantum pinning in strongly-interacting bosonic fluids [617] and Meissner-to-vortex transition in bosonic ladders [27, 618].

The behavior of the system is highly dependent on whether the two lengths, $d$ and $a$, are commensurate (C) with a rational ratio or incommensurate (IC) with an irrational ratio. The way in which the system accommodates the incommensuration gives rise to various interesting phenomena. For example, one can have situations where incommensurations float within the commensurate phase [619] or else appear as interstitial phases separating the commensurate ones [4]. Generally enough, a C-IC transition can be characterized by the pinning of the system to an otherwise floating and translationally invariant phase. The translational invariance of the incommensurate phase manifests as a superlubricity and corresponds to the presence of gapless excitations, called phasons. Instead, an energy gap opens up in the excitation spectrum when in the commensurate phase, with the breaking of translational invariance. This gap amounts to the energy needed for a classical particle to surmount the energy barrier. Finally, defects can significantly contribute to the complex behavior and rich dynamics of the system in the pinned regime, their nature being in the form of walls, dislocations, or vortices [615].

In the following, we first discuss the model that has been extensively used to describe these physical phenomena, focusing on its realization as an Aubry-like transition [620] with atomic technologies, and the possibility of identifying emerging quantum effects. Then, we discuss one peculiar example falling in the C-IC universality class, the Meissner to vortex transition.

*Aubry-like Transitions* The paradigmatic model used to explain the physics of competing length scales consists of a one-dimensional (1D) chain of particles connected by harmonic springs. While at equilibrium, the particles are initially positioned at a distance $d$ from each other, when subjected to a periodic substrate lattice with a period $a$ a rich physical phenomena takes place. This model, originally proposed by Frenkel and Kontorova and by Frank and Van der Merwe model (FKVdM) [621, 622], indeed captures the essence of the competition between two key potentials.

In the FKVdM model, the competition in fact arises between the elastic potential, which favors a periodic structure with a characteristic period of $d$, and the lattice potential, which seeks to confine the particle positions to integer multiples of $a$. Consequently, when the lattice potential $V$ is weak, the particles positions are largely unaffected by the ratio $w \equiv d/a$, tending to float on the lattice. This corresponds to the incommensurate (IC) phase. On the other hand, above a critical threshold for the strength of the lattice potential, the particles tend to localize near its minima. This is the commensurate phase. In essence, they form a structured arrangement where the average spacing between particles is a rational multiple of $a$, where the lattice potential dominates over the elastic potential, causing the particles to align with the lattice in a commensurate manner.

To delve into the transition in more detail, we can introduce the concept of average spacing $\overline{d}$ between atoms in the chain and the so-called winding number $\tilde{w} = \overline{d}/a$. These quantities help providing further insight into the system behavior. At a lattice potential depth of $V = 0$, the winding number $\tilde{w}$ is equal to the ratio $w$, and the particles float without significant localization. However, as the lattice potential becomes non-zero, it tends to localize the particles. For non-rational values of $w$, the system accommodates this by allowing an increase in energy while maintaining the winding number $\tilde{w} = w$. This energy increase persists until it reaches a critical point where there is enough energy to form a discommensuration.

The C-IC transition, as predicted by Aubry [620], occurs as a second-order transition where the winding number $\tilde{w}$ abruptly jumps to the next rational number. A critical lattice strength $V_c$ exists, below which there are intervals of non-rational values of $w$ where $\tilde{w}$ remains irrational, that form a Cantor set with non-integral fractal dimension [4]. As the lattice strength exceeds the threshold $V > V_c$, the size of these intervals reduces to zero measure.

It is worth noting that the critical lattice depth $V_c$ depends on the specific irrational value of the length ratio $d/a$, and it attains its largest value at the golden ratio $(\sqrt{5} - 1)/2$. Remarkably, this transition persists even when the number of particles in the chain is finite [623].

The quantum version of the FKVdM model has been as well subject of extensive investigation, as reported in the study by Borgonovi [624]. Quantum Monte Carlo (QMC) studies, extending to systems with up to 144 particles, have provided insights into the impact of quantum effects on the Aubry transition [625]. Indicators analogous to those used in the classical transition, such as the hull function, particle position variance, and density-density correlation function, have been defined to identify the transition, along with an effective Planck constant measuring the degree of quantum behavior, and essentially dictated by the elastic energy and the lattice spacing [625, 626].

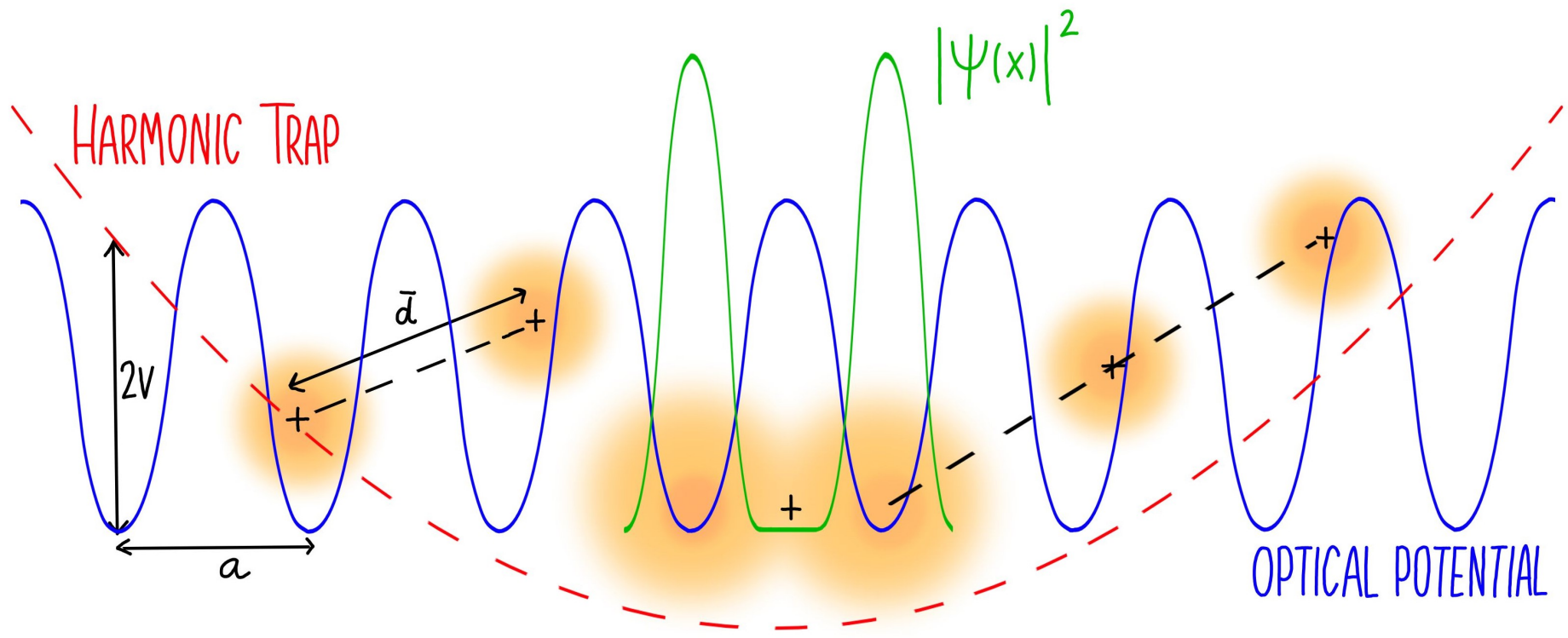

Figure 14: Aubry-like transition in ion platforms. Concept of the experiment in [28]. Laser-cooled trapped ions with charge $+|e|$ are arranged at an average distance $d$, determined by the interplay between Coulomb repulsion and external (harmonic) confinement. Superimposed is a periodic optical lattice with period $a$ and potential height $V$. The lengths $a$ and $d$ are chosen to have an irrational, incommensurate, ratio. In the commensurate phase, whenever $V > V_c$ exceeds a critical value $V_c$, ions pin into the bottom of the lattice wells forming commensurate segments separated by incommensurate regions. With increasing the level of incommensuration, pinning occurs at progressively larger $V_c$ values. Quantum effects arise from tunneling between lattice sites. When this happens, the quantum probability density $|\psi(x)|^2$ (here depicted for the central particle) becomes bimodal, and can be detected. Image inspired from [643] .

Density-Matrix Renormalization Group (DMRG) and Path-Integral Molecular Dynamics methods have been employed to investigate correlations in the C phase and the IC phase, respectively [627–629]. The excitation dynamics across the transition have been discussed, revealing a transformation from a pinned instanton glass to a sliding phonon gas [626]. The influence of long-range interactions has been explored in a study by Pokrovsky [630]. In a nonlocal FKVdM model considering long-range power-law interactions, the formation of kinks (topological solitons) and kink-antikink pairs has been characterized, with considerations for finite-size effects [631].

The relevance of the FKVdM model extends beyond simulations. It has been noted that the model relates to the structural zigzag instabilities of ion strings in optical resonators [632], as highlighted in studies by Cormick, Gangloff, and Fogarty [633–635]. Theoretical approaches also exist, including the description of the C-IC transition using an Ising model, as demonstrated in the influential work by P. Bak and R. Bruinsma [636].

Bylinskii et al. [28] have observed an Aubry-like transition in a small chain of trapped ultracold $Yb^+$ ions, investigating the dependence on temperature [637] and commensurability [638]. In the confined system, the competition between the period of the applied optical potential and the average interparticle spacing is seen to drive a transition between pinned and sliding arrangements. By employing precise atom-by-atom control and observation techniques, the transition has been explored from superlubricity to stick-slip behavior as a function of the optical potential height and the commensurability of the length scales. This experimental setup, depicted in Figure 14, provides a versatile and controllable platform to explore the fundamental aspects of friction at the nanoscale over a wide temperature range [639–641]. A similar ion-trap experiment has been conducted in 2D [642], providing insights on nanofriction and transport processes with atomic resolution, and enabling the observation of a soft vibrational mode during the transition.

The 1D experiment with ions is especially interesting because the power law $\alpha = 1$ of the Coulomb interaction equals the dimensionality $d = 1$, and it is well known that under such conditions the interactions are long-range in nature. One should therefore ask whether the detailed knowledge developed on the FKVdM model, the Aubry transition, and the quantum effects remain unaltered and, in particular, whether the quantum effects might be observable or washed out. An answer in the case of a finite-size system comes from a Path-Integral Monte Carlo (PIMC) simulation study [643] that replicates the finite temperature and other conditions of the experiment [28]. Here, the signatures of quantum effects have been identified from the analysis of the Binder cumulant $B$, an additional indicator not previously considered. $B$ indeed, continuously evolves from zero for monomodal to 2/3 for bimodal distributions, in this case signaling the existence of a superposition state in adjacent lattice sites. resulting phase diagram is depicted in Figure 15 from [643]. A feasible experimental strategy has been envisioned to detect the emergence of these quantum effects, which in the simulation have demonstrated to remain robust in the presence of thermal fluctuations at temperatures already achieved in experiments, and distinguishable even in the relatively small systems that are currently available [643]. Notwithstanding this wide range of existing studies, a number of crucial questions regarding the quantum Aubry transition remain open, among which is the relation between the C-IC transition and the paradigm of many-body localization (see Section III B 1), the degree of universality across different dimensions and length scales, also depending on the range of the interactions among the particles, or the implications of the presence of disorder.

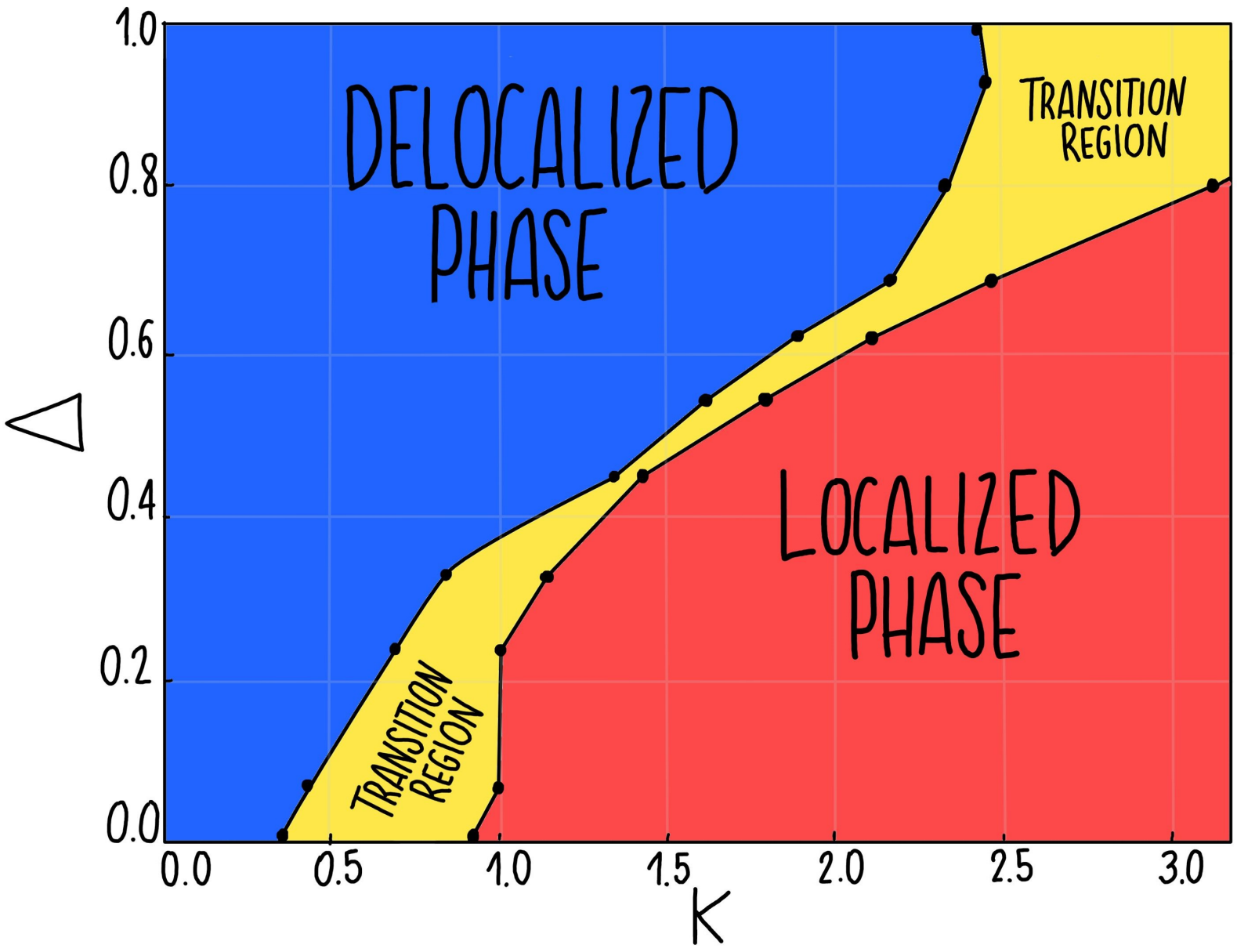

Figure 15: Commensurate-incommensurate physics. Phase diagram of the Aubry transition for the experimental realisation in Figure 14. The variance of the central-particle density probability is used as an indicator for the transition in the space of the governing parameters, that are the incommensuration degree $\Delta$ and the normalised height $K$ of the lattice potential. Red region: localized phase. Blue region: delocalized phase. Yellow region: intermediate phase. Figure inspired from [643] (Permission to use this content granted by Creative Commons Licence `https://creativecommons.org/licenses/by/4.0/`).

*Meissner to Vortex Transition* It is known that superconductors subjected to an external magnetic field $H < H_{c1}$ exhibit the Meissner-Ochsenfeld effect, where surface currents effectively shield the magnetic field within the bulk, resulting in perfect diamagnetism [644]. In type-I superconductors, the normal state is restored when $H > H_{c1}$. However, in type-II superconductors, a vortex phase is formed for $H_{c1} < H < H_{c2}$, where the magnetic field partially penetrates the system in the form of flux lines surrounded by screening currents. The transition from a Meissner to a vortex state is an example that falls in the commensurate-incommensurate universality class [645, 646]. Indeed, in the commensurate Meissner state, the flux lines are homogeneously distributed throughout the material due to the expulsion of magnetic fields via the Meissner effect. However, as external magnetic fields increase beyond a critical threshold, the material undergoes a phase transition characterized by the formation of vortex lattice structures. These vortices arise due to the penetration of magnetic flux lines, leading to the breakdown of commensurability between the vortex lattice and the underlying crystal lattice. The incommensurate vortex state exhibits a spatially modulated arrangement of vortices, resulting in a complex magnetic field distribution, accompanied by diverse physical phenomena, such as vortex pinning, flux creep, and Josephson coupling. In material science, a number of factors can affect the detailed manner with which the transition occurs, such as temperature, magnetic field strength, material composition, and crystal symmetry, and their study can be relevant to unveil the interplay between superconductivity, lattice structure, and magnetic field dynamics.

A striking demonstration of the Meissner to vortex transition has been realized in quasi-one dimensional noninteracting bosonic ladders of ultracold atoms [27], subjected to external synthetic gauge fields as predicted to occur theoretically [647, 648] and experimentally engineered [649]. The type of experimental concept is also depicted in Section III B 2 and the corresponding figure  In this ladder arrangement, atoms sit in two different (pseudospin) states, with the possibility of tunneling in one-dimensional optical lattices, and are coupled by light fields across the ladder rungs. The use of synthetic dimensions can overcome the Mermin-Wagner-Hohenberg theorem, otherwise prohibiting the $U(1)$ spontaneous symmetry breaking needed for the superfluid state to appear [650, 651].

In particular, an analogous Meissner phase is predicted to exist in the ground state for low flux of the synthetic gauge field, while a Tomonaga-Luttinger liquid (TLL) of vortices is expected for higher flux [652–654]. Density-Matrix Renormalization Group (DMRG) studies combined with bosonization methods have been conducted with increasing flux at different fillings, finding that in the presence of on-site hard-core repulsion among the bosons determines a persistence of the Meissner phase over the vortex state at any flux above a minimum interchain hopping [655]. When the applied flux is $\rho\pi$ or close to it, with $\rho$ the filling per rung, a second incommensuration is found in the vortex state [618] (see Figure 16).

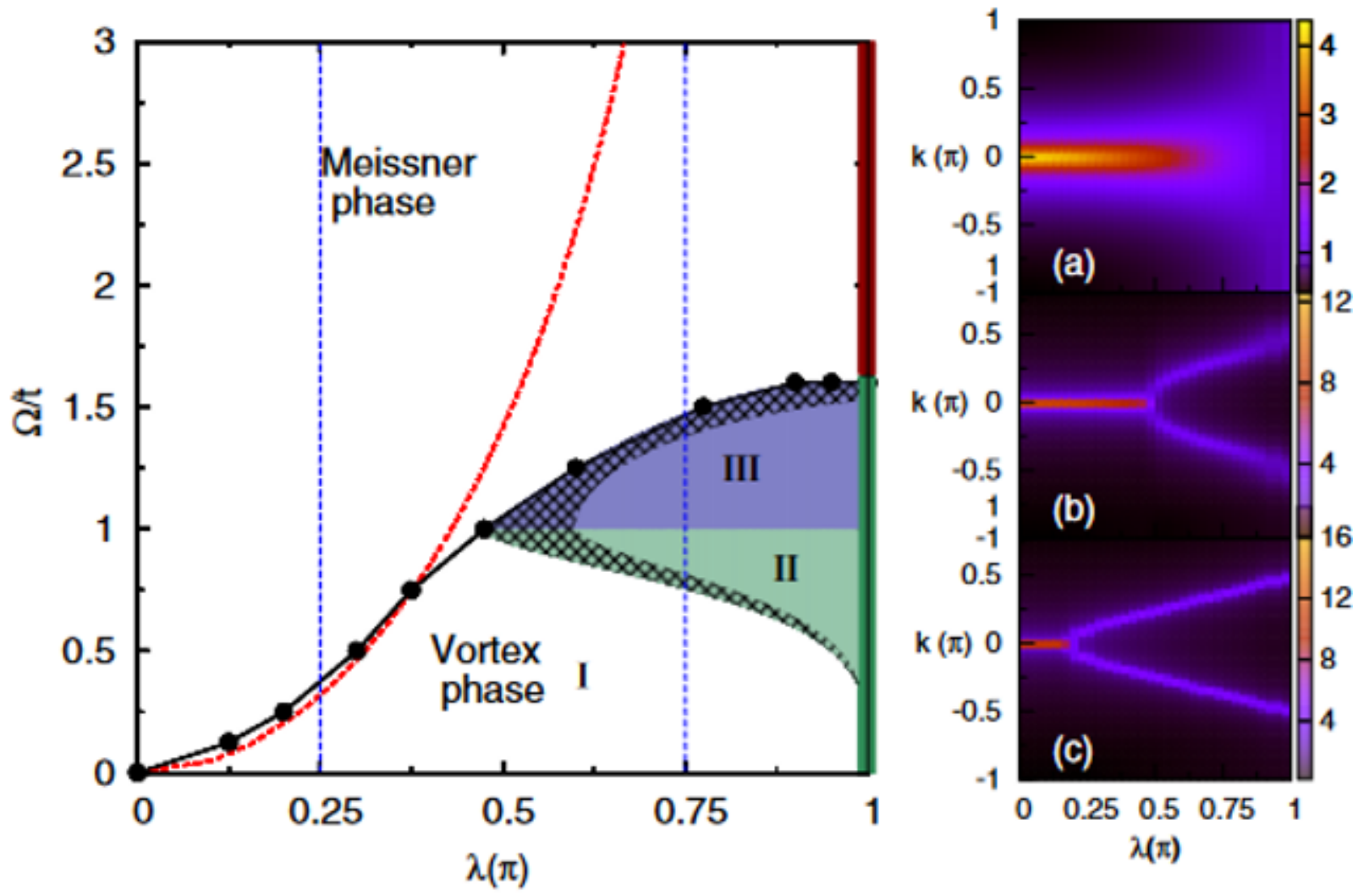

Figure 16: Commensurate-incommensurate physics in the Meissner−to−vortex transition. DMRG prediction of the phase diagram for experiments of the type of Atala et al. [27] on a two−leg ladder (see also Figure 18c), but in the case of hard−core spinless bosons case, i.e., infinitely repulsive on-site bosons. Left panel: phase diagram in the space of the governing parameters, that are the flux $\lambda$ of the synthetic gauge field per plaquette, and the coupling $\Omega/t$ between across the rungs in units of the tunneling $t$ along the legs. Black solid line: phase boundary between the Meissner and vortex phases. Compared to the non−interacting case (red dashed line), the hard-core nature of the bosons favors the persistence of the Meissner phase above the threshold $\Omega > \Omega_c$ for all fluxes $\lambda$, except at $\lambda = \pi$. Shaded area: a second incommensuration appears. In the green region (region II), additional peaks at $k = \pi$ emerge in the Fourier transform of the rung current correlations, which dominate in the blue region (region III). The double line (green vs dark red) at $\lambda = \pi$ represents the transition to a localized phase. Right panel: Intensity plots of the momentum distribution $n(k)$ versus $\lambda$ and $k$. (**a**) $\Omega/t = 1.75$ in the Meissner phase, characterised by one single maximum at $k = 0$ for all $\lambda$. At $\lambda = \pi$, $n(k) = 1$, corresponding to the formation of a fully localized state (dark-red solid line). (**b**,**c**) for $\Omega/t = 1$ and 0.25, respectively, showing the transition from the Meissner phase to the vortex phase, characterised by two maxima of $n(k)$ symmetrically located around $k = 0$. Figure from [655].

Additional predicted orderings include chiral superfluid order at half a flux quantum per plaquette [652, 656, 657] and a chiral Mott insulating phase [658–662], which exhibits both chiral currents and a spin-density-wave phase. Diagonal interchain hopping in ladder structures has also been studied using Density Matrix Renormalization Group (DMRG) methods [663–668].

## B. Condensed Matter: Sprouting Paradigms

### 1. *Eigenstate Thermalisation Hypothesis and Anderson/Many-Body Localisation*

In this section, we focus on the emergence of thermal behaviour in isolated quantum many-body systems and scenarios where this behaviour is violated, as it is the case of localised states. As it is well-known, an isolated quantum system undergoes unitary, hence reversible, evolution and should retain information of its initial configuration but instead the quantum system is observed to thermalise. Quantum thermalisation is often described in literature in terms of quantum chaos and ergodicity (Two elusively different concepts as an ergodic system can present non-divergent dependence to its initial conditions.) [669–672]. For our brief description, we focus on the concept of ergodicity: an ergodic quantum many-body system is that explore all configurations permitted by its conservation laws, as a consequence it is also one whose local d.o.f. entangle with each other leading to the decoherence of local quantum correlations and to macroscopic observables becoming stationary after a slow classical hydrodynamical evolution [673]. These observables become thermal and can then be described in terms of statistical mechanics and loose any dependence on the initial condition. Importantly, this process should occur for any state of the system, even those forming its eigenbasis which are stationary. Thus, this process should already occur at the level of individual eigenstates, this is denoted as the

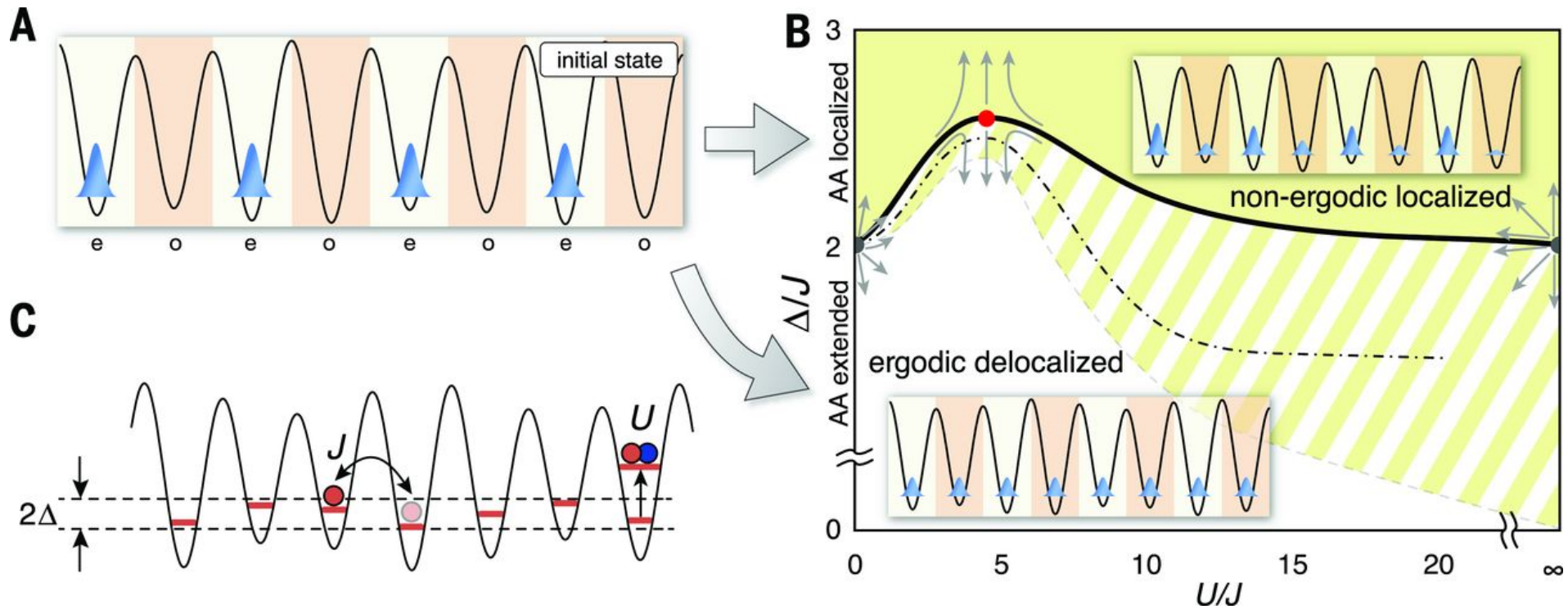

Figure 17: Schematic experimental detection protocol of Many-Body Localization (MBL). (**A**) The system is initially prepared in a charge-density state with half-filling, where a particle/doublon is present in every second lattice site. Over time, particles can tunnel with rate $J$ to neighbouring sites and doublons experience a local energy offset $U$ due to particle interactions and local random disorder of order *Delta*. (**B**) A phenomenological phase diagram illustrates the MBL transition, reflecting the ergodicity properties of the system. In the localization phase (yellow region), the initial configuration persists for long periods of time, in contrast to the ergodic delocalized phase (white region), where the particles evenly distribute across the lattice. The stripped region represents the variation in the transition point based on the initial number of doublons, highlighting the interplay between disorder $\Delta$ and interaction $U$, which are factors included in the Hamiltonian contributions schematically depicted in (**C**). Figure from [33].

Eigenstate Thermalisation Hypothesis (ETH) [672].

While AMO platforms have allowed for the understanding of many aspect of quantum thermalisation given their ability to measure local and global observables and hold conditions of quasi-isolation for long system timescales allowing to observe thermalisation in controlled quantum system, in cold atoms [674] and more recently in trapped ions [675], we would like to focus in the class of systems that fail to exhibit this behaviour due to localization.

Localization was first predicted by Anderson as a consequence of quantum interference at the level of single non-interacting particles in the presence of random potentials or tunneling rates [676] already for infinitesimal disorder rates. It took close to 40 years for this to be observed experimentally [677], since then in a large variety of systems [678, 679], including cold atom platforms [680–682]. The localization mechanism was also extended and observed in the case of interacting particles, a phenomenon known as many-body localization (MBL) [683–685]. In contrast to the Anderson localization case, with interacting particles all eigenstates exhibit localization for high disorder values [686], while for smaller disorders rates current theoretical and numerical evidence predicts the localization of a finite subset of the eigenstates in one-dimensional systems [687]. Thus, in the interacting MBL systems the violation of ETH and non-ergodic behaviour occurs at a finite disorder rate with the system thermalising below the critical value. This was predicted [688] and experimentally observed in random and quasi-periodic models [689] initially in 1D [33, 690], and later on in quasi-2D [691] and 2D [692]. In these experiments, it is actually the degree of ergodicity that is tested in order to detect an MBL phase by measuring the degree of survival at long times of an initial space-imbalanced configuration, see Figure 17. Since then, MBL has been studied in a large range of non-disordered scenarios, such as dynamically constraint systems [693–695]. Moreover, the long-time scales required to reach the steady configuration [696] together with the technological advances making longer experimental running times possible, makes it necessary to consider scenarios where the system is coupled to its environment [697] where the signatures of the transition or the long-time steady states are modified by the rate of dephasing [698–700] and particle losses [701], leading in certain regimes to the breaking of the localised phases. From the theoretical perspective, MBL has been characterize in terms of the entanglement growth rate [683, 684], by the appearance of local conserved quantities [702] or by the information spreading of system subregions [703].

While MBL has been discussed in literature for a relatively long period, this paradigm offers an ideal platform for the testing and understanding of thermalisation in quantum systems, leaving a number of open questions. In particular, challenges remain on the understanding of the mechanism beyond 1D [704] and particularly in parameter regions near the phase transition [705], where thermal rare regions appear and can lead to eventual thermalisation [706–709]. Only very recent experiments tackle the characterization of the critical behaviour [710] or have been able to measure the entanglement content in a vicinity of them [526]. Other relevant directions include its combinations with topological phases [711, 712], its relationship with glass transitions [693–695] and with periodically-driven systems and time crystals [713].

Notably, in the recent years, the community's attention has focused on the understanding and generation of another example of non-ergodic behaviour, that are the so-called quantum many-body scars. These are examples of weak-breaking of ergodicity, where abnormally long relaxation times linked to non-thermal high energy eigenstates—and thus non-thermal behaviour—can be detected for certain initial conditions. These states, first observed in Rydberg-atoms simulators [56], do not only challenge the general applicability of ETH and our understanding of thermalization and chaos in quantum systems, but have also been predicted to be relevant in describing the emergence of glassy dynamics, certain lattice gauge theories or fast scrambling [714]. Moreover, their naturally long coherence times can be seen as a useful resource in QT for state preparation, with some proposals already been posed towards their applicability in quantum sensing [715]. While a deeper understanding of the systematic preparation of many-body scars, their theoretical classification or the discovery of a wider range of systems that exhibit them is still under development we expect useful applications in the coming years. For details on the current state of these phenomena we refer to the recent topical reviews [714, 716].

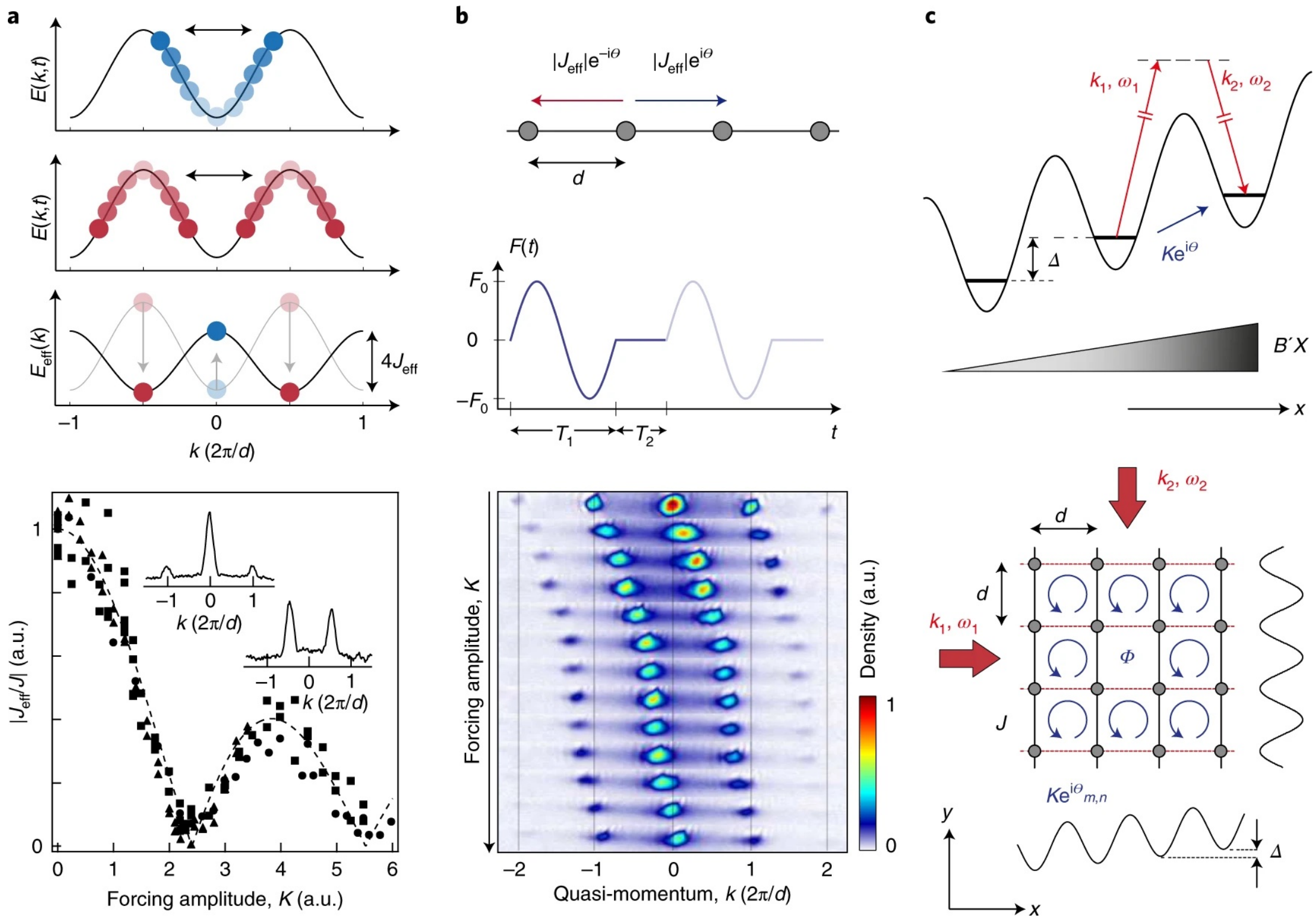

Figure 18: Floquet engineering: summary of the main protocols. (**a**) The behavior of atoms in a shaken optical lattice can be described by an effective tunneling constant $J_{\text{eff}}$, which can take negative values within certain parameter ranges. When $J_{\text{eff}} < 0$, condensation occurs at new minima (shown in red) in the distribution instead of the previous minima centered at zero quasi-momentum (shown in blue). The effective tunneling is experimentally estimated in a Bose-Einstein condensate (BEC) by measuring the suppression of diffusion. (**b**) If the driving force is asymmetric in time, this lattice distortion leads to a change in the energy bands, which amounts to the introduction of Peierls phases [717] to the effective tunneling elements. By controlling the amplitude of the drive, the positions of the minima in the quasi-momentum distribution can be adjusted, allowing the creation of condensates with finite momenta, as observed in experimental observations. (**c**) By combining field gradients with the imprinting of Peierls phases, it becomes possible to engineer intriguing plaquette models with controlled effective flux. Figure from [718].

### *2. Periodically-Driven Systems: Floquet Engineering*

When describing some of the potential uses of cold atomic platforms in Section II A 3 we already highlighted the possibility of engineering the system into the generation of synthetic dimensions as a versatile tool for quantum simulation, e.g., in designing synthetic bosonic ladders [649] for the studies in Section III A 2, or for applications to fundamental interactions, see Section IV B. In this section, we describe Floquet engineering, an already well-established field, which uses periodic coherent driving to modify the behaviour of a quantum system. We include a description of the basic concept of the protocol in the case of a BEC in Figure 18, including condensation at finite momenta and the induction of complex tunneling phases.

Periodically driving a cold atomic platform allows for the generation of synthetic gauge fields that appear in the form of additional—generally complex—phases in an effective Hamiltonian picture, mimicking the effects of the desired static magnetic field in a controllable way [719, 720]. The continuous driving of the system generates non-equilibrium steady states for simulation [718, 721], complementing other approaches of material and solid-state design. We include in Figure 18 a summary of the main protocols accessible through Floquet engineering, that constitute a well-established tool in current quantum technologies. These mechanisms have allowed to access unexplored parameter regimes and configurations. Among other early examples, a partial list includes: the creation of effective spin-orbit coupling for the study of paradigmatic condensed matter magnetism phenomena such as the Quantum Hall effect [722], the study of frustrated magnetic systems [723], the creation of space-dependent tunneling [724], the study of phenomena in ladder systems such as Vortex superfluidity [725] or the preparation of topological insulators in driven superconductors [726, 727]. Another advantage of using Floquet theory is the fact that, despite it applies to the description of systems out of equilibrium, its theoretical formulation remains numerically treatable, see e.g., [721].

Finally, there is a subset of problems in driven systems that has attracted particular attention in recent years. That is the case for instance of accelerated counter-adiabatic protocols for quantum control and information ([728] and references therein), and of boundary-driven systems [729]. While the drive has been introduced in a variety of frameworks, both dissipatively [730, 731] and coherently-periodic [732], these model systems constitute an important test-bed for the understanding of transport phenomena not only in quantum matter and technologies [733] but also in biology [734, 735], as we discuss in Section VII. Moreover, these tools have been applied in the simulation of a set of models included in this review. In particular in IC-C transitions in condensed matter, see Section III A 2 and in the simulation of fundamental interactions, see Section IV B.

#### *3. Measurement, Control and Feedback in Open Quantum Systems*

In Section II B 1, we discussed how the measurement of a quantum system can be treated theoretically, both when the measurement is due to some apparatus or instead via some non-monitored degrees of freedom of the environment. The measurement process can lead to dephasing and decoherence, heating [36], freezing of the system's dynamics due to the quantum Zeno effect [736], and even induce a phase transition as we present here and in Section III B 4.

Here, we highlight how the measurement process, typically in a weakly continuous manner as described in Section II B 1, can also be combined with control methods (see Section II C 3) so that the measurement outcome is classically processed and fed back to perform a conditional action onto the physical system steering it into a desired state (see also Figure 19). This framework is often referred to as quantum measurement and feedback control [304]. This is similar to what occurs in error correction (Section II C 1), where the classical outcome of the ancillary qubit is processed in order to act on the rest of the system to mitigate some operational errors. The quantum feedback protocol has been successfully implemented to: improve cooling schemes in both trapped ions [737] and neutral atoms [738] by coherently, and not dissipatively, coupling the system to a radiation mode; generate strongly correlated states [739]; induce phase transitions [740]; generate novel non-equilibrium steady states [741]; prepare self-organised phases [742] or, recently, cool BECs in 2D preventing the effects of measurement back-action [743]. In addition, optimal control theory has also been applied in quantum devices in the presence of dissipation [744].

While these protocols are rather system-dependent, we believe that current technological developments will make their use more frequent as a framework for quantum simulation [745, 746], particularly for AMO systems in optical cavities (Section II A 7). These mechanisms for the preparation of programmable out-of-equilibrium systems can be combined with the existing coherent and dissipative approaches, some of which are described in Sections II B 1 and III B 2, composing an evergrowing toolbox for the canvas of fundamental and applied use cases of quantum technologies that we discuss in subsequent sections.

#### *4. Out-of-Equilibrium Systems, Dynamical Phase Transitions and Dissipation*

Before discussing the applications of quantum simulators in a different fields, we would like to bring the attention to a final set of tools that have proven useful in the recent years when discussing driven-dissipative quantum systems. In Sections III B 1 and III B 2, we have discussed instances of non-thermalising quantum systems due to disorder or driving. This is also true in more general scenarios like in integrable models or after performing specific quantum quenches, where long-lasting quasi-stationary states can be observed [747]. In this context, the first examples of dynamical phase transitions in driven quantum systems have been found [748].

Dynamical phase transitions are well understood and characterised in classical systems [749] where the universal behaviour arises—both in time and space—after a quench were parameters cross an equilibrium phase transition point. Similarly, we find instances of phase transitions of dissipative nature related to the gap closing of the evolution superoperator in open system [750, 751] with the associated change in the steady state properties. Consequently, efforts for the characterisation of these different phenomena in quantum systems have seen a considerable rise in recent years in order to extend existing approaches (as those based on Green functions [300]) and create a generalised description of dissipative non-thermal systems with their related phase transitions. While the lack of an equilibrium statistical description could seem a problem, it also allows to engineer states and phenomena that are not possible in such scenarios. Moreover, there has been progress in their formalisation particularly in terms of Keldysh approaches [290, 430] and some of the transitions have already experimentally observed, with the first examples here [752, 753]. For a detailed description, we refer to the following excellent topical reviews [430, 748].

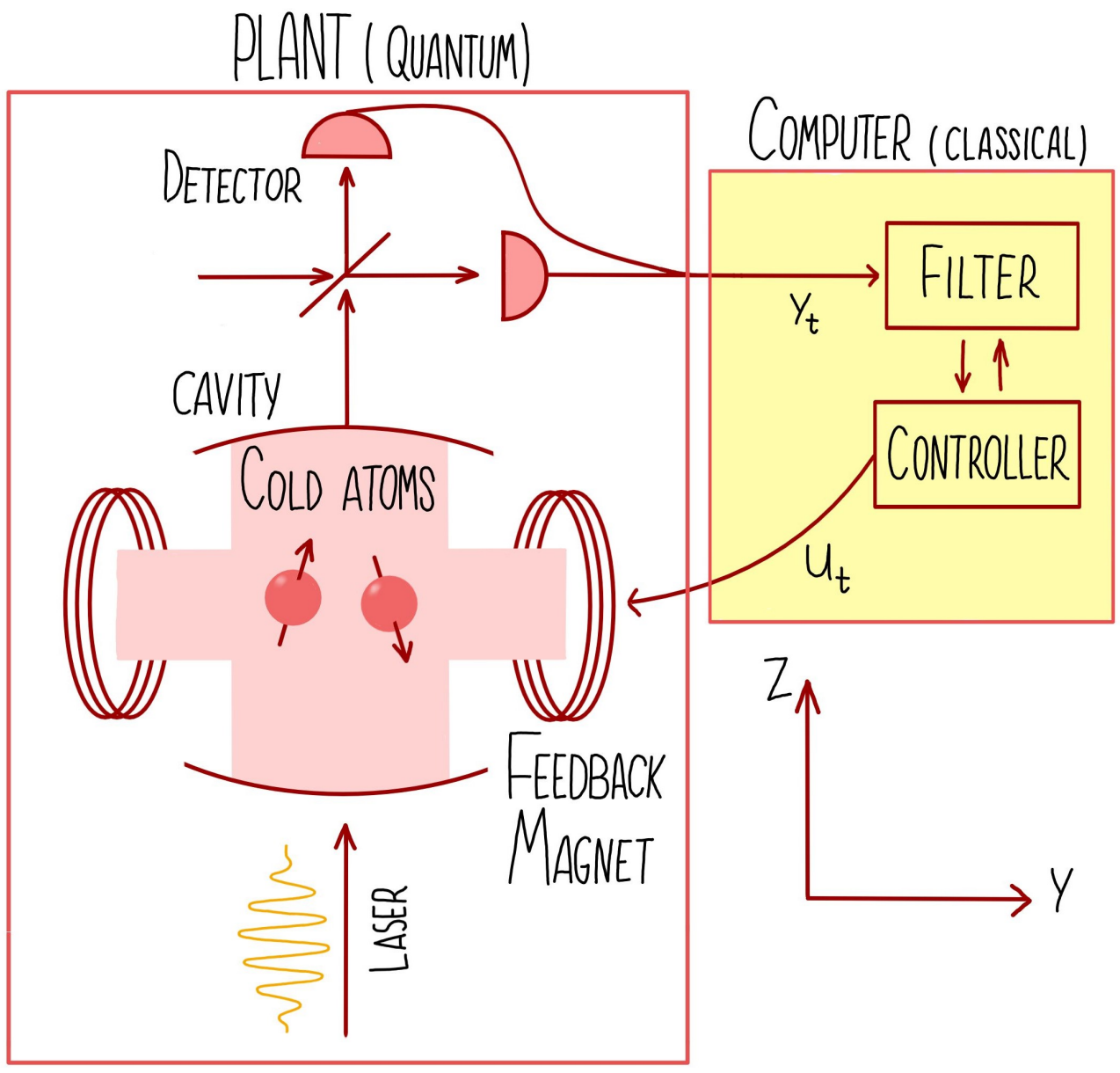

Figure 19: Diagram of a quantum measurement and feedback protocol. The AMO platform, in this case atoms in an optical cavity, is weakly probed via some homodyne-like measurement. The outcome is classically processed by a computer that modifies the feedback controller accordingly in order to steer the state to the desired phase.

One class of dynamical phase transitions that has attracted a large interest is that of measurement-induced phase transitions (MIPT). These have been observed both in gate-based [292] and in continuous [754] systems. MIPT arise from the competition between the creation of entanglement by the coherent system dynamics and its destruction due to the projective nature of measurements. This interplay can lead to drastically different properties in the individual realisations of the dynamics, while their average density operator at long times is a trivial infinite-temperature state due to the heating induced by the measures, regardless of their rate. The signatures of these transition, which are only present in non-linear quantities of the state, are then elusive to experimental probing – as that would typically require to measure a system twice. Since then, several proposals have appeared using ancillary degrees of freedom, establishing links with purification phase transitions [755, 756]. Importantly, links with quantum error correction and information scrambling have been established [757, 758], leading to links with relevant questions in cosmology that can be proved via quantum simulation [115].

## IV. FUNDAMENTAL PHYSICS

Section III has highlighted several relevant new paradigms, or old paradigms with new implications, related to many-body quantum matter physics, that can be explored by means of cold and ultracold atomic quantum platforms in perspective of their current and potential technological applications. However, many-body quantum matter technologies can be applied in other sectors of physics. We dedicate this section to applications in fundamental physics.

In this field, our understanding relies on four fundamental theories, Quantum Field Theory (QFT) and the Standard Model of particle physics (SMPP) on the one side, and General Relativity (GR) and the Standard Model of cosmology (SMC) on the other. While each of them has resisted to a wide range of testing over decades up to more than a century, a unified description has yet to be found. A number of questions seek a definite answer. Examples are the unification of GR with other SMPP interactions, the account for Dark energy (DE) and Dark matter (DM) in the SMPP, and the discrepancy of the vacuum energy density expected from the SMPP, $\sim$ 40–120 orders of magnitude larger than observed DE.

The observation of the Higgs boson at the LHC confirmed the final piece predicted by the standard model of particle physics. While it has been tested in detail over the years, the Standard Model of particle physics cannot explain certain universe observations. Therefore, new particles, fields, and forces must be introduced in order to have a more comprehensive understanding of the behavior of elementary particles and forces from shortly after the Big Bang through to the creation of atoms, molecules, stars, and galaxies. Experimental particle physics has focused on the discovery of these new particles and forces beyond those in the SMPP for decades, through very-high-energy particle colliders, and direct evidence searches for dark matter. To date, no new particles have been found. However, with advances in atomic, molecular, and optical physics techniques and quantum-limited measurement devices, it is now possible to explore these questions with table-top experiments based on a completely different concept, that is the sensitive measurement of extremely small energy shifts in quantum-mechanical resonances, caused by the existence of such new SMPP fields. It is worth noting that these novel types of experiments allow to push the frontiers of fundamental physics as a complementary approach, but also in certain occasions to outperform traditional methods [47].

In this section, we thus overview the main problems in fundamental physics that have been or potentially can be addressed by means of cold and ultracold-atomic platforms, that are linked to tests of the Standard Model of particle physics. We then refer to Sections V and VI for the discussion of foundational problems in General Relativity and Quantum Mechanics, respectively.

We notice that dark matter and dark energy are preeminent examples of unknown systems in physics, which do not fit the SMPP for particle physics, and which have deep connections with cosmology. As a result, we choose to discuss them in Section V. Here, we first necessarily focus on a third unanswered question, that is how the atoms making the visible universe managed to survive the Big Bang, which is connected to the matter-antimatter asymmetry. We then dive into the possibilities offered by atomic quantum technologies to make quantum simulators for lattice-gauge field theories and implement analogue models to investigate nuclear and quark matter.

### A. Electric Dipole Moment

The generation of a matter-antimatter asymmetry in the universe requires forces that violate the combination of charge conjugation (C) and parity transformation (P) symmetries, i.e., CP symmetry. At the same time, CP violation is always accompanied by a violation of time-reversal (T) symmetry, while the combined symmetry CPT is preserved, as for every theory with Poincaré invariance, as it is the case for the SMPP. The CP violation also results in the fact that ordinary particles having a magnetic dipole moment, also develop an electric dipole moment (EDM). This is the case of the neutron, whose possible (but not observed so far) EDM would signal CP violation in the strong (baryonic) sector of the SM.

Contrary to the strong sector, the CP symmetry is violated in weak (leptonic) sectors of the SMPP, however the associated effects cannot suffice to account for the observed matter-antimatter asymmetry. Beyond the SMPP, if new particles and forces are present, these can induce CP violation, leading for instance to larger EDMs compared to those predicted by the SMPP [759].

Experimental efforts have focused on detecting EDMs. The experiment concept is to orient the spin of an electron or a nucleus in a given direction, by means of polarized laser light resonant with a transition between atomic or molecular energy levels. If the particle were to possess an EDM, application of an electric field perpendicular to the spin would result into a torque, which in turn would be detected in the form of a spin precession about the direction of the field [47].

These experiments enhance their sensitivity by maximizing the strength of the electric field, the time of the spin-field interaction, and the number of particles observed.

In the ACME experiment in Harvard [46], electrons bound inside an ultracold, slow, and intense beam of polar molecules, experience an intense effective electric field of $\approx 10^{11}$ V/cm for about 1 ms. In the JILA experiment [45], molecular ions have been trapped by a rotating electric field, so that despite the lower number of detected molecules, the observation time for each of them could reach nearly 1 s. A third type of experiment uses $^{199}$Hg nuclei bound in Hg atoms: these are exposed to fields of $10^4$ V/cm and confined in small transparent cells, the nuclear spins remaining polarized for more than 100 s [760].

Though no finite EDM has been detected so far by any of these experiments, these three experiments are characterized by remarkable sensitivities, in fact $|d_e| < 9.3 \cdot 10^{-29}$ $e\cdot$ cm, $|d_e| < 1.3 \cdot 10^{-28} e\cdot$ cm with 90% confidence and $|d_e| < 7.4 \cdot 10^{-30} e\cdot$ cm with 95% confidence, respectively. A more recent experiment at JILA improves on the previous best upper bound by a factor of about 2.4, providing constraints on new physics above electron volts, beyond the direct reach of the current particle colliders [761].

### B. Fundamental Equations and Symmetries

Ultracold atoms and trapped ions reveal to be ideal platforms for the simulation of analogous phenomena of relativistic quantum mechanics and quantum field theory.

In the framework of relativistic quantum mechanics, the first notable experimental example has been the realization of the honeycomb lattice [32], known to host at half filling low-energy excitations effectively described as (3+1)D-Weyl (that is massless) fermions, moving in a (2+1)-dimensional space [762–764]. A similar spectrum has been proved on a square lattice pierced by a magnetic $\pi$-flux, $\Phi = \pi = \mathbf{B} \cdot \mathbf{S}$ per plaquette, $\mathbf{S}$ being the plaquette area vector [765]. Indeed, a continuous map exists between the two lattice schemes, see e.g., [766].

Later on, Dirac mass terms have been conceived for Weyl fermions [767] also via the Wilson mechanism (a method to give mass to lattice fermions avoiding ambiguities from different lattice momenta [768]) [769] In the latter approach, a direct effective realization of axion electrodynamics (possibly from a Peccei-Quinn scheme, a proposed method for the solution of the CP problem in the strong sector, exactly based on a similar (pseudo)-scalar boson [770]) is obtained [769, 771]. Following similar ideas, proposals have been conceived for the Majorana equation [772] (designed for trapped ions platforms) and for Majorana fermions, that means massive fermions with Majorana mass terms [773]. Finally, still concerning relativistic quantum mechanics, we report the theoretical proposal for Nambu-Jona-Lasinio models [774, 775]) in [767], and low-dimensional fermionic models, as the Thirring and Gross-Neveu models [776]. All these examples are pure fermionic interacting models, originally proposed as toy models for QCD and related phenomena, as chiral symmetry breaking.

These notable achievements become even more interesting by the possibility, theoretical and partly experimental, of synthesizing static gauge potentials and dynamical (gauge) fields, both Abelian and non-Abelian.

More in detail, Abelian gauge potentials, i.e., gauge configurations fixed from outside without their own dynamics, have been realized experimentally on two-dimensional systems, both in the continuous space and for lattice systems. Indeed, Abelian electric potentials are realized relatively easily via space-dependent energy offsets. Concerning the achievement of Abelian magnetic potentials, the more successful approach exploits optical transitions between atomic hyperfine levels to accumulate a suitable Berry phase along close spatial loops [117, 777]. A similar approach can be designed for non-Abelian potentials, accumulating the product of matrix unitary operators, instead of phases, acting on multiplets representing the chosen non-Abelian group symmetry. In the Abelian case, magnetic potentials with moderate intensity—typically strong enough to probe the Landau levels regime and related phases, as in probing the integer/fractional quantum Hall phase—can be obtained via lattice rotations, also in the presence of parabolic trapping [777–779].

More involved, even theoretically, and experimentally challenging, is the synthesis of dynamical gauge fields with proper dynamics. Various schemes have been proposed so far. The scheme by [780, 781], that first triggered the recent efforts in ultracold atoms, is a direct approximation of lattice gauge theory [768], valid both for Abelian and non-Abelian theories, and uses the so-called quantum-link models [782–785]. The latter are based on the use of link operators organized in discrete sets, allowing to approximate the unitary gauge transformations on the links of a lattice formulation. For a $SU(2)$ theory, the same operators can be chosen as angular momentum representations. Quantum link models have been proven to be realizable via gauge invariant quantum link models, both Abelian and non-Abelian, and can be exactly described in terms of tensor networks [784, 786–788]. This fundamental ingredient, yielding a direct link with Section II B 2, allowed the direct simulation on classical computers.

Other proposals for the simulation of gauge fields, equally recent, involve specific dynamics of Bose-Einstein condensates [789, 790], long-range correlated honeycomb lattices [791] exploiting the fact that gauge dynamics characterizes the low-energy behaviour of systems hosting topological order [792], Floquet dynamics [727] as described in III B 2, or internal degrees of freedom of single atoms [793]. Besides these analog quantum simulators proposals, digital proposals also appeared, mimicking the gauge-dynamics time evolution via stroboscopic approaches [172, 794].

The mentioned lattices setups with effective gauge fields have been used to describe and partly simulate numerically a number of remarkable physical phenomena, ranging from color confinement and baryons formation, to confining string breaking and vacuum polarization, i.e., production of lepton-antilepton pairs in the presence of ultra-intense electromagnetic fields (see e.g., [784] and references therein).

Next to gauge symmetries, flavour symmetries play a relevant role in fundamental physics [795]. While the experimental realization of color dynamics, as described above, is still lacking on cold and ultracold-atomic setups, fermionic mixtures with effective flavour dynamics are realized using earth-alkaline or lanthanide atoms [117, 777, 796, 797]. This is because these atoms are characterized by interactions that do not depend on the hyperfine levels of the multiplet they belong to. Identifying these levels with flavour degrees of freedom, an effective flavour-invariant interaction can be modelled and, being also the atomic mass not dependent on the same levels, effective $U(N)$ or $SU(N)$ invariance flavour groups are realized. The particular features of earth-alkaline or lanthanide atoms also allowed to simulate synthetic compact dimensions, using the levels of these atoms as effective sites of an hyperlattice along the synthetic dimension [798]. In addition, synthetic gauge fields involving synthetic dimensions have also been realized [799]. Notably, similar techniques also offered an alternative route towards experimental realizations of topological phases, as quantum Hall phases [800].

The issue of symmetries simulation involves unavoidably their spontaneous breaking [801, 802]. When concerning global symmetries, this phenomenon is ubiquitous in the context of ultracold atoms physics.

The first notable example is the presence of Goldstone bosons in superfluids, due to the spontaneous breaking of the global $U(1)$-symmetry related to the conservation of the number of bosons. Goldstone bosons mostly appear in condensates as propagating density excitations, i.e., sounds, see e.g. [802]. A similar phenomenology can be observed in superconductive phases of ultracold fermions [117, 777], possible also in the presence of gauge potentials. It is important to note that gauge potentials do not allow Higgs dynamics, but only Goldstone dynamics, due to the lack of dynamics for them. In turn, this results in the lack of dynamics for photons, that are not present in the theory as propagating excitations [801].

As we discuss below, the problem is substantially harder for the Higgs mechanism. Higgs effective excitations have been realized first in fermionic setups, at the transition between superfluid and Mott phases [26, 803–805]. In perspective, in ultracold setups of fermions with colour degrees of freedom and corresponding gauge dynamics, as of the types described above, a Higgs realization can be expected from a di-fermion condensate (a non-vanishing vacuum expectation value for a proper fermionic bilinear operator $\psi_\alpha \, \psi_\beta$, $\alpha$ and $\beta$ labeling generic internal degrees of freedom). This condensate would give rise to the so-called dynamical symmetry breaking [774, 775]. In particle physics, this condensate has been proposed as an alternative to the scalar Higgs boson of the Standard Model, possibly useful for open problems left by the Standard Model itself, as the hierarchy problem [806]. In short, this problem concerns the huge differences between the coupling constants of the four fundamental forces of Nature, and often referred as between weak force and gravity.

The simultaneous presence of independent symmetry groups, as involving flavour and gauge degrees of freedom, enriches further the relevance of spontaneous symmetry breaking. Indeed, even if at Lagrangian or Hamiltonian levels these systems display $G_c \times G_f$ group invariance (with $G_c$ and $G_f$ the colour and flavour symmetry groups, respectively), a certain condensate (as of a fermion bilinear) can induce a pattern of spontaneous symmetry breaking to a subgroup $G_{cf} \in G_c \times G_f$, composed by simultaneous transformations of $G_c$ and $G_f$. This is the invariance group of the condensate, such that colour and flavour degrees of freedom are not any longer independent of each other. The phenomenon is called colour-flavour locking [5, 807–810], and plays an important role in the physics of deconfined quarks, as we will comment in the Section IV C. Locking of two flavour (global, in general) groups is also possible while locking of gauge groups is forbidden by the Elitzur's theorem [801, 802].

Still in the wide framework of symmetry phenomenology, up to condensed matter physics, a special role is played by anomalous symmetries, aka anomalies. An anomalous symmetry is exact at the level of classical Hamiltonian or (more often) Lagrangian formulation, and instead explicitly broken by quantum corrections, even at the perturbative level [801, 811]. In principle, every symmetry can be anomalous, but only global ones are accepted in consistent theories. Among them, the most important ones are axial anomalies, involving chiral (massless, in even space-time dimensions) fermions coupled to gauge fields, even external (static). In particle physics, axial anomalies have very important physical consequences, the first known one being the decay of the pion $\pi_0$ in a pair of photons [812] or a finite mass for the $\eta$ particle. Realizations of axial anomalies have also been developed in condensed matter, as argued long ago by Nielsen and Ninomyia [813], on the basis of theoretical arguments only. More recently [814], these realizations have been described in Weyl semimetals, which means three-dimensional systems hosting zero-dimensional Fermi surfaces, consisting of pairs of isolated points, topologically protected. Dispersion is linear around these points, eventually allowing a direct link with chiral fermions. Various condensed matter realizations have been discovered later on, as well as various generalizations [815, 816], as with more involved dispersions. At the same time, various ultracold atoms realizations of those have been proposed, see [817] and references therein. Along these lines, above introduced honeycomb lattices have been recognized as a suitable framework for the quantum simulation of axial anomalies [818]. This progress, together with developments in the synthesis of gauge potentials and fields, pave the way for the simulation of axial anomalies in ultracold atoms setups. In this respect, we stress that the appearance of axial anomalies does not require coupling with gauge fields, gauge potentials being sufficient. This fact holds both for axial anomalies in particle physics and in condensed matter/ultracold atoms devices. Moreover, axial anomalies turn out to be realizable in not topologically protected semimetals, as Dirac semimetals [818]. The first notable example is the honeycomb lattice (as for the graphene), already simulated by ultracold atoms in [32].

Analogous models for group locking of global and gauge symmetries can also be obtained in ultracold atom setups. Indeed, fermionic mixtures with effective flavour dynamics are realized using certain earth-alkaline or lanthanide atoms [117, 777, 796, 797]. There, suitable di-fermion condensates result in group-flavour lockings [797]. Further possibilities are in perspective in ultracold setups of fermion with colour degrees of freedom and corresponding gauge dynamics, as of the types described in the next Section IV C.

Still concerning analogies and analogous models in ultracold atom setups with interest in fundamental physics, it is important to mention Bose-Einstein condensates, possibly multi-species, around unitarity. Indeed, these systems display interesting analogies with meson condensates in ultracompact stars, in a deconfined regime characterized by sufficiently high barionic and/or isospin chemical potential [548, 819]. Similarly, fermionic quantum gases at unitarity, i.e., in the middle of the BEC-BCS crossover, are known to lose their characterization in terms of any intrinsic length-scale, then effectively displaying scale-invariance. This property is approximately shared by quark-gluon plasma [820] or by certain non-Abelian gauge theories with a large number of colours [821] and/or flavours [822].

So far, we mainly discussed ultracold atoms in their ground state. However, also the Rydberg atoms platforms discussed in Section II A 5, are gaining a steadily increasing attention [172]. Among other notable achievements, they allowed recent progresses on the quantum simulation of gauge theories [823, 824]. These results stand in a more general program for quantum simulations, that already started to be experimentally developed. More in detail, simulations with devices composed by up to 256 (Rydberg) atoms have been realized so far [56, 57, 177].

Beyond ultracold atoms, a separate approach makes use of trapped ions, a platform already mentioned concerning the simulation of the Majorana equation [772]. As discussed in Section II A 4, due to the long-range interaction that can be encoded in such devices, ions, organized in suitable arrays, proved to be effective as quantum simulators for various relevant quantum systems and models [825–827]. A first notable application is given by long-range systems, as long-range generalizations of the standard quantum Ising chain [828, 829] or of the quantum Heisenberg chain [157]. Other relevant proposals exploiting ultracold atoms and trapped ions concern long-range topological superconductors [830], see [831]. Beyond their intrinsic importance as strong and long-range correlated systems and as potential quantum simulators [832], tunable interactions allow for variable spreading of correlations on the system, effectively without a finite maximum speed. This makes these platforms ideal toy models as interacting systems with (even strong) violation of Lorentz invariance. Connected with this aspect, recently peculiar phenomena of dynamical mass generation, due to long-range couplings, have been shown. This is the case for instance of some Goldstone excitations from the spontaneous symmetry breaking of continuous symmetries, see e.g., [833–835].

Finally, we highlight that the quantum Ising chain with vacancies, often referred to as the tricritical Ising model, is known to host a global $N = 1$-supersymmetry [836, 837] at criticality, hunted since long-ago in particle accelerators. This symmetry can be spontaneously broken by continuously lowering the vacancy density, ending up in the critical quantum Ising model. As customary in supersymmetric theories, this breakdown is paralleled by the appearance of Goldstino excitations, the fermionic counterparts of bosonic Goldstone excitations for global ordinary symmetries. Actually, the Majorana fermion, describing the low-energy behaviour of the critical quantum Ising model, can be shown to be nothing but a Goldstino mode from the pattern described above [836, 837]. Other notable proposals for emergent supersymmetry in ultracold atoms setups involve Bose-Fermi mixtures [838], also spin-orbit coupled [839], and atom-molecule mixing [840]. In the same references, the spontaneous breaking of supersymmetry, even thermal [840], has been also inferred.

A separate treatment will be devoted to the analogous description of QCD in some of the deconfined regimes, a topic treated in the following section, Section IV C.

### C. Ultracompact Stars and Quark Matter

Ultracold atoms proved to work as an ideal platform also for quantum simulation of selected nuclear physics phenomena, relevant as well for ultra-compact objects like neutrons stars [5, 841]. As customary, this approach exploits simplified toy models, including solely essential properties of much more involved dynamics. Most prominent examples involve the multispecies Bose-Einstein condensates (BEC), BEC-BCS crossover [552, 566]—as described in Section III A 1—and superfluid fermionic mixtures [796, 797, 842], as well as the proposal for gauge fields included in Section IV B.

In order to contextualize the role of such a modeling, let us focus at the beginning on the description of the core of certain ultracompact stars. In this region of the star, it is expected that sufficiently high densities can be reached so to induce deconfinement of quarks. In this condition, various deconfined regimes, typical of the QCD phase diagram, are expected to arise. The expected interior of a neutron star is depicted in Figure 20, while a qualitative drawing of the QCD phase diagram is reported in Figure 21 [819].

In Figure 21, the yellow zone denotes the regime where quarks are confined in the nucleons, while the main other deconfined reported regimes are denoted by typical order parameters [819]. We stress that even if the yellow region appears as well-defined, a deconfined crossover is nowadays expected, at least in temperature. Moreover, the existence of a single critical temperature for deconfinement and for chiral symmetry is still debated, see e.g., [844].

A widely studied [5, 807–810] regime concerns the region of the QCD phase diagram at large-enough baryonic chemical potential $\mu_B$. There, due to an attractive channel between quarks (possible to be obtained even at a fully perturbative level, and well described by effective Nambu-Jona-Lasinio models [774, 775]), the formation of coloured counterparts of the Cooper pairs can be favoured, also depending on the relative chemical potentials of the different flavours. These pairs are described by the anomalous difermion condensate $\langle\psi_{a,\alpha}(\mathbf{r})\psi_{b,\beta}(\mathbf{r}')\rangle$, $\{a, b\}$ denoting color and $\{\alpha, \beta\}$ the flavour indexes. Provided that formation of these pairs occurs coherently, it can give rise to colored superfluid states of quarks, possibly belonging to different flavour multiplets. In this way, color-flavour locking, see Section IV B, is generally obtained. Moreover, due to imbalances between the number densities of the pairing species (colour and flavour), pairs with nonzero total momentum can be stabilized, resulting in spatial modulations and crystalline patterns of the pairings [809, 810, 819, 845, 846]. Peculiar astrophysics signatures of colour-flavour locked phases have been described in the emission spectrum of the stars hosting these phases [807, 809, 810, 819], deserving present and future investigation. Also for this purpose, ultracold atoms can still be useful, as suggested by the discussions in the same references.

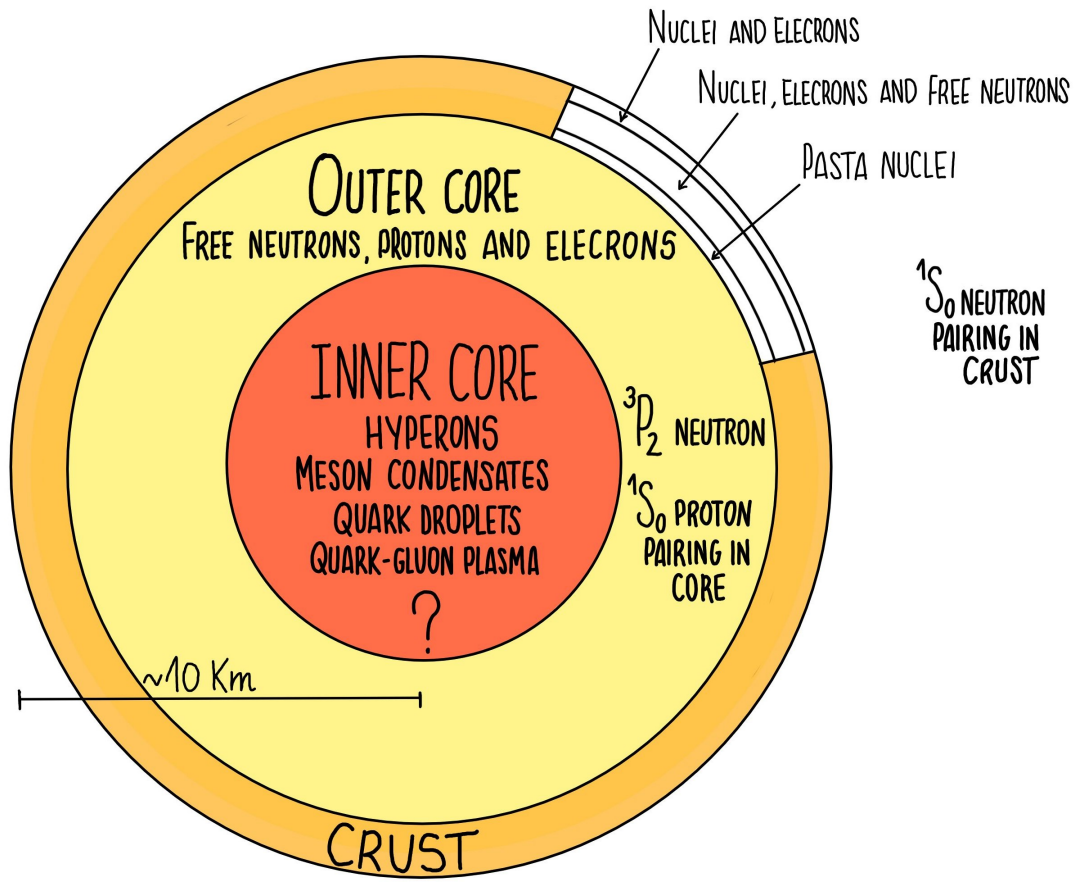

Figure 20: Cross-section of a neutron star. The neutron superfluid in the crust and the proton superfluid in the core are expected to be paired in $^1S_0$ states, and the neutron superfluid in the core in $^3P_2$ states. The electrons are instead expected to be in a normal state. A quark superfluid in the core is also expected to be BCS paired in $^1S_0$ states. Figure re-drawn from [843].

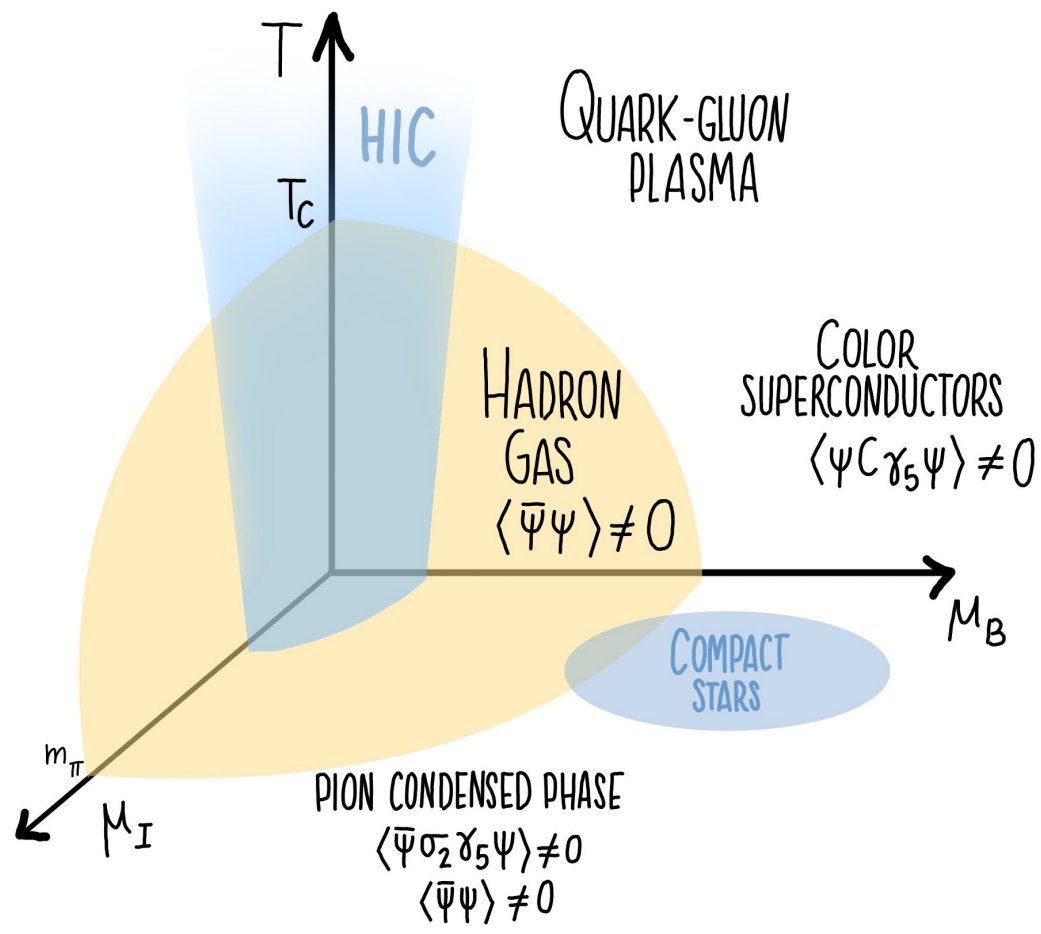

Figure 21: Qualitative QCD phase diagram for varying temperature $T$, baryonic chemical potential $\mu$, and isospin chemical potential $\mu_I$ (see text). Reported are the order parameters identifying the different phases. They are expressed as vacuum expectation values of bilinears of the field $\Psi = (\psi_u, \psi_d)$, where $\psi_u$ and $\psi_d$ are fields for the up and down quarks, $C$ denotes the charge-conjugation unitary matrix acting on each of them, $\sigma_2$ is the second Pauli matrix acting on the color (u, d) indexes, and $\gamma_5$ denotes the fifth Dirac matrix. The bar over the field operators indicates the usual Lorentz inverse conjugation. For a detailed description of the order parameters, see [807, 809, 810, 819].

Still concerning the QCD phase diagram in Figure 21 and its effective simulation, it is also interesting to focus on the opposite, deconfined, regime where the isospin chemical potential $\mu_I$ is sufficiently large (we remind that $\mu_I$ is the difference between the chemical potentials of different quarks in the same flavour multiplet). There, also based on Monte-Carlo simulations (see e.g., [847–851]) it has been inferred [852] the formation of a BEC of scalar (spinless) particle called mesons, primarily involving the pion doublet $\pi_\pm$, at least at not too large baryonic chemical potential $\mu_B$. Probably, due to the mesons interactions, a nonvanishing superfluid fraction parallels the BEC. The further role of the strange quarks and consequently of the k-mesons (kaon) is also considered, at larger strange-quarks chemical potential. Focusing on the pions dynamics, a BEC regime arises when the isospin chemical potential approaches the pion mass, $m_{\pi_\pm} \approx 139.57$ MeV. Increasing further the same potential, up to $\approx 1.7\, m_{\pi_\pm}$, theoretical and numerical studies indicate the continuous evolution of the BEC into a BCS state, à priori not linked to the colour-flavour phases described above. We stress that this crossover does not involve a colour deconfinement transition, indeed deconfinement is a common feature of all the regimes discussed here. More at a phenomenological level, the described phenomenology suggests the existence of meson stars [852], an issue still under investigation. Finally, for even larger $\mu_I$, the BCS regime can display again spatial modulations and crystalline patterns of the pairings [809, 810, 819, 845, 846] (equivalently, with nonzero momentum), linking with the large $\mu_B$-regime, still reported in Figure 21.

This physics can in principle been investigated with ultracold atoms. The simulation of the physics of di-fermion condensates and colour-flavour locking has been already discussed in Section IV B. Concerning instead the BEC-BCS crossover, a similar phenomenology can be obtained in current experiments, even in the presence of additional degrees of freedom, as mimicking color or other elementary quantum numbers. Indeed, the same phenomenology is mainly driven by the difference in the chemical potentials between the fermions forming the BEC molecules, which constitute central controllable parameters in ultracold atoms setups [117, 777]. This is the case also for the related regime of superconductivity with non-zero momentum. Detailed studies, also at finite temperature, of unbalanced ultracold fermions, indicating similar pairings, are included in the review [853, 854] and references therein.

## V. COSMOLOGY AND ASTROPHYSICS

Section IV has highlighted the main problems that have been or potentially can be addressed by means of cold and ultracold-atomic platforms, to test the Standard Model of particle physics. In this section, we proceed with our journey along selected aspects of the four fundamental theories, Quantum Field Theory and the Standard Model of particle physics on the one side, and General Relativity and the Standard Model for cosmology on the other, and discuss how atomic quantum technologies can be exploited to address foundational problems in General Relativity.

In particular, General Relativity (GR), our so far best theory of gravity, lies on the following principles: relativity principle, stating that no preferred inertial frames and that all (accelerated or not) frames can be equally considered; general covariance, stating that field equations must be in covariant form; causality, stating that each spacetime point has past, present, and future; equivalence principle, essentially stating that locally, inertial and gravitational effect are indistinguishable.

In turn, the equivalence principle encompasses a number of different aspects. One is related to the universality of free fall, often referred to as weak equivalence principle (WEP), stating that inertial and gravitational mass are equivalent. Thus, (a) locally, one cannot distinguish inertial and gravitational effects using the straightforward observation of free-fall of physical objects, and (b) objects with different internal composition are subject to the same acceleration when moving in a gravitational field. The generalized Einstein equivalence principle (EEP) embodies (i) WEP through the fact that special relativity is locally valid; it additionally encompasses (ii) local Lorentz invariance (LLI), stating that the outcome of any "local non-gravitational" test is independent of the frame velocity of the measuring apparatus, and (iii) local position invariance (LPI), that adds independence of the position and time where the test is performed. In this perspective, gravity is indeed a curvature of spacetime with the trait of a universal property, and the EEP is indeed crucial for all metric theories of gravity.

Despite its many successes, a number of issues arise with GR, both at infrared—cosmological—scales and at the ultraviolet scales of quantum field theory. In the former case, the Standard Model of particle physics and the standard model of cosmology appear to be inadequate at extreme energy curvatures regimes. In the latter, the problem is that GR is hardly made to work as a fundamental theory of gravity with a quantum spacetime. Thus, solving such controversies requires to either act on the sources of Einstein field equations and introduce exotic forms of matter (Dark Matter, DM) and energy (Dark Energy, DE), or on the geometric view. Acting on the geometric view implies, e.g., building on effective theories with GR recovered to some limit, or extended theories of gravity with additional higher-order curvature invariants and (non)minimally coupled scalar fields, possibly (not necessarily) due to field quantization of spacetime [52, 855, 856].

Interestingly enough, EEP breaking is invoked in a number of scenarios. For example, several dark matter and dark energy models break EP due to the introduction of new fields that are expected to non-universally couple to Standard Model particles, and the same occurs with other unification theories. Also, attempts to develop a quantum theory of gravitation lead to a breaking of Lorentz symmetry. In addition, EEP breaking can explain the values of some arbitrary Standard Model constants. On other hand, WEP and LLI can be violated in extended SMPP theories, e.g., due to coupling to generalized charges, the strong EP encompassing also gravitational energy, can be violated in extended GR theories, e.g., in presence of a scalar field non-minimally coupled to geometry. Finally, EEP can be violated at finite temperature $T$ since a fraction of particle mass might arise via finite-$T$ radiative corrections spoiling Lorentz-invariance of vacuum [856, 857].

Consistent efforts have been performed to address all these intriguing questions by employing highly-sensitive quantum sensors based on cold atoms, proposed to make unprecedented advancements at the interface of general relativity, gravity, atomic physics, and quantum mechanics [48, 494, 858–862], and fostering the reach of technology readiness also for practical applications in Earth observation, geodesy, time-keeping, and navigation.

One essential ingredient in this quest necessarily are precise and accurate frequency standards and clocks [489], with steadily improving stability [863, 864] up to record precisions of $5 \times 10^{-19}$ s, i.e., failing approximately one second in the age of this universe [865, 866]. The demonstration that precision is not necessarily limited by the local-oscillator stability, with the reach of a $8.9 \times 10^{-20}$ precision after 3.3 h of averaging [860, 867], suggests that we may expect even further steady improvements. This exceptional advancements are enabling table-top experiments relevant for fundamental physics beyond the Standard Model, as demonstrated in achieving the capability of resolving gravitational redshift across mm distances in table-top experiments [51], and the applications in the search for dark matter [868–871].

One second ingredient is atom interferometry [494, 499, 500, 872] that, along with atomic clocks, is having a tremendous development as an accurately controllable tool characterised by even long coherence times [873]. Atom interferometry is exploited for experimental gravitation, including precision measurements of gravitational acceleration [874–876], also via schemes using Bloch oscillations in optical lattices [224, 873, 877–879], curvature gradients [880, 881], geodesy [882], gyroscopes based on the Sagnac effect [883, 884], testing the Newtonian $1/r^2$ law [885], general relativity tests [264] and investigation of quantum gravity models [886], geophysics [887], and measurements of the gravitational constant $G$ [42, 43].

In the following, we step on some of these more recent developments especially relevant for the spirit of the present review.

## A. Gravity and General Relativity Tests

### 1. *Measurement of Big G*

While the Newtonian gravitational constant $G$ has no definitive relationship to other fundamental constants and its value is not predicted theoretically, improving the precision with which $G$ is known is of paramount theoretical importance in gravitation, cosmology, particle physics and astrophysics, and for geophysical models [43]. About 300 are the experiments aimed at determining the value of $G$, however resulting into discrepancies too large to lead to sufficiently precise conclusions. In the experiment performed at LENS with the use of ultracold atoms and quantum interferometry, in the presence of macroscopic tungsten masses [43], $G$ has been determined to be $G = 6.67191(99) \cdot 10^{-11}$ $\mathrm{m^3Kg^{-1}s^{-2}}$, with a relative uncertainty of 150 parts per million, allowing to identify systematic errors from previous experiments, and improving confidence in the value of $G$. The gravity gradiometer consisted of two vertically separated atom interferometers operated in differential mode with any spurious acceleration induced by vibrations or seismic noise rejected. The double-differential configuration has been designed to cancel out common-mode spurious signals and maximize the signal. The (two sets of) well-characterized tungsten masses were placed in two different positions to modulate the relevant gravitational signal, and designing for additional cancellation of common-mode spurious effects [43]. The atom interferometer was realized using light pulses to stimulate $^{87}$Rb atoms at the two-photon Raman transition between the hyperfine levels $F = 1$ and $F = 2$ of the ground state.

This experiment has improved by one order of magnitude the measurement previously performed at Stanford [42].

### 2. *Variation of Fundamental Constants*

One use of atomic clocks is in disentangling the fundamental problem of possible variations of fundamental constants in nature, such as the fine-structure constant $\alpha$ and the electron-proton mass ratio $\mu$. These are typically assumed to be fixed, but some forces-unification theories predict their variation in space and time. Atomic and molecular transitions, which determine the clock frequencies, depend on them. Atomic clocks can therefore constrain these fundamental constant variation over the period during which they are locally operated, typically years, thus complementing astronomical measurements spanning instead larger fractions of the universe's time history [489].

Optical clocks, which have higher precision than traditional cesium (Cs) microwave clocks, are particularly useful in studying the variation of the fine-structure constant, given that they are instead insensitive to the electron-proton mass ratio $\mu$. Then, variations in $\alpha$ and $\mu$ can be investigated by comparing the frequency of different atomic species. Additionally, frequency measurements can be analyzed to search for couplings between $\alpha$, $\mu$, and gravitational potentials due to the elliptical Earth orbit that introduces an annually varying solar gravitational term. Currently, constraints of $\delta\alpha/\alpha = -2.0(2.0)10^{-17}/\text{yr}$ and $\delta\mu/\mu = -0.5(1.6)10^{-1}/\text{yr}$ are extracted from these studies [489].

Similar goals can be pursued by means of atom interferometry [495], in fact after setting up two closed interferometers by extending the laser-pulse sequence. The phase shift experienced by the two closed interferometers is directly related to the recoil energy acquired during the $\pi/2$ beamsplitter pulses. Precise measurement of this phase difference, with the accurate determination of the laser wavelength, gives access to the ratio $h/m$ of the Planck's constant to the mass of the involved atomic species. Combining this measurement with the extremely accurate determinations of the Rydberg constant, yields the most accurate results for the fine-structure constant to date, $\alpha = 1/137.035999046(27)$ with relative uncertainty $\simeq 2 \cdot 10^{-10}$. These results surpass the precision achieved through measurements of the electron's anomalous magnetic moment. In fact, a comparison of the two results can serve as a high-precision test of quantum electrodynamics.

### 3. *Equivalence Principle Tests*

As highlighted in the introduction to this section, a number of reasons has shifted some of the efforts performed to design accurate tests of the equivalence principle. Three motivations are especially relevant. First, several theories developed to unify gravity with the other fundamental forces suggest that yet undetected fifth forces would violate the Universality of Free Fall (UFF) and the Equivalence Principle (EEP) [857]. Then, quantum interactions of dark matter with Standard Matter particles could show signatures leading to violations of UFF and EEP [888–891]. Finally, though still controversial, contradictions are proposed to exist between general relativity and quantum mechanics, as e.g., with possible gravitational-induced collapses of the wavefunction and the understanding of the measurement problem in quantum mechanics [892–895]. Needless to say, EEP remains a crucial point for any self-consistent theory of gravity and can discriminate between competing theories. Moreover, EEP might hold at a classical level and be violated at quantum level. In fact, given that EEP is not based on any fundamental symmetry and can be seen more as an heuristic hypothesis, instead of asking whether EEP is violated, one might ask its validity extent and why a violation has not been observed yet [52].

The concepts underlying the design of EEP tests are the encoding of (i) local Lorentz invariance, so that clock rates are independent of the clocks velocities, (ii) local-position invariance, i.e., the universality of red-shift, and (iii) the universality of free fall, so that all free-falling point particles follow the same trajectories independently of their internal structure and composition. To this end, atom interferometers based on multi-photon Raman (involving internal states of the atoms) or Bragg (for external, momentum states) transitions and Bloch oscillations have been developed. In particular, (i) and (ii) can be tested by means of atomic clocks, and (iii) by atom interferometers using momentum states. For example, absolute redshift measurements can be performed in which a terrestrial clock is compared to a clock in a spacecraft, and null redshift measurements -i.e. tests of the universality of the redshift–in which two different types of clocks on board of the same spacecraft are compared, with any GR deviation would manifest in a modulation of the frequency ratio between the clocks. The UFF can be for example tested by atomic Mach-Zender like Bragg interferometers like those illustrated in Section II D 1.

As to the cases (i)–(ii), optical clocks in space have indeed the potential to greatly improve tests of fundamental physics such as Einstein's theory of relativity, time and frequency transfer, and the accurate determination and monitoring of the geoid. Mission scenarios like SAGAS [896] and EGE [897] propose to use an optical clock in space that could surpass terrestrial tests by several orders of magnitude in terms of redshift measurements, Local Lorentz Invariance (LLI) tests, and parametrized post-Newtonian gravity measurements. Additionally, optical clocks in space can act as stable time and frequency servers for time/frequency transfer between continents, and relativistic geodesy applications. For example, ACES is a mission designed to test Einstein's General Relativity from the International Space Station, using a laser-cooled Cesium atom clock and an active H-maser. It will measure the gravitational redshift, search for time variations of fundamental constants, and perform Standard Model Extension tests. In currently undergoing qualification tests, ACES states a fractional frequency stability of $10^{-16}$ and accuracy of 1–2 $\times 10^{-16}$ after 10 days of integration [898, 899].

Turning to the case (iii), we begin by noticing that historically tests of the UFF have been conducted since the 16th century by Galileo after measuring the time of "free falling" balls of different compositions under gravity with a $10^{-3}$ precision, this being ingeniously reached with thousands of about 1 second oscillation periods of a pendulum, having no more than his heartbeats as a timekeeper. Ground tests of EP with macroscopic torsion balances have been performed reaching $\eta = (0.3 \pm 1.8) \times 10^{-13}$ [900] for the Eötvös parameter $\eta \equiv 2(a_A - a_B)/(a_A + a_B)$ built from the accelerations $a_{A,B}$ of the two macroscopic test bodies, and other proposals have been pushed forward [901, 902] until the state-of-the-art space experiment MICROSCOPE [903–905].

Table I: State-of-the-art in UFF/EEP tests. Numbers in brackets are results expected in the future. Table reproduced from [52] (Permission to use this content granted by Creative Commons Licence https://creativecommons.org/licenses/by/4.0/) and adapted from [53].

| **Class** | **Elements** | $\eta$ | **Year** | **Comments** |
|---|---|---|---|---|
| | Be–Ti | $2 \times 10^{-13}$ | 2008 | Torsion balance |
| Classical | Pt–Ti | $1 \times 10^{-14}$ | 2017 | MICROSCOPE first results |
| | Pt–Ti | $3 \times 10^{-15}$ | 2022 | MICROSCOPE full data |
| Hybrid | $^{133}$Cs–CC | $7 \times 10^{-9}$ | 2001 | Atom Interferometry |
| | $^{87}$Rb–CC | $7 \times 10^{-9}$ | 2010 | and macroscopic corner cube (CC) |
| | $^{39}$K–$^{87}$Rb | $3 \times 10^{-7}$ | 2020 | different elements |
| | $^{87}$Sr–$^{88}$Sr | $2 \times 10^{-7}$ | 2014 | same element, fermion vs. boson |
| Quantum | $^{85}$Rb–$^{87}$Rb | $3 \times 10^{-8}$ | 2015 | same element, different isotopes |
| | $^{85}$Rb–$^{87}$Rb | $3.8 \times 10^{-12}$ | 2020 | 10 m drop tower |
| | $^{41}$K–$^{87}$Rb | $(10^{-17})$ | 2037 | STE-QUEST |
| Antimatter | $\bar{H}$–$H$ | $(10^{-2})$ | 2023+ | under construction at CERN |

Ultracold-atom interferometry has been developed for UFF and EEP tests since the 1990s [499, 874], reaching now $\eta = (1.6 \pm 1.8(stat) \pm 3.4(syst)) \times 10^{-12}$ [906]. In white papers submitted to the ESA Voyage-2050 call [50, 907], the growing AEDGE (Atomic Experiments for Dark Matter and Gravity Exploration in Space) community and the STEQUEST (Space Time Explorer and QUantum Equivalence principle Space Test) team [52, 907] have structured proposals to test the Universality of Free Fall (UFF) and the Einstein Equivalence Principle (EEP) using (ultracold) atom interferometry, beyond the best existing results achieved by MICROSCOPE (see Table I). In general, these atom-interferometry proposals are also constructed to be multipurpose under different usages and/or data analysis, allowing also for Lorentz invariance tests, gravitational waves detection and (ultra-light type) dark-matter quests (see next sections).

The STEQUEST proposal for example [52], aims at $\eta \approx 10^{-17}$ after 18 months of operation, by using a double atom interferometer on a satellite with $^{87}$Rb and $^{41}$K test masses in a quantum-state superposition (see Table I). Interestingly, this proposals build on the technological advancements achieved in MICROSCOPE and LISA-Pathfinder missions, besides the Technical Readiness Level of the payloads reached by ground-based and microgravity experiments, that include drop-tower [908, 909], zero-gravity flights [910], sounding rockets [911], and the International Space Station [912], special efforts devoted to controlling the main systematic effects [913].

At the same time, the AEDGE community is building up a roadmap for terrestrial very-long-baseline Atom (VLB) Interferometry (AI) based on underground or tower infrastructures [914]. Five proposals are being pushed forward: the 150 m deep array of three Rb Bragg AI MIGA-ELGAR in France [915, 916], the 240 m deep array of Rb and Sr Raman/Bragg AI ZAIGA (Zhaoshan long-baseline Atom Interferometer Gravitation Antenna) in China [917], the 10 m tall tower with Rb and Yb VLBAI in Germany [918], the clock AI 10-m tall tower AION in UK [919], and the 100 m deep clock/Bragg AI MAGIS at Fermilab in USA [920]. These large facilities are at various stages of planning and realization, while others possibilities are flourishing such as those in the underground laboratories SURF (USA), CallionLab (Finland), and an underground VLB-AI at CERN.

### *4. Detection of Gravitational Waves*

The first direct evidence for gravitational waves (GWs) was provided by LIGO (Laser Interferometer Gravitational-wave Observatory)/Virgo observations of black hole and neutron star mergers [921]. LIGO and Virgo are interferometers essentially composed by 4 km and 3 km long perpendicular arms, respectively. In the basic operating concept, a passing gravitational wave stretches one arm while shrinking the other, so that the light bouncing between the arms mirrors takes different amounts of time to travel, and interference fringes manifest and can be detected. Extraordinary sensitivities have been reached, corresponding to arm-length changes of the order of $10^{-19}$ m: the longer the arms, the bigger being the arm-length change (see Figure 22). The GW frequency range which LISA/Virgo are sensitive to is 10–$10^3$ Hz in the high portion of the spectrum, corresponding to objects with relatively small masses up to several tens of solar masses. This discovery has opened the door to new explorations in fundamental physics, astrophysics and cosmology and, on the methodology side, fostered the era of so-called multimessenger astronomy. Other GW experiments are currently being prepared or proposed, such as KAGRA (Kamioka Gravitational Wave Detector) in Japan, INDIGO (Indian Initiative in Gravitational-wave Observations) in India, and the Einstein Telescope (ET) in Europe, which will focus on the same frequency range of detectable gravitational waves as LIGO/Virgo while aiming at higher sensitivities.

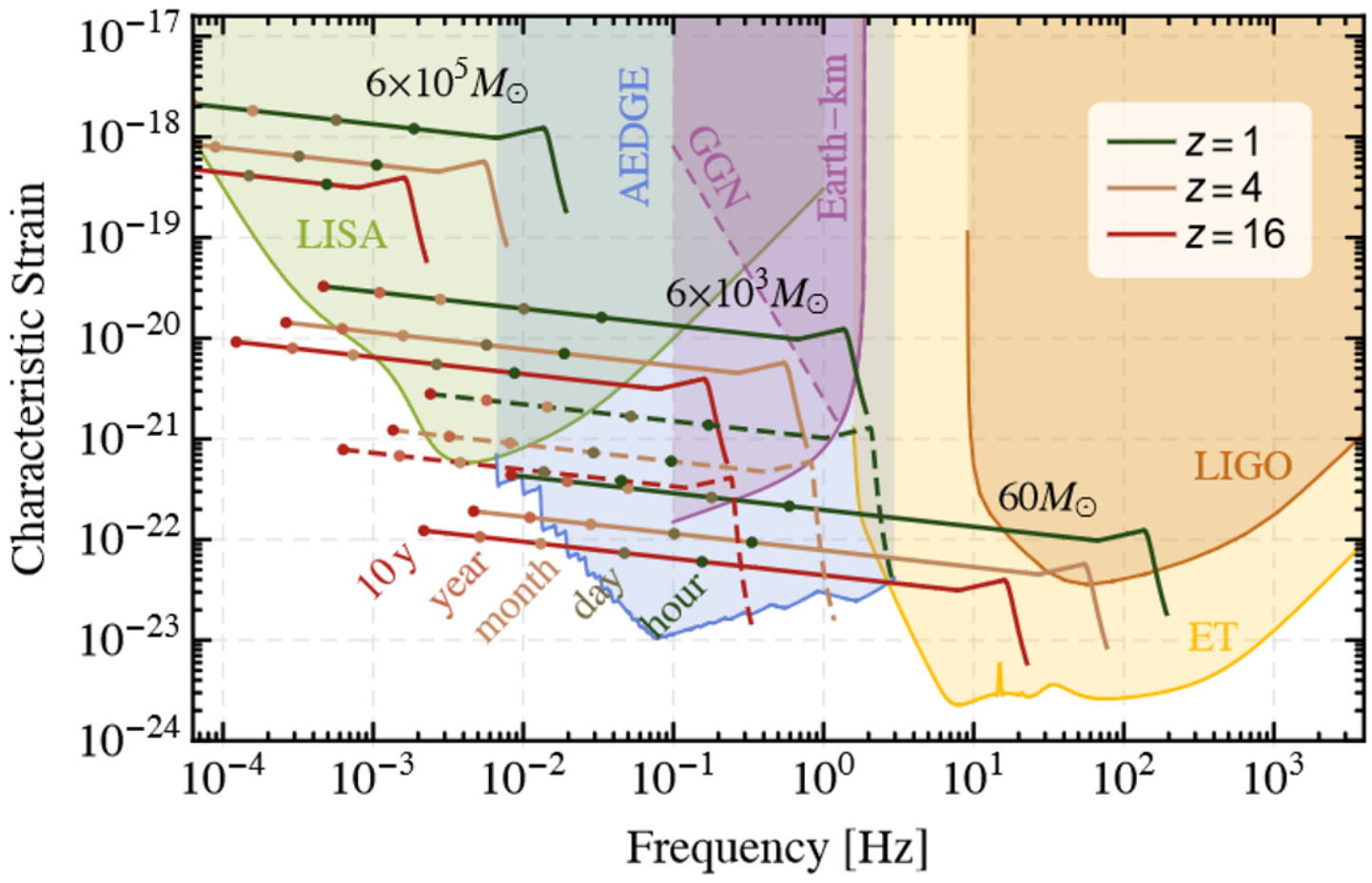

Figure 22: Gravitational Waves detection with atom technologies. Comparative analysis of strain measurements proposed by AEDGE with those proposed or realised by light-interferometry experiments as in the legend. Sensitivities to black hole (BH) mergers at different redshifts $z$ and with varying total masses are showcased. Lines: predicted strain signals for binary BH mergers, with equal mass (solid lines) and significantly different masses (dashed lines), specifically 3000 $M_\odot$ and 30 $M_\odot$. Shown is also the estimated level of gravitational gradient noise (GGN), that could potentially arise in a terrestrial detector of kilometer scale: this emphasizes the need for effective mitigation strategies. Notice that potential synergistic collaborations are possible between AEDGE and other detectors investigating different stages and histories of BH mergers. Image from [50] (Permission to use this content is granted by Creative Commons Licence https://creativecommons.org/licenses/by/4.0/).

On the other hand, supermassive black holes with masses larger than $\sim 10^6$ solar masses are known to exist, being relevant to cosmological structure formation. For this reason, designing lower-frequency interferometers is crucial. This concept is a big focus in the forthcoming LISA (Laser Interferometer Space Antenna), the ESA/NASA mission led by ESA, and building on the results of the LISA pathfinder mission. LISA will be a space interferometer using three satellites arranged at the vertices of an equilateral triangle with about 5 million km side [922] and orbiting the Sun behind the Earth, the long interferometric arms opening to a (best) sensitivity in the low-frequency range ($10^{-4}$–1 Hz), and thus corresponding to GW caused by heavier objects or with wider orbits (see Figure 22).

The AEDGE proposal using atom interferometry [50, 53], is aimed at sensitivities in a frequency range suited to observe mergers of intermediate-mass black holes with masses between $10^2$ and $10^5$ solar masses, thus intermediate between the LIGO/Virgo-like and the forthcoming LISA capabilities, possibly allowing also for complementary observations (see Figure 22). For example, it could detect mergers of $\sim 10^4$ solar-mass black holes with signal-to-noise (SNR) higher than $\approx 1000$ out to 10 years redshift, and mergers of $\sim 10^3$ solar-mass black holes with SNR higher than $\approx 100$ out to 100 years redshift. In so doing, the AEDGE proposal can pursue the goal of providing evidence of how some of the most massive stellar black holes eventually grow into super-massive black holes, to observe the inspiral stages of lower-mass black holes, and investigate whether a gap exists in the spectrum of black holes masses around 200 solar masses [50, 53].

We highlight that some of the same EEP-test aimed at very-long-baseline terrestrial experiments are under design, listed in the previous section, are also aimed at GW detection in such intermediate frequency range. These are, e.g., MIGA, ZAIGA, and MAGIS.

Finally, experiments like AEDGE and LISA can be helpful also to link GW detection to cosmological sources, whereby e.g., the Standard Model of particle physics predicts first-order phase transitions in the early Universe, including cosmic strings [50, 53].

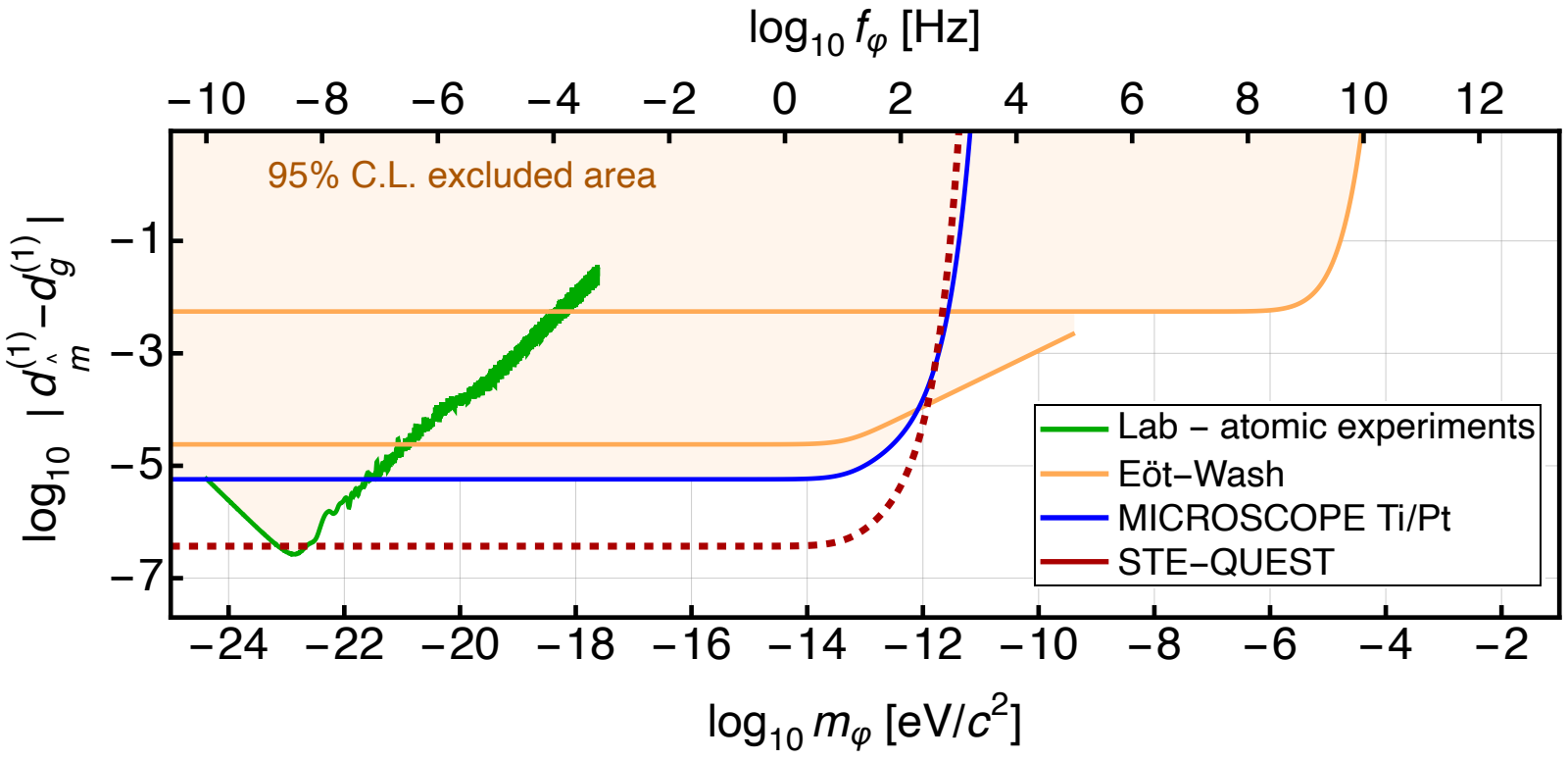

Figure 23: Quest for Dark Matter with atom technologies. Comparison between current experiment sensitivity to a linear coupling of scalar ULDM to quarks (shaded region) and the sensitivity proposed by STE−QUEST mission (dashed line). Experiments include atomic clocks [927], the MICROSCOPE experiment obtained from [904] and torsion balances [928], as in the legend. Image from [52] (Permission to use this content granted by Creative Commons Licence https://creativecommons.org/licenses/by/4.0/).

### 5. *Quest for Ultra-Light Dark Matter*

Dark matter (DM) and dark energy (DE) are preeminent examples of unknowns remaining to be unveiled at the verge between the standard model of particle physics and of cosmology. It is estimated that DM comprises around 84% of the matter density in the Universe. However, nothing about the nature of the corresponding particles is known: neither its mass, nor energy, spin, parity, lifetime, not even the type of interaction (if any) with normal—e.g., SMPP-matter, to start with. DE is theorized to be related to the cosmological constant $\Lambda$ of general relativity, a yet unknown form of energy, in fact the dominant form of energy in the universe: while the vacuum energy of quantum fields can potentially account for dark energy, a correct prediction of its value requires a viable theory of quantum gravity [889]. Searches for DM in a wide range of energies between zeV and TeV/PeV have been performed [923]. One of the most popular high-energy candidates, the weakly interacting massive particles (WIMP), have however not been directly detected at the CERN Large Hadron Collider. Therefore, alternative dark matter candidates have more recently gained interested in research, within the so-called ultralight dark matter (ULDM) scenario, e.g., DM in the sub-eV mass range. Different hypothesis include QCD axions and axion-like-particles (ALPs), dark vector bosons, and light scalar particles [924–926].

Atom interferometers can be sensitive to these low-energy particles, i.e., in the zeV-keV range (see Figure 23). The concept underlying the detection of scalar ULDM by atom interferometry exploits a temporal variation of atomic transition frequencies, due to scalar DM-induced oscillations in time of fundamental constants like the electron mass and the electromagnetic fine-structure constant, at frequencies set by the mass of the scalar DM and amplitudes driven by the DM mass and local density [870]. The sensitivity increases with increasing the interrogation times, the number of atoms in the beam, the amount of momentum transfer, and the fringes contrast, and with decreasing the sampling time [50, 53]. Massive vector ULDM fields would be detected also because of a composition-dependent Yukawa-type modification of the two-body interaction, leading to a differential acceleration of different isotopes in dual-species atomic interferometers [52].

The AEDGE roadmap aims at probing extensive new regions of parameter space in the largely unexplored mass range of $\sim 10^{-2}$ eV to $\sim 10^{-16}$ eV, for DM with linear and quadratic couplings to electrons and photons [50, 53]. Using atomic isotopes with different nuclear spins, detection of axion-like DM lighter than $10^{-14}$ eV [929] could become accessible, while running two coupled interferometers would become sensitive to dark vector bosons below $10^{-15}$ eV coupled to the difference between baryon and lepton number [930].

Limits to possible oscillations of a linear combination of constants like the hyperfine-structure constant, the $\alpha$ quark mass, and the QCD mass scale, have also been set after performing approximately 6 years of highly accurate hyperfine frequency comparison of atomic $^{87}$Rb and $^{133}$Cs clocks [927], making more stringent a previous Dy-spectroscopy based experiment assuming only $\alpha$ variations [869].

While the present open debate continues on whether one day it will be possible to detect gravitational waves or dark matter by atom interferometry, this field of research is rapidly growing and extending, always involving a cross-disciplinary community of condensed-matter and AMO physicists, astrophysicists, and high-energy physicists [50, 931].

The topic has entered The 2021 ECFA detector research and development roadmap (DRDR) (see e.g., [932]) and two fully dedicated work packages (WP1 and WP2) are being conceived as input for the ongoing Quantum Detector initiative started at CERN. A dedicated location at CERN has been more recently announced to host a ground atom interferometer at the more recent Terrestrial Very-Long-Baseline Atom Interferometry Workshop (13–14 March 2023 at CERN) [933].

We finalise this section by highlighting one aspect of the ULDM scalar boson hypothesis, that represent one more example of integration between ultracold-atoms physics and cosmology. In fact, it is known that the conventional cold-particle interpretation of dark matter (so-called cold dark matter) misses basic properties of the density profiles for common dwarf galaxies. While this outcome is fostering increasing interest, remarkably complex simulations of cold, wavelike dark matter composed of Bose–Einstein condensates are being performed. Indeed, it is found that when Bose-Einstein condensed into what is named fuzzy dark matter, dwarf galaxies profiles are reproduced, with the observed uniform central masses and shallow density [934]. These simulations are computationally extremely challenging because of the very different length scales involved, which require large and adaptive simulation meshes. Ideally, one might also want to embed the simulation of the small-scale galaxy-size BEC DM into the large-scale structure dark matter simulations [935], leading to even more demanding computations. In the spirit of this review, one may still envision implementing at least the fuzzy-DM simulation in real BEC experiments, using BEC platforms as a quantum simulator.

#### *6. A Special Perspective*

We conclude this section by highlighting that these forefront enterprises are signing a new era in the way quantum technologies are used for fundamental science. Some of the illustrated proposals ride along the—so-far unexplored—path of integrating microscopic quantum sensors in large-scale facilities. Thus, practicing a cross-disciplinary approach to one common search, sourcing from high-energy physics, condensed matter and atomic-molecular-optical physics, astrophysics, general relativity and cosmology. All of them operate within tightly-joined theoretical and experimental efforts, implementing techniques and technologies across different scientific communities, and fostering the advancement of technology readiness to stimulate dedicated companies and ignite sustainable economical development.

### B. Quantum Simulators for Gravity and Cosmology Problems

We have discussed how atomic clocks and atom interferometry can be exploited for fundamental tests and to detect gravitational waves and dark matter. On the other hand, in the spirit of this review we highlight that atom-based quantum technologies can also be designed to work as quantum simulators for specific phenomena. In this section, we discuss one such example interesting to gravity and general relativity, that is analogue gravity. This is attracting increasing interest on both the theoretical and experimental side and proposes to encode gravity-related phenomena in ultracold-atom experiments.

#### *1. What Is Meant by Analogue Gravity*

Analogue gravity is a paradigm built on the concept that a physical phenomenon can be described using an *emergent* or *analogue* metric [936], thus enabling to simulate the dynamics of objects living on a spacetime and described by classical or quantum field theory in curved spacetime. While depending on the specific matter model, this emergent metric can act as an effective background quantity governing field fluctuations. Along these lines, analogue models essentially work as effective-field theories, with their dynamics being determined by matter equations, aiming to replicate aspects of general relativity and field theory in a curved background.

In essence, the analogy works as follows: the fields propagating through other gravitational fields correspond to the perturbations propagating in continuous systems, that are the limiting form of a microscopic underlying structure, thus *emerging* from a more fundamental theory. In this view, general relativity would be the analogue to hydrodynamics, whereby gravity is simulated by the model kinematics and the dynamics through the system's microphysics: the metric experienced by the quantum excitations would emerge from the fluid-dynamics equations with parameters determined by a thermodynamic equation of state. Of course, the possibility of describing emergent phenomena within an analogue gravity approach, by no means implies that general relativity must emerge from a more fundamental theory of quantum gravity. This notwithstanding, the analogy can be quite useful to construct different cosmological metrics, as it has been the case for example of the Friedmann-Lemaitre-Robertson-Walker [937].

Playing with analogue gravity offers three main advantages. First, it encourages creative thinking by associating concepts from seemingly distinct systems, potentially leading to new understanding and ideas that, in this case, might lead to completely different perspectives on gravity as a quantum or as an emergent phenomenon. Second, the knowledge of the matter model—in our case an ultra-cold atomic fluid, can be exploited to gain insight on classical general relativity and viceversa, or yet on curved-space quantum field theory. Third, selected concepts from gravitational theory can be encoded into real table-top experiments, allowing for practical investigations, albeit with limitations [37, 38]. In this respect, two caveats are relevant in this discussion. Firstly, the matter stress-energy tensor does not obey the Einstein equations, so that no correspondence directly exists with the analogue metric. Second, at the scales where the system's microscopic properties become relevant, the equations describing particles and waves in the gravitational field can deviate from those of gravitation: for example, local Lorentz invariance can break, whereas at larger length scales it is intrinsically encoded along with local-position invariance. Nonetheless, these analogies serve as a source of inspiration, opening up new perspectives and generating novel ideas. Turning to the specific models, a number of classical and quantum systems have been explored in the last couple of decades, including water and classical fluids with on-purpose designed background velocities [938–941], electrodynamical systems [942–945], liquid Helium [38], and quantum gases.

One common tool is the formation of horizons. In a moving fluid, this is realised by arranging the fluid to move with a speed below and above the speed of sound in different regions: in the subsonic portion, the fluid drags the sound waves along, while in the supersonic region sound waves cannot go back upstream, so that the limiting border between the supersonic and the subsonic regions plays the role of an event horizon. In fact, analogs of horizons with the corresponding Hawking radiation, and rotating black holes have been simulated in classical fluids. However, in classical systems the field analogues must be simulated, while in quantum systems they spontaneously pop up. For this reason, quantum models result to be especially appealing and useful. Among them, those of interest for the present review, which are physically realized by Bose-Einstein condensates (BEC) of ultracold atoms [37, 946, 947].

Here, the acoustic metric is fully determined by the BEC wavefunction and the analogue particles are the Bogoliubov quasi-particles spontaneously appearing in the condensate as a quantum perturbation of the condensate wavefunction. The BEC healing length, governed by the system density and the interatomic interaction strength, plays the role of the Planck scale and sets a natural cutoff preventing the ultraviolet divergences and solving the transplanckian problem [948–950]. In fact, the connections between gravity and quantum hydrodynamics via thermodynamics [951] and collective quantum fields [952, 953] is sometimes speculatively used to develop a theory of quantum gravity. The cosmological particle creation can be simulated after dynamical tuning of the atomic interactions through time variations of the low-energy scattering length operated by means of the Fano-Feshbach resonance mechanism [954, 955], see Section II A 1.

As a field of research, analogue gravity started flourishing after the pioneering work of Unruh [956], building up an analogue model based on fluid flow to investigate aspects of Hawking radiation in real black holes, years later developed in another seminal paper by Jacobson [948]. In playing the analogy, the lesson was learnt that Hawking radiation should be thought of as a fundamental phenomenon of curved spacetime, occurring whenever a horizon is equipped with an effective geometry, thus not necessarily specific of general relativity.

While we refer to the detailed review by Barcelò et al. [37] for an extensive account of this subject, we discuss below the main ideas and selected recent results implying the use of quantum fluids of atoms, and highlight the possible perspectives. In particular, we overview three significant outcomes of the joint experimental and theoretical research performed in analogue acoustic (or dumb) holes realized in these analogue systems: the creation of acoustic horizons with the non-trivial observation and understanding of (analogue) Hawking radiation [957–961], the perspective of investigating the information-paradox problem, and of testing the Kovtun, Son and Starinets (KSS) conjecture about the existence of a universal bound for the shear viscosity-to-entropy density ratio $\eta/s \geq 1/(4\pi)$ at the black-hole horizon [962]. Other ongoing theoretical and numerical studies are aimed at mimicking rotating black holes, the occurrence of superradiance, and the Penrose effect.

### 2. *Analogue Horizons and Hawking Radiation*

In the acoustic (AH) analogue of black holes (BHs), also named *dumb* holes, long-wavelength sound waves cannot propagate upstream against the supersonic fluid flow across the acoustic horizon, that works as a trapping surface. Hawking radiation and temperature $T_H$ can be defined in analogy to photons spontaneously emitted at the BH horizon [37, 956, 963–968]. As first derived by Unruh [956], $T_H$ results to be determined by the gradient of the velocity field for a transonic flow in a background with spatially uniform sound speed:

$$T_H = \frac{1}{2\pi} \left( \frac{\partial |c_s - v|}{\partial n} \right) |_H, \tag{12}$$

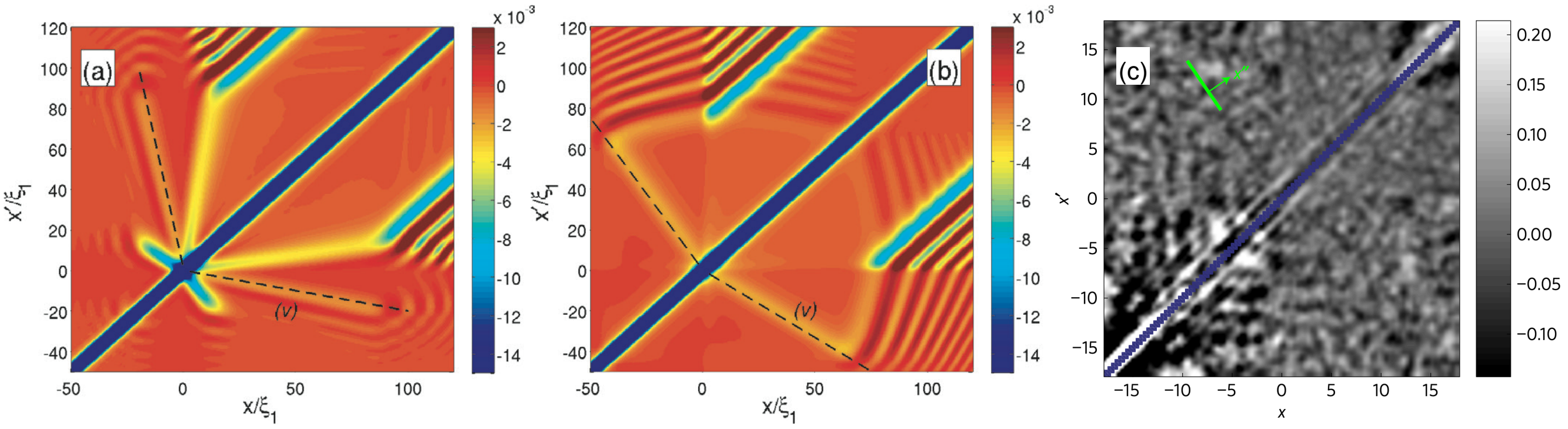

Figure 24: Analogue gravity simulation with atom technologies of Hawking radiation in acoustic holes. (**a**) predicted density−density correlation function $(n\xi1) \times [G^{(2)}(x, x') - 1]$ between Hawking particles and their partners, rescaled to the diluteness parameter $n\xi_1$, with $n$ the density and $\xi_1 = \sqrt{\hbar^2/(m\mu_1)}$ the healing length in terms of the atom mass $m$ and the chemical potential $\mu_1$. The temperature is $k_BT_0/\mu_1 = 0.1$. (**b**) the same quantity as in (**a**), in the absence of a black−hole horizon, where the flow remains sub-sonic everywhere. Image from [959] (Permission to use this content granted by Creative Commons Licence `https://creativecommons.org/licenses/by/4.0/`). (**c**) Experimental observation of the predicted two−body correlation function in (**a**) panel, performed in [39]. The origin represents the location of the horizon, and the dark bands emerging from the horizon indicate correlations. Image from [39].

with $v$ and $c_s$ the space-dependent fluid velocity and sound speed profiles, respectively, the normal derivative being evaluated at the event horizon [37]. Following the theoretical suggestion in [959] of identifying the signatures of Hawking radiation in the density-density correlation function between analogue Hawking particles (see Figure 24a), analogue Hawking radiation has been observed in the laboratory [39] with quasi-one-dimensional BEC (see Figure 24b). Here, an imprinted density profile created a space-dependent speed of sound and thus an analogue horizon. In a more recent theoretical work [41], the expression (12) for $T_H$ has been derived within a covariant kinetic approach, with steps that lead to learn two special lessons. First, the analogy between hydrodynamic flow and gravity results in a covariant expression for the distribution function of phonons [969, 970], that highlights the geometric nature of $T_H$. Second, the derivation is performed by equating the entropy and energy losses of the acoustic horizon and the entropy and energy gains of the spontaneously emitted phonons, that highlights the connection between Hawking radiation and the horizon area [971] known from real black holes. In particular, this result supports the conjecture that the entropy of acoustic horizons is proportional to their horizon area, consistently with what is expected from considerations about the information loss problem [972] and the entanglement of quantum fluctuations on opposite sides of the horizon [973] (see below).

### *3. Information-Loss Paradox*

While radiating, the black hole evaporates until it becomes a thermal bath. This, however, would be in contrast with a unitary evolution of the quantum state from the initial to the final state. In that case, it is easily understood that different initial states may lead to the same thermal state and information would be lost. This paradox is a major open problem in (analogue) gravity, connected to the link between geometry and quantum matter, whereas a semiclassical theory of gravity essentially separates them: thus, information transfer among the two is not possible [974]. Restoring such an information transfer among matter and geometry is the solution, that is what quantum theory of gravity should be capable of. Atom technologies can be an effective platform where to investigate this concept of information transfer between matter and geometry.

More recent studies [974] have suggested that during evaporation the information might remain encoded in the entanglement between condensed and excited particles, which would then preserve unitarity. This can be viewed by studying the corresponding many-point correlation functions in a beyond-Bogoliubov approach that accounts also for the quantum behaviour of the condensate correlations between quantum degrees of freedom of matter and geometry build-up during evaporation.

### 4. Viscosity to Entropy Density Ratio

We finally turn to one of the most striking predictions of the universal behaviour of black holes, that is the Kovtun, Son and Starinets (KSS) lower bound conjectured to exist for the ratio between the shear viscosity and the entropy density ratio $\eta/s \geq 1/(4\pi)$ at the BH horizon [962]. This is quite a robust result. Indeed, though KSS originally derived the bound within the AdS/CFT correspondence, completely independent derivations have been performed instead in Rindler causal horizon in flat spacetime [975, 976]. In addition, since the speed of light does not appear in the bound, the KSS conjecture seems to be valid for all real fluids, relativistic and not, and amenable to be extended to classical fluids. Given such a degree of universality, one may ask whether are there necessary conditions that the fluid should fulfill, that have a quantum mechanical nature [552]. This question presents a pathway to investigating the interplay of general relativity and quantum mechanics at the black hole horizon. Experimental studies of the KSS ratio in BECs have been performed [977], that however show discrepancies with microscopic theoretical predictions based on the theory of strongly interacting Fermi gases [552]. In fact, KSS had already highlighted that one microscopic condition for the bound validity is that the fluid be strongly interacting without well-defined quasi-particles [552]. Interestingly, this condition can be met in relativistic heavy-ion collisions close to the deconfinement transition temperature [978], in pure gauge numerical simulations [979], in $O(N)$ [980] and hadronic [981] models. In both the microscopic [552] and quantum hydrodynamic [976, 982] treatments of the KSS bound, the bound calculation has been based on the determination of the Kubo transport coefficients formulas, involving finite temperature Green's functions of conserved currents. An entirely different, simpler and particularly transparent derivation has been proposed [983], where the KSS bound is derived within a kinetic theory approach, where interaction-driven and geometric effects can be easily disentangled, the latter being enhanced, and is open to further developments also based on the quantum hydrodynamic microscopic formulation for superfluids, that is based on the time-dependent density functional theory [984–986].

## VI. FOUNDATIONS OF QUANTUM MECHANICS

In Sections IV and V we have explored the extent to which atomic quantum technologies can be exploited to explore unsolved questions about the connections between Quantum Field Theory (QFT) and the Standard Model of particle physics (SMPP) on the one side, and General Relativity (GR) and the Standard Model of cosmology (SMC) on the other. We have highlighted that while these theories have resisted to all sort of testing over decades up to more than a century, a unified description has yet to be found. In particular, Quantum mechanics is our best fundamental theory to describe the microscopic building blocks of Nature, and it has not been disproved so far by any of the increasingly sophisticated and precise experiments that have been conceived in more the 100 years, scaling up in size from the behavior of subatomic particles, chemical reactions, the functioning of electronic devices, up to matter-wave interferometry with tests masses of nearly $10^5$ atomic mass units (amu) [987, 988]. In fact, the possibility that macroscopic quantum-mechanical behavior be relevant in the realm of biological systems is attracting considerable interest as well, and will be the subject of the next Section VII.

However, even before questioning its connections with the other theories, a number of fundamental questions remain to be unveiled. One especially intriguing mystery is the concept of time, unavoidable and crucial question in the search of how to reconcile QM and general relativity [989]. In this domain, a seminal paradigm shift has been initiated by Page and Wootters [990], considering time as a quantum degree of freedom with a corresponding Hilbert space, the time flow being then provided by the entanglement between the time degree of freedom and the system itself, the global state remaining time-independent. For an observer internal to the system, the clocking would reduce to a normal time evolution resulting from projecting the global quantum state on the given time $t$, thus a conditioned state. However, the mechanism fails to provide the correct quantum propagators and correct quantum statistics of the measurements performed at different times, an issue that has been subsequently fixed by formalizing measurements according to the von-Neumann prescription [991]. The presence of gravitational interactions among the clocks and the definition of consistent quantum reference frames has been also addressed [992, 993] and a well-defined classical limit recovered through Generalized Coherent States [994]. Needless to say, the remarkable progress being achieved with atomic quantum technologies and specifically with quantum clocks [51] potentially leads to explorations and experimental testings of these new paradigms.

In the following, we focus on a second fundamental and foundational problem, that is the classical-to-quantum crossover. This question revolves around the observability of quantum phenomena in macroscopic objects and how quantum theory retrieves the predictions of classical physics in the limit of large quantum number, a topic that is also intimately connected with the measurement problem in quantum mechanics and that has generated considerable debate [987, 995–997]. While decoherence can explain the emergence of classical behavior [998] and quantum physics does not pose any fundamental limitation to the system sizes at which quantum behaviors—like the observability of superposition states—can be observed, many proposals have been pushed forward to explain how classical behavior emerges from quantum physics at macroscopic scales, and current quantum technologies can provide accurate test-beds for these theories. Among them is the Bohmian interpretation [999, 1000], decoherence histories [1001], the many-world interpretation [1002], and collapse models [893, 894, 1003–1005]. Alternatively, one could consider that quantum mechanics emerges as the effective framework of a more fundamental theory [1006].

In the last two decades, systems of increasing size have been realized, whose behavior requires quantum mechanics to be understood. These include picogram mechanical oscillators [1007, 1008], optomechanical systems [1009], superconducting quantum interference devices [1010], Bose–Einstein condensates [1011, 1012], matter-wave interferometers [872, 1013], and massive molecules [988, 1014].

In particular, quantum superposition has been demonstrated in the experiment performed in the Arndt group using a Long-Baseline Universal Matter-Wave Interferometer comprising a three-grating Talbot–Lau interferometer with a 2 m baseline [988], where quantum interference of single molecules beyond 25kDa has been observed, corresponding to the record macroscopicity parameter [1015] $\mu \equiv \log_1 0[(m/m_e)^2(\tau/1s)/\ln(\eta)] = 14.1$. Here, the mass $m = 26.777$ Da is scaled with the electron mass $m_e$, the coherence time is $\tau = 7.5$ ms, the fidelity $\eta = (0.93 \pm 0.06)$, and a de Broglie wavelength of about 53 fm. Since 1980, an approximately linear scaling-up of the macroscopicity parameter has been achieved in quantum experiments, from $\mu \simeq 6$ to the present $\simeq$14.

In general, these tests are classifiable under two main strategies, depending on whether interferometric methods are used or not [895]. Interferometric experiments rely on quantum superposition and are therefore more fragile with respect to decoherence issues, though they provide access to a wider landscape of fundamental physics tests. Non-interferometric experiments can be more stringent for collapse models on the ground. In this scenario, atom technologies can indeed provide an effective platform to explore the boundaries of the quantum superposition principle when applied to larger-scale systems, in fact testing the wavefunction collapse models. The idea behind the collapse models is that collapse terms enter the otherwise conventional Schröedinger dynamics, that induce the localization of the wavefunction within a preferred basis [1005, 1006]. In fact, collapse models are invoked also in the context of cosmology to explain the formation of cosmic structures and mechanisms for the cosmological constant [895].

Two such models are currently attracting much interest to design ultracold atoms experiments: the phenomenological Continuous Spontaneous Localization (CSL) and the Diosi-Penrose (DP) model. The CSL [1016, 1017] considers that a fundamentally quantum system be subjected to weak and continuous measurement-like dynamics, in the form of white-noise terms.

A conceptual map of the CSL model is displayed in Figure 25, again from the STEQUEST proposal [52], in terms of the two relevant free parameters: the collapse rate $\lambda$ related to the collapse strength and the correlation length of the collapse noise $r_c$, defining the spatial resolution of the collapse. The parameter values denoted as GRW (Ghirardi, Rimini, and Weber) correspond to $\lambda = 10^{-16}s^{-1}$ and $r_C = 10^{-7}$ m, and were theoretically proposed [1018] to ensure the effective collapse of macroscopic systems. The values of $\lambda = 4\times10^{-8\pm2}s^{-1}$ and $r_C = 10^{-7}$ m were instead proposed by Adler [1019] to enable the collapse to occur at the mesoscopic scale instead. While the latter have already been experimentally excluded, GRW has not yet been ruled out.

Alternatively, the model independently conceived by Diosi and Penrose (DP) [893, 894, 1003, 1004], postulates that quantum superposition states collapse due to gravitationally-induced decoherence. In Penrose's argument in fact, this mechanism would at once unveil the quantum-to-classical crossover transition and re-connect the principle of general covariance of general relativity with the superposition principle of quantum mechanics. According to the argument, spontaneous collapse into localized states occurs, which would be enhanced as the system's mass increases. Diosi in turn has formalized the concept by analyzing a master equation where Newtonian gravity enters the non-unitary system-environment coupling. Formalization of the idea requires to avoid the divergences of the Newtonian potential at small distances with consequent diverging collapse rate for a point-like particle regardless of its mass, implying an unphysical instantaneous collapse. Thus, the Diosi-Penrose model needs to be formalized with including an extended mass distribution characterized by a minimum length $R_0$, in fact the sole free model parameter. Experiments have been performed to test the lower bounds on $R_0$ [1029, 1030], the most stringent having been performed by X-ray measurements, exploiting the fact that the collapse mechanism makes charged particles emit radiation [1026]. This notwithstanding, more recent proposals are being pushed forward [1031] also envisioning to exploit current atomic technologies.

In the STE-QUEST concept, CSL would be tested by measuring the variance in position of a non-interacting Bose-Einstein condensate (BEC) in free fall, moving to space an experiment implemented on the ground [1027], predicting the red-line result in Figure 25. When related to the CSL parameters, this variance can be expressed as $\sigma_t^2 = \sigma_{\mathrm{QM},t}^2 + (\hbar^2/6m_0^2r_c^2)\lambda t^3$. The idea is that the presence of collapse models enhances the variance of the BEC compared to the prediction of quantum mechanics, whereas $\sigma_{\mathrm{QM},t}^2 \propto t^2$. The scaling induced by collapse models is thus distinguished to be proportional to the cube of the free evolution time. Needless to say, operating the experiment in space offers the usual significant advantage of longer free-fall evolution times [895, 923, 1032, 1033]. By measuring the BEC expansion over times of the order of 50 s of free-fall and with an accuracy of a position variance in the order of micrometers (both accessible in the STEQUEST space-design), the expected sensitivities to the CSL parameters are approximately four orders of magnitude stronger than those in ground-based experiments (see Figure 25). Similarly, the bounds on the DP model can be improved by more than an order of magnitude.

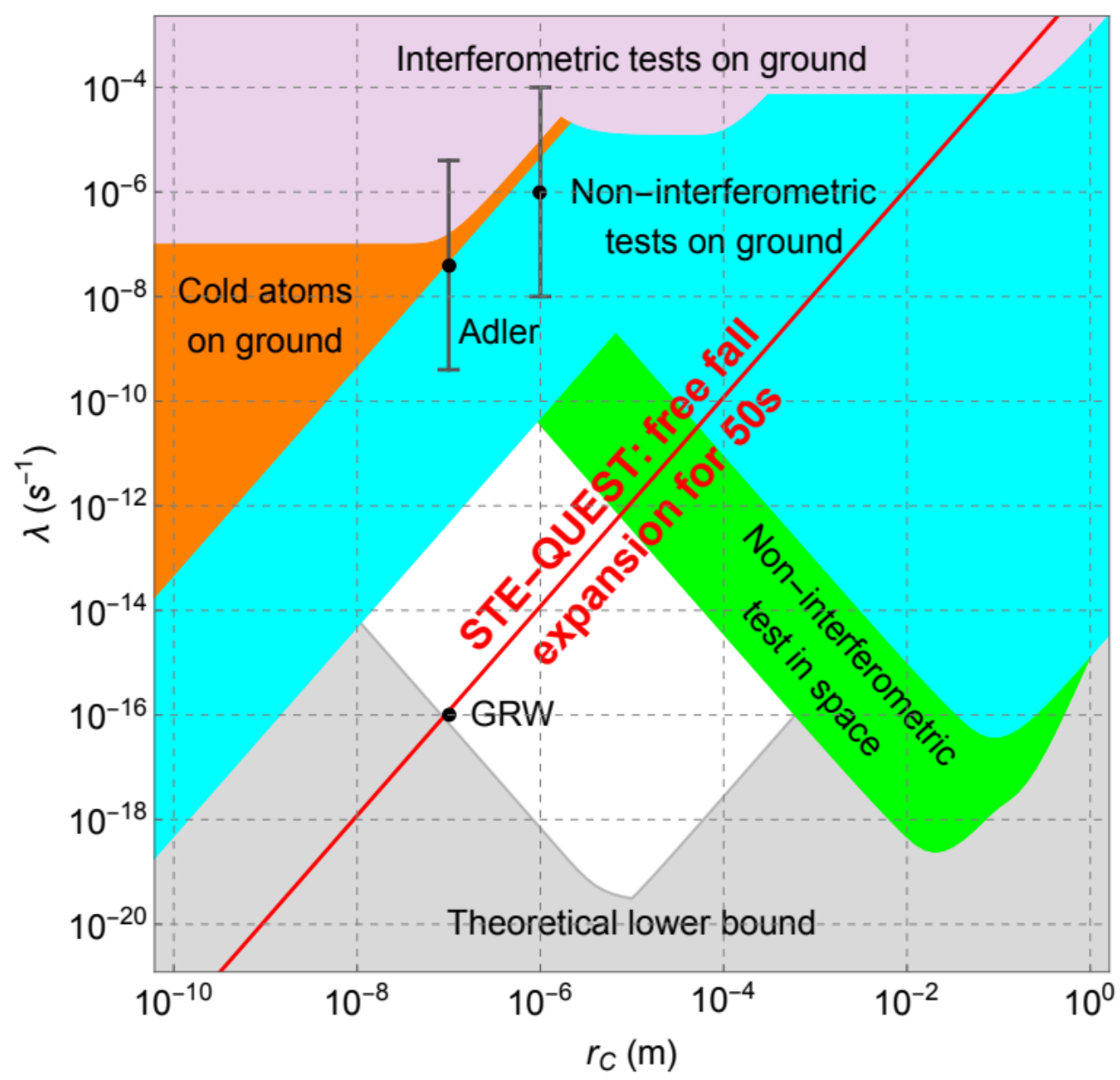

Figure 25: Testing the foundations of quantum mechanics with atom technologies. Test of the phenomenological Continuous Spontaneous Localization (CSL) model for the collapse of the wavefunction, in the governing parameters space: the collapse rate $\lambda$ and the correlation length $r_c$ of the collapse noise (see text). Diversely colored areas indicate the excluded regions based on results from different experiments: ground-based interferometric experiments (pink) [988, 1020, 1021], non-interferometric experiments (blue) [1022–1026], cold-atoms on the ground (orange) [1027], and non-interferometric experiments in space (green) [1022, 1028]. Red line: the STE-QUEST prediction. Grey area: theoretically excluded region assuming a collapse at the macroscopic scale, which is a fundamental requirement of the model [1021]. Black dot: prediction of the Ghirardi-Rimini-Weber (GRW) model [1018]. Black interval: Adler model [1019]. Image from [52] (Permission to use this content granted by Creative Commons Licence `https://creativecommons.org/licenses/by/4.0/`).

## VII. QUANTUM SIMULATORS FOR BIOLOGY AND CHEMISTRY

In this review, we have already introduced in Section II C the applicability of certain QT tools and algorithms to problems of relevance in chemistry and biology. In particular, when discussing quantum control protocols in Section II C 3 or more generally in quantum variational algorithms in Section II C 2, since the description of microscopic complex structures like those of relevant biological molecules is one of the heralded applications of large-scale QC, e.g., in drug-design [436]. Thus, a considerable effort has been put forward in recent years towards possible applications already in the NISQ era, see e.g., [435, 436, 1034], combining gate-based approaches, Hamiltonian simulation and hybrid protocols [425, 1035], even with some initial experimental implementations [162]. Moreover, the developed of NISQ protocols has shifted the approach with the combination of classical and quantum algorithms or classical processing of the quantum results; moving towards an approach where only the suitable hard problem is solved via the quantum co-processor while the rest is solved classically.

This route towards the description of applied physical phenomena has indeed been pursued in depth by a large community and it is steadily advancing, in parallel to quantum technological (not exclusively atom-based) and algorithmic development. For an account of this vast subject we refer the reader to the relevant reviews, see [3, 427, 436], and here we do not pursue it further. Instead, here we would like to take the opportunity to highlight a complementary approach, that is the quantum-like paradigm.

### A. Quantum-like Paradigm and the Brain

The quantum-like paradigm relies on a simple idea: using quantum mechanics for the mathematical description of non-linear, complex phenomena that do not require to be inherently quantum [1036, 1037]. This approach uses the mathematical toolbox of quantum systems exploiting: (i) the linearity of quantum information processing, mapping the problem's complexity into the intricate build-up of correlations through superposition and entanglement that appears due to the coherent evolution; and (ii) the innate non-linearity of the measurement process. Mapping a problem into a quantum one can take advantage of the large set of tools that we have discussed in this review allowing in certain scenarios to simplify the description in a controllable way, particularly via the approaches in Sections II B 1 and II B 2, and making it possible to save computational resources by the use of these well-understood approximations. In particular, those tools of non-Markovian open quantum systems [334, 345] or those in periodically and boundary-driven systems [729, 732] that allow for the description of transport processes [733] particularly in biological matter [734, 735]. In addition, certain properties of the quantum counterpart can be exploited to prevent computational bottlenecks as, e.g., in quantum annealing where quantum dynamics have been shown to be more robust in avoiding self-trapping in metastable states [1038]. Thus, finding the right mapping allows to exploit the quantum theoretical and numerical toolbox, utilizes any inherent advantages of the quantum system, and complements existing classical methods in a synergistic manner. Finally, while quantum simulators technologies progress, we can envision that this mapping can be performed not simply theoretically but directly on the relevant QT platforms surpassing the limits of what can be classically simulated.

In literature there exist numerous examples of these coarse-grained quantum models to applied physics. Here, we consider the quantum-like paradigm as applied to the description of certain brain functions. Existing examples include models to describe long-lived quantum-chaotic patterns generation [1039], information processing and consciousness measures as in the the quantum versions of the Integrated Information Theory (IIT) [1040, 1041], dissipative quantum models of the brain [1042], information processing in more general biosystems [1043], quantum-inspired techniques in psychology defining the field of quantum cognition [1044], and comparing the brain to a quantum Bayesian inference machine [1045, 1046].

In order to provide a flavour of how the quantum-like paradigm may work, in the following we illustrate it through one specific example related to the description of a visual, perceptual function of the human brain, that is the perception of number or *numerosity*. Most existing approaches to this problem rely on the use of artificial and deep convoluted classical neural networks [1047, 1048] to reproduce the observed phenomenology [1049–1051]. In particular, the hallmark of perception is given by Weber's law, an observation that the error in number perception of a set of visual, auditory or tactile cues is linear with the number of stimuli for a wide range of numbers up to several hundreds [1049]. While the classical neural network approaches manage to reproduce the phenomenology, the ability to perceive numerosity is only related to a small subset of the network nodes and not as a global property. In addition, the neural mechanisms underlying the relationship between the error rate and the number of perceived items across such a broad range of conditions remain unknown. Finally, this partial description of the problem requires rather complicated architectures for the underlying artificial neural network which could be considered as more complex that the network they try to describe.

Here instead, we introduce a paradigmatic example of the quantum-like paradigm, that reveals to be capable of accounting for this phenomenology in a relatively simpler manner. In a truly interdisciplinary research reported in [1052], an open quantum spin model—a dissipative XXZ model—maps the information processing of a network of neurons, modeled as 1/2-spin particles with varying connectivity (Figure 26a). Excitations, produced via spin flips (Figure 26e), can be transferred through exchange coupling (Figure 26b); they are subject to an energy offset when multiple excitations are in close proximity (Figure 26c), and they can decay to a resting state due to interactions with the environment, including losses and dephasing effects (Figure 26d). It was observed that certain time-dependent observables (Figure 26f–h) in the all-to-all connected network carried in their spectrum information about the number of previous excitations. A robust feature that remains consistent regardless of the location, time or amplitude of the stimuli and in a wide parameter range. By employing an ideal-observer decoding procedure [1052], the uncertainty $\sigma(N)$ associated with number estimation can be retrieved, which remarkably follows Weber's law, i.e., $\sigma(N)/N \sim \text{const}$ (Figure 26i). All in all, this model shows that applying the quantum statistical toolbox can reproduce complex phenomenology with a minimal model that can be efficiently simulated. We remark that the use of the quantum toolbox does not necessarily imply the claim of quantum effects in perception, having instead only a descriptive power.

Besides being relevant to the specific visual neuroscience problem, this example connects to the emerging field of research aimed at exploring the effects of long-range interactions in open quantum systems dynamics. Indeed, the dynamical coordination necessary to create this number-dependent spectral signal can be understood as a manifestation of the concept of cooperative shielding [1053]. This phenomenon observed in long-range models describes how certain initial states, locked in Hamiltonian subspaces, can be protected from evolution for long time periods [1054–1056] or even dissipation [1057], with applications to describe efficient biological phenomena [1058, 1059].

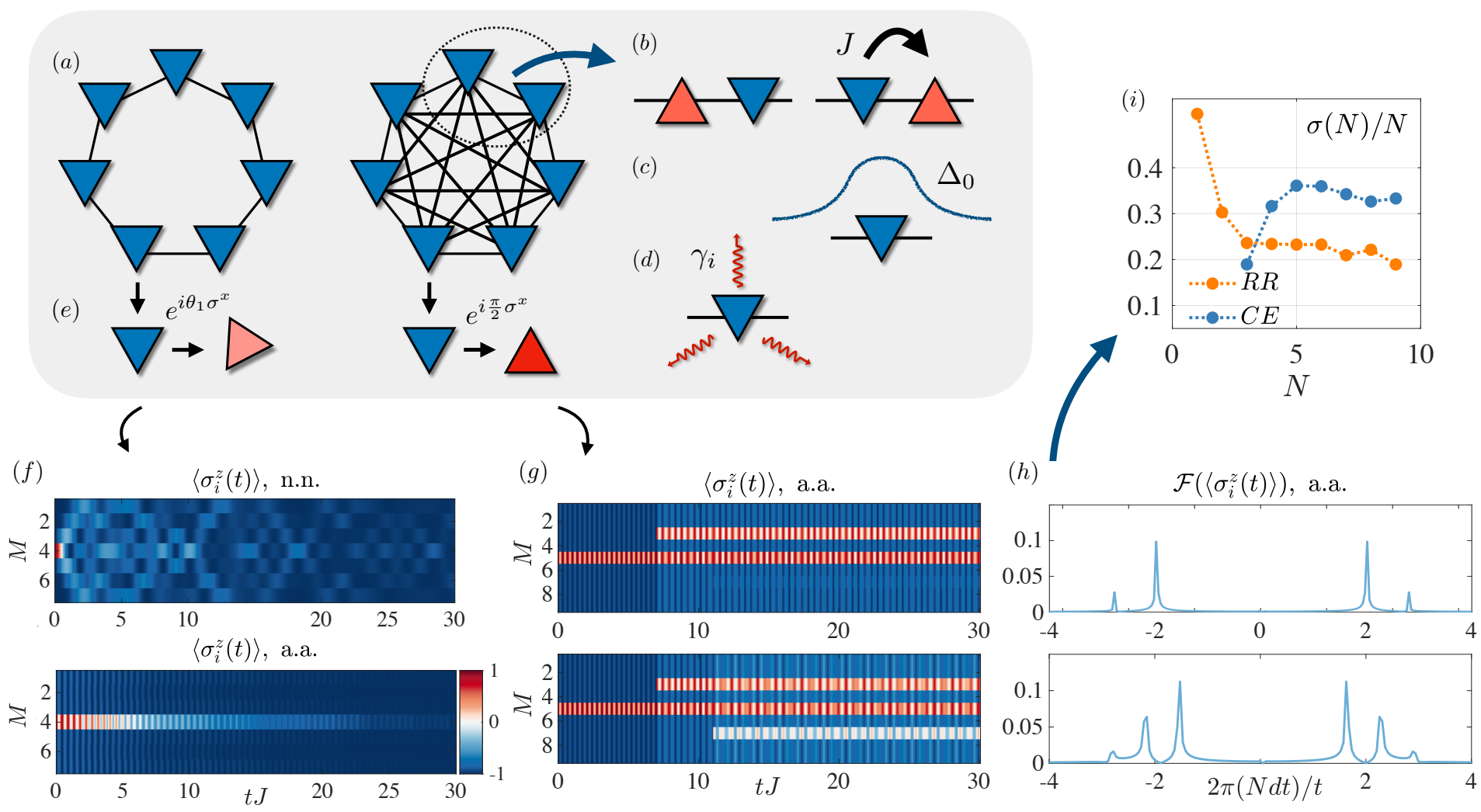

Figure 26: Quantum-like paradigm and the brain. Quantum spin model for visual neurosciences: a network of neurons or groups of neurons is mapped into an open quantum chain of spins with given connectivity. (**a**) Different connectivities: nearest-neighbour (left) and all−to−all (right); (**b**) Excitations propagate via exchange $J$; (**c**) nearby excitations experience an energy offset $\Delta_0$; and (**d**) can be subject to different dissipative channels at rate $\gamma_i$; (**e**) Excitations of variable amplitude are injected via spin flip operations. (**f**) Magnetisation profile of one single excitation in nearest−neighbor (n.n., top) and all−to−all (a.a., bottom) connectivity. (**g**) Magnetisation propagations in the a.a. case for $N = 2$ (top) and $N = 3$ (bottom) excitations, the regular interference generates clear patterns with additional frequencies as more spin flips are introduced. This can be understood in the power spectrum of the time signals (**h**). (**i**) Weber's law for numerosity is recovered using an ideal−observer decoding procedure independently of whether the excitation is injected with random amplitude (RR) or random amplitudes constrained to constant total energy (CE). Adapted from [1052] (Permission to use this content granted by Creative Commons Licence `https://creativecommons.org/licenses/by/4.0/`).

We close this section by evidencing in perspective other approaches to the description of biological and neuronal systems through the quantum-like paradigm, that make use of concepts typical of non-linear dynamics and quantum chaos. Examples include: chaotic behaviour for the amplification of emergent quantum macroscopic effects [1039], dissipative driving for the enhancement of chaotic behaviour [1060] or models for long-lasting coherence in warm noisy media due to near-criticality from quantum chaos to order, discussed e.g., in FMO complexes [1061] or in descriptions of nuclear physics via random Hamiltonians [1062]. Additionally, models for the brain as a resonating biosystem in a dissipative environment [1063], connect with ideas of time crystals [1064] and synchronisation [1065], as those discussed in Section III B 2, providing avenues for the description of the brain in terms of out-of-equilibrium systems [1066].

All in all, the quantum-like paradigm is providing means to use the quantum mechanical statistics toolbox to describe complex dynamics that are not-necessarily quantum, and to establish educated models for non-equilibrium dissipative driven phenomena that could explain the emergence of quantum effects in warm macroscopic biological media. These models can be benchmarked in experiments with atom-based quantum technologies and, on the way back, might be exploited to make a bio-imspired quantum technology.

### B. Transport and Quantum Effects in Biology

Another important problem in biology is the study of transport in noisy media, relevant in processes such as photosynthesis. Here, it is also possible to derive quantum models to describe the potential role of quantum effects in such tasks. After the experimental observations of relatively long-coherence times even at room temperature in biological complexes involved in photosynthesis [1067], it is important to understand whether these quantum effects, even if already observed, play a role in the transport efficiency and whether nature has evolve to use them for its advantage.

The understanding of photosynthesis is progressing in hand with advances in theory and experiments in quantum biology [7, 8, 1068] that build, among other tools, on the use of femtosecond two-dimensional spectroscopy methods [8, 59, 1069, 1070] to investigate energy transport in light-harvesting complexes. The crucial question is how can these coherent effects survive, given that the system's typical frequencies provide an estimate for the quantum-to-classical crossover temperature of approximately 0.1 µK, which is far from room temperature. Some studies suggest that optimized evolutionary conditions would steer away from fully coherent or incoherent processes, exploiting also the available dissipative couplings [59]. A known example in biology is Fröhlich condensation [1071, 1072], where externally driven polar molecules condense into a narrow low-frequency state and exhibit coherent collective motion [1073]. And, following on the tools mentioned in Section II B 1, particularly those oriented towards the description of non-Markovian systems constitute a suitable platform to describe these transport problems—given that biological baths are known to have non-trivial spectral densities [734, 735].

In addition, when considering transport problems in large systems these tools can be combined with ideas of quantum complex networks [1074] to create a flexible framework for both the understanding of the optimized microscopic transport phenomena and the emergence of certain properties at larger scales. Complex network theory is a powerful classical tool that was developed after the fundamental understanding that despite their different scale or origin, many networks in nature or society share some general properties [1075–1077]. The study and characterisation of a set of network classes allowed for the universal description of a large set of problems. Once the right network properties are identified for a given problem, basing on a series of quantifiers such as network topology, clustering or disparity, it is possible to predict the network behaviour, its dynamics or its response to perturbations based on this theory. One of the ubiquitous network classes is so-called *"small-world"* appearing in systems as varied as the brain or the internet [1075]. Then, it is natural to consider if this emergent universality and its related tools can be applied to networks of quantum elements both in theory [1078] and experiment [1079]. Moreover, the quantum information description of such systems can then be combined and applied to general network theory creating a generalised theory inspired both by quantum and classical tools [1080]. In the same way, network tools can be used to describe quantum phenomena [1081] or to improve certain QC protocols [1082]. For further details on the topic the authors refer to the topical review [1074] and the references therein.

While these networks are used to describe a large category of problems, we highlight the discussion on the pertinent topic of transport in such systems. A large amount of research has been conducted in recent years in order to compare the information and energy transport properties of a given network, in their classical and quantum version, in what is often denoted as quantum walks [1083], also in their stochastic account [1084]. This is opening the avenue to connect to dissipative scenarios [1085, 1086], with potential applications in biological systems.

All in all, in this section we have shown how the mapping of applied physical phenomena into quantum Hamiltonians or algorithms can benefit from the large theoretical and experimental toolbox to describe quantum matter and from emergent quantum technologies. This change of description framework can offer deeper understanding of the system and foster combinations of protocols (classical and quantum or even hybrid), that may help developing quantum-inspired classical methods and further synergistic applications.

## VIII. RESPONSIBLE RESEARCH AND INNOVATION, RESEARCH-BASED EDUCATION AND OUTREACH IN QUANTUM TECHNOLOGIES

The scientific blossoming in quantum technologies that we have described so far presents new challenges for society, even in the narrower perspective of atom technologies. In this section, we briefly present our scientific perspective on the subject, that stems from more recent intense research.

Responsible Research and Innovation (RRI) is a concept that emphasizes the need for scientific research to be conducted in a way that is socially responsible, inclusive, and sustainable [1087]. The European Commission defines RRI as "an approach that anticipates and assesses potential implications and societal expectations with regard to research and innovation, and the aim of fostering the design of inclusive and sustainable research and innovation" [1088]. In fact, RRI emerges as a response to concerns about the potential negative impacts of scientific research. To do so, RRI encompasses six dimensions, i.e., open science, public engagement, gender equality actions, science education and outreach and ethics [1089]. The principles of RRI are increasingly being adopted by research funding agencies and institutions around the world. For example, the European Union has made RRI a key component of its research and innovation strategy and has established a framework for promoting RRI across all stages of the research process [1090]. In the United States, the National Science Foundation has adopted a similar approach, emphasizing the importance of RRI in its funding priorities [1091].

When dealing with rapidly developing areas like quantum science and technologies, which have the potential to significantly impact society in the next decade, public awareness, education and outreach activities play an important role in the development of the society itself [1092, 1093]. In fact, quantum technologies are expected to have a profound impact on our daily lives, and produce a shift to entirely novel economies and job markets [62]. As we have discussed from the perspective of atom technologies, examples include the environmentally safe and sustainable development of batteries exhibiting increased efficiency, the development of materials to absorb and convert carbon dioxide in efficient manner, the development of fertilizers with increased efficiency and reduced environmental impact, sustainable food production, engineering of precision medicine, quantum communications and criptography, and the mitigation or solution of complex-network problems in the domain of logistics, finance, artificial intelligence, and even the brain. With such a list, hard sciences, as well as philosophical, economical, juridical, social, and policy-making sciences are expected to be affected by these developments, and potentially interested by the emerging quantum industry. In addition, enormous amounts of financial resources are being invested by public bodies and companies. It is then apparent how all the six Responsible Research and Innovation (RRI) dimensions should be adopted as glasses through which the research in quantum technologies should be viewed.

Effective public communication is crucial in achieving this goal and is already recognized as an important aspect of research and study programs: scientists worldwide are beginning to acknowledge their responsibility in this regard [1094]. However, educating the general public in quantum science presents a formidable challenge. Though similar to any other arisen in response to other technological advancements throughout history, this challenge for quantum technologies pose the additional problem of severe limitations in the experimental, creative, and mathematical literacies that are the pillars of scientific thinking [61].

Indeed, even basic forms of understanding of quantum science and technology concepts, require forms of experimental and mathematical literacy that are not readily available in everyday life, with quantum-physics experiments conducted in specialized labs and involving math that is out of reach for most people. Furthermore, quantum physics deals with intangible objects and phenomena, which challenge our creativity and imagination to levels which are not usually well-developed [1095]. Despite the importance of quantum technologies, quantum-matter physics is also usually excluded from school curricula, or only present at a descriptive level. In an outreach context, quantum physics is often presented via analogies or historical perspectives, which provide only factoids instead of explanations or understanding through a scientific-thinking process, leading to potential misunderstandings. Thus, even more than with classical physics [1096, 1097], education and outreach in quantum science and technologies would risk to become a bare transmission of historical facts and scientific factoids often using misleading analogies [1098, 1099], with no fundamental understanding nor education to scientific thinking. Without grounding in experiments and a certain level of formal structure, educators run the risk of offering transient narratives that are often misleading, relying on familiar analogies as a substitute. Therefore, the development of a research-based approach to outreach is required, that has been introduced in [61] and given the name of Physics Outreach Research (POR). Though paralleling the well established field of Physics Education Research (PER), and the development of PER and POR for quantum physics can be beneficial to each other, profound differences are in order [1100]. For example, the outreach context is different from classroom by definition: the public may not be uniform in age and background, the assessment of new learning tools and techniques usually cannot be obtained through tests and exams, outreach events are usually a short-period activity, and more importantly, the participation to an outreach activity is voluntary. This poses additional questions when the design of learning paths and especially the assessment of their efficiency and effectiveness are addressed.

One significant step has been pushed forward by the QTEdu-CSA (Quantum Technology Education—Coordination and Support Action) under the European Quantum Flagships initiative [1101]. This has been established to foster a robust competence ecosystem for the quantum-technologies workforce and to amplify outreach and education in the realm of quantum science. By bridging the gap between academic and industrial quantum communities, QTEdu reinforces its dual objective. Over a period of one and a half years spanning from 2021 to 2022, QTEdu has successfully executed 11 pilot projects in 25 European countries, diligently addressing educational needs and enlightening society about the remarkable potential of quantum technologies [1101]. These QTEdu pilot projects have addressed diverse segments of the public, including citizens of all ages, policymakers, teachers, students, media personnel, and industry workforce. Design of didactic material has been tailored for high schools and universities, the latter finally leading to the EU-funded Digitally Enhanced Quantum Technology Master (DIGIQ) [1102]. Specifically interesting in the context of RRI in quantum techonologies is the pilot QUTE4E on outreach and education [1103], that has produced the research-based approach Culturo-Scientific Storytelling (CSS) [61], evolving the proposal pushed forward in [1104].

As depicted in Figure 27a, CSS essentially considers that a discipline has a nucleus of concepts with their relationships, a body of knowledge about how these concepts are applied into practice, and a periphery. The latter is the space where the most exciting things happen from both the scientists and the citizens viewpoints. For scientists, the periphery is the space where the boundaries of knowledge are pushed to reach new understanding. For citizens, the periphery is the space where their beliefs are developed. Therefore, the periphery is a highly transformative space. In this perspective, the CSS narrative follows a unique approach that explores the intersection of discipline and culture, thereby naturally suited to RRI. For quantum science and technologies, story-telling of the nucleus and of the body is performed through the process of scientific thinking depicted in Figure 27b. The elements entering the CSS approach mirror those characterising the five minds for future introduced by Howard Gardner, including the disciplined, synthetic, creative, ethic and respectful minds [1105].

In fact, sustaining creative, formal, and experimental literacies for quantum science and technology storytelling is not an easy task. The proposed way out to overcome this challenge is envisioned in the design of resources and tools that can work to compensate the corresponding language limitations. Within the QUTE4E pilot, a survey of available and desirable resources has been performed [1106], that are in terms of quantum games and interactive tools, and conceptual and experimental virtual labs, working as effective mediators between the mathematical or experimental technical tasks—demanded to a computer—and their management by users. In particular, quantum games have been proven to work as effective tools to support high-school students in grasping elementary concepts of quantum mechanics [1107], and are being developed as outreach resources to leverage on the awareness to cultural heritage [1108].

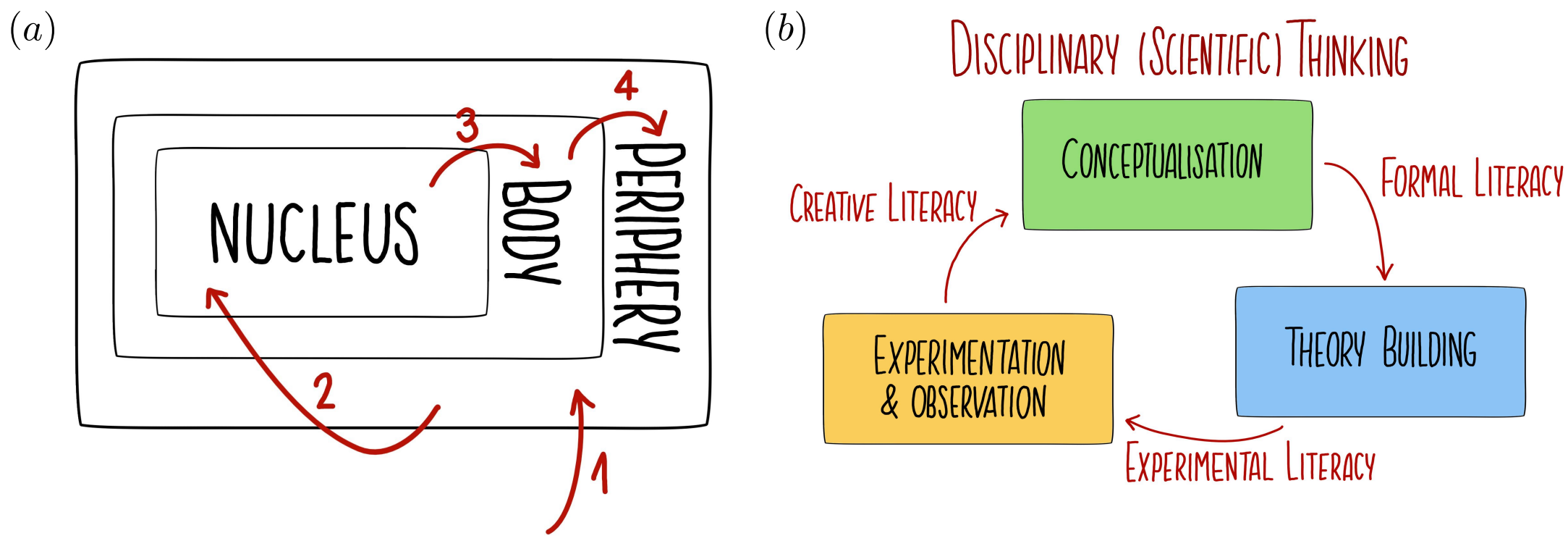

Figure 27: Culturo-scientific storytelling (CSS) approach to Physics Outreach Research. (**a**) CSS consists in an experiential journey that mirrors the encounters of scientists and citizens, with each subject being approached through the process of scientific thinking, as depicted in (**b**). The journey begins by acquainting oneself with peripheral knowledge (arrow 1) and gradually delves deeper into the consolidation of fundamental concepts at the core (arrow 2). Practical applications are then explored and comprehended in the main body (arrow 3), followed by a return to the periphery to emphasize the ever-evolving and unfinished nature of the discipline-culture (arrow 4). Image inspired from [61].

The idea of language compensation is the core in the development of the QPlayLearn platform [1095] to educate K12 and general public to quantum physics and technologies, and inspired by the theory of multiple intelligences proposed by Howard Gardner in the 1980s. In a nutshell, a dictionary of quantum physics and technologies concepts is proposed, the storytelling of each item being offered through the use of different languages, suited to our diverse intelligences. The sections play, discover, learn, apply use the language of games, experiments and animations in the form of quantum pills, mathematics, and quantum information and logic, respectively. In addition, an integrated art and science approach is offered, like the one developed with the Quantum Jungle, an interactive and immersive art installation created by the artist Robin Baumgarten, supported by a team of quantum physicists from Helsinki and Pisa Universities for the conceptualisation and the coding side. As shown in Figure 28, the Quantum Jungle is capable of visualising the behaviour of a single quantum particle from a given initial state, providing a scientifically effective and artistically fascinating tool to grasp the essential quantum concepts of quantum state, superposition, tunneling, and measurement. The analysis of didactic experiments is under way, wih promising results involving students from the kindergarten to the high school, as well as general public during the six months of exhibition in Pisa in the Palazzo Blu museums.

## IX. DISCUSSION

In this perspective review we have illustrated the state-of-the-art quantum technology platforms based on ultracold atoms and molecules that are currently leading in terms of their control in system size, coherence time, level of noise, and in terms of number of applications. In particular, we have focused on those based on neutral atoms in optical lattices, optical cavities or optical traps, trapped ions in linear traps, Rydberg atoms, and dipolar gases. For completeness, we also very briefly discussed solid-state platforms, including the most relevant superconducting qubits on which current quantum computers technology is often based. We have discussed how the diverse atom-technologies provide a flexible framework of experimental platforms. While these still exhibit certain limitations in terms of operation time or noise effects and are generally better suited for specific tasks, the degree of control and the variety of applications is remarkable, especially if one considers the development of AMO as a whole. Moreover, the consistent improvement over an extended period of time that spans several decades, heralds on the technological and theoretical sides the continuation of a healthy pure and applied research landscape.

The development of these platforms then requires an adequate level of advancement in their theoretical characterization. In this review, we have discussed the theory of open quantum systems providing the tools for the description of coherent and dissipative couplings in many-body physics platforms. We have also shown how this set of dissipative tools can access novel phenomena in quantum matter and be used as a resource, in fact beyond being considered as a technical limitation in what is denoted as reservoir engineering. Moreover, we have summarized the main approximations for a resource-affordable description and discussed selected approximate methods based on stochastic unravelling, to ensure that these problems remain treatable classically at least for moderate system sizes. In this direction, we have provided the reader with a summary of tensor network methods that can complement the tools provided by open quantum systems theory, to tackle current relevant system sizes. There, we have discussed their success in one dimension and their moderate progress in higher dimensions, while also connecting with their applicability in quantum simulation in other fields: from their links with the simulation of lattice gauge theories to their synergies with neural network approaches.

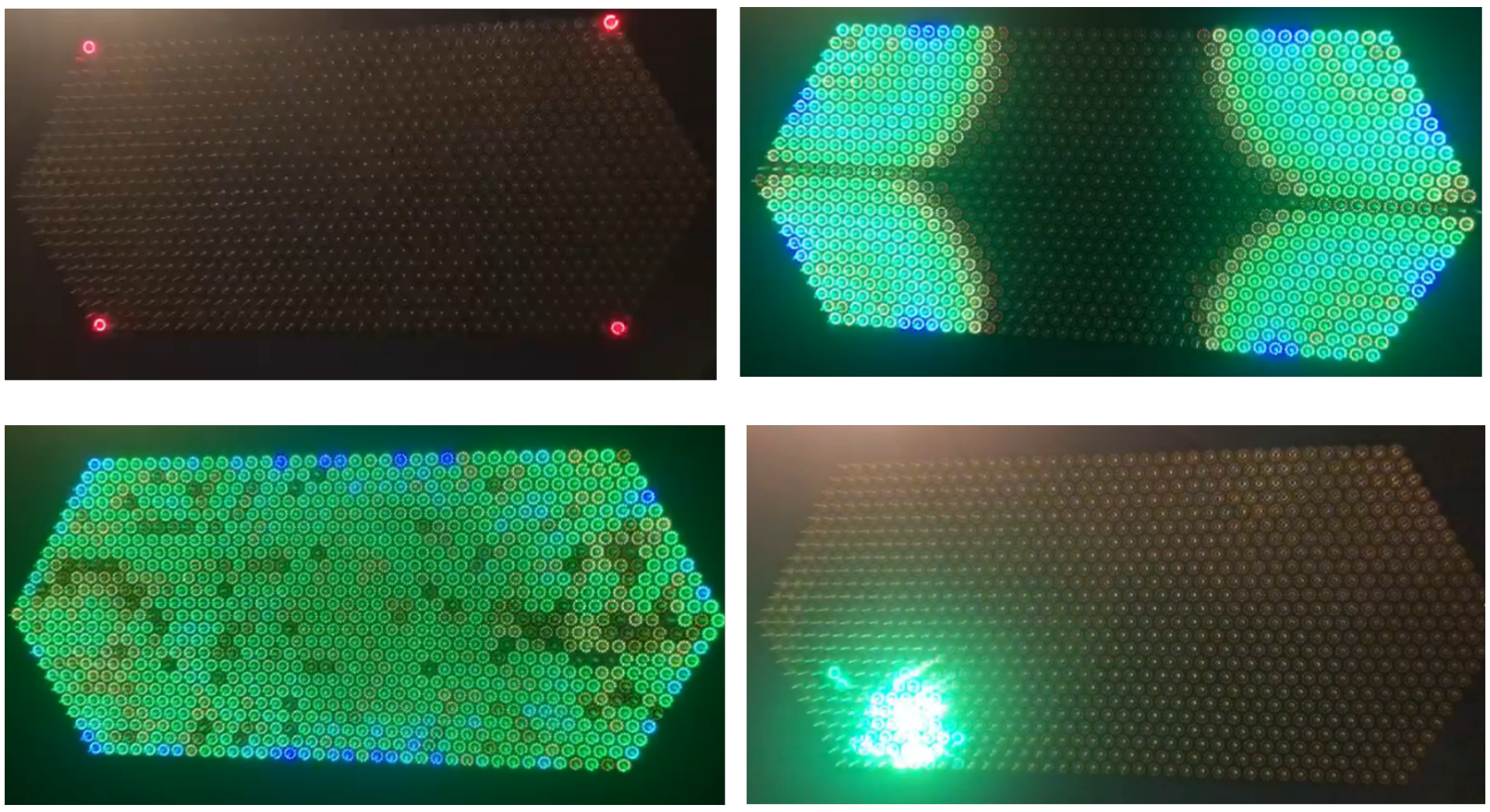

Figure 28: Interactive tools for CSS. Screenshots from the art and science 6 sqm installation Quantum Jungle by Robin Baumgarten, visualizing the time evolution of one quantum particle. After touching the springs, the quantum particle is created in a superposition state (**top left**). The probability from computer simulations of the Schrödinger equation evolves according to a quantum walk (**top right**, **bottom left**), and is visualized via switching on of LEDs with proportional intensity. Later spring touch is interpreted as a measurement action, visualizing the collapse (**bottom right**).

Once the description of the platforms and the theoretical description of quantum simulation was laid down, we have presented that the applications of AMO platforms reach far beyond their use as quantum simulators. In particular, we have discussed the potential use of some platforms as NISQ computers and discussed their potential utility in view of resource-conscious, optimization and variational approaches, where in certain scenarios these atomic platforms can complement and outperform existing solid-state approaches.

Then, we discussed their applicability as high-sensitivity sensors, which is essential for their applications e.g., for fundamental physics tests and for cosmology problems. There, we described strategies for the production and characterization of highly entangled states as a new paradigmatic resource for high-sensitivity performance and the general measurement strategies to exploit these advantages. Using quantum entanglement for quantum metrology and sensing still requires various directions of development. An important limitation is the consistent preparation of highly entangled states, characterized by a higher number of degrees of freedom and/or by a longer coherence time. The second challenge is the realization of more efficient protocols, for single-state and collective manipulation, useful for applications on a growing number of physical systems. However, as we discuss in the text, we believe that variational and optimization approaches derived from classical systems, together with dissipative driving methods, could provide useful solutions to some of these technical limitations.

Once that the tools and platforms were described in detail, we focused on the wide range of applications of atom technologies, starting from those originally motivating their development: condensed-matter physics. Given the wide variety of interesting possibilities, we adopted as selection criterion their potential applications in engineering quantum states of matter or their potential coding in atom technologies to simulate phenomena otherwise hard to control in their original context. We have addressed the description of superconductivity/superfluidity, the effect of incommensurations or disorder leading to non-thermal states, the role of driving and of dissipation in state preparation or the characterisation of topological phases of matter. Importantly, we highlight the relevance of these frameworks—typically based on well-understood microscopic rules—to describe complex problems in other fields and establish connections with cosmology, information theory or physics foundations.

In this direction, we have delved into the application of atom technologies to address open questions in fundamental physics, cosmology or even physics foundations. We have discussed the use of ultra-sensitive atomic clock technologies and quantum metrology to test the equivalence principle or the microscopic effects of general relativity in molecule ensembles; the possibility of coding into atom technologies Hamiltonians relevant for the description of quark-gluon matter in neutron stars, and the use of analog-gravity models for the simulation of cosmological problems via cold-atomic ensembles in table-top experiments. While some limitations still exists, in terms of e.g., systematic generation of gauge fields in dimensions larger than 1 spatial plus 1 temporal, or in the required system sizes, one can expect that these applications will progressively be better at hand with technology improvements, and that new ones will come to life.

The final applications that we have covered in this review are related to their applications in biology and chemistry. We started from their microscopic descriptions as an emergent field where the role of quantum phenomena is still being discussed. Then, we introduced recent coarse-grained approaches connected with quantum complex networks for the benchmarking of transport phenomenology in noisy and warm media. This cross-fertilization can work in both ways: learning from biological systems efficient ways to harness quantum advantage in quantum technologies and quantum computers, and exploit the latter to understand biological processes. The latter two discussed applications convey the message that the degree of control and sensitivity reached in atom-based quantum technologies, is pushing the limits of fields well beyond condensed matter and atomic-molecular physics.

In this direction, as the transformative effect of quantum technologies impacts society, the need for the scientific community to reflect on their role in educating the general public and in understanding the implications of their research and the value this could bring to the world. Consequently, it is important that quantum technology development comes in hand with a Responsible-Research and Innovation perspective, providing both scientists and stakeholders with the tools to assess the importance of these innovative advances in quantum matter.

## X. CONCLUSIONS

Physics is living an era of unprecedented cross-fertilization among the different areas of science. In this perspective review, we have discussed the manifold impact that state-of-the-art cold and ultracold-atomic platforms can have in fundamental and applied science through the development of platforms for quantum simulation, computation, metrology and sensing. Engineering of table-top experiments with atom technologies is engendering applications aimed at understanding problems in condensed matter and fundamental physics, cosmology and astrophysics, to unveil foundational aspects of quantum mechanics, and advance quantum chemistry and the emerging field of quantum biology.

This has become possible via two main approaches: creating quantum analogues and building quantum simulators. The former fosters learning from analogies between two quantum systems via associative creativity, the latter implies the encoding of specific microscopic system Hamiltonians in a highly controllable quantum platform, in fact a specific type of quantum computation. At the boundary between the two approaches are paradigmatic model Hamiltonians, that can serve to create analogue simulations or code microscopic quantum problems. Needless to say, quantum computing would represent a universal answer to all these questions, though not necessarily to be adopted alone.

The first message from this review is that indeed this forefront enterprise is marking a new era in the way quantum technologies are used for fundamental science even beyond the advancement in content knowledge: this is characterised by truly cross-disciplinary research, extended interplay between theoretical and experimental thinking, and intersectoral approach. Cross-disciplinary approach to common research is manifest in the connections that we have illustrated existing among condensed matter, atomic-molecular-optical (AMO) physics and quantum information, foundations of quantum mechanics, high-energy physics, astrophysics, general relativity and cosmology, with new branches sprouting across quantum chemistry and quantum biology, and neuroscience. In fact, the steady and rapid progress in quantum-gases platforms capabilities has since the very beginning resulted into the emergence of a new scientific community from the originating disciplines, fostering cooperation, contamination, and cross-disciplinary thinking as values for the way of practising research. The seeding and care in building up such a community has been one of the roots for its success with fast and extended growth, a lesson that should be valued. On the other hand, the development of atom technologies has led to a different cooperation between theory and experiment, going beyond the conventional pillars of scientific thinking, where phenomena understanding is formalized into a theory and/or theoretical predictions are tested in experiments. In the new environment, theory is also part of the experiment, in that serves to efficiently map the problems in a quantum simulator, and the experiment works as an alternative or complementary simulation for the theory. Pushing the boundaries even further, systematic intersectoral connections are being developed between quantum science and technology developed in academic contexts, and their implementations in everyday life operated by small to large companies, the small ones not rarely stemming from academic experiences. Based on science history, such an outcome could in fact be expected when considering the impact of condensed matter physics and material science in the human-kind progress. What is instead less obvious, is how this is happening also for fundamental science. This is apparent for example going back to more recent enterprises encountered in this journey, like the AEDGE community [50, 53] and the efforts to advance gravity exploration in terrestrial very-long-baseline or else in space experiments. In these advanced cross-disciplinary searches, the – so-far unexplored – path is being pursued of using microscopic quantum sensors in large-scale facilities. Besides requiring tightly-joined theoretical and experimental efforts and the implementation of techniques and technologies across different scientific communities, this route is also fostering the advancement of technology readiness to stimulate dedicated companies and ignite sustainable economical development.

One second message emerging from this review is that quantum many-body physics is taking center stage in frontier's science. In fact, it addresses phenomena on a size scale and complexity level that stand closer to the emergent boundary between classical physics and quantum mechanics, as the microscopic theory accounting for more than 40 orders of magnitude in length scales from the cosmology of the universe down to the quarks as elementary constituents of matter. While the classical-to-quantum crossover is moved by the advancement of quantum technologies, quantum many-body physics and its contemporary evolution in terms of quantum complex systems can provide useful reading glasses to understand whether and where this border should be placed. In fact, quantum technologies are playing a vital role in the investigation of the emergence of macroscopic quantum-coherence in biological systems, leading to the establishment of the new field of quantum biology. While the effective function of macroscopic quantum coherence in optimizing or even solving biological functions is at the center of live debates, this field of research is seeding remarkable knowledge advancements: first, to develop refined experimental and theoretical methods suited to address this question, and, second, to explore whether one can learn from Nature hints useful for the design of quantum technologies, and viceversa. In parallel, quantum technologies are reinforcing the use of the quantum-like paradigm to address complex systems, in fact exploiting the mathematical machinery of quantum many-body physics to describe aspects of their behavior.

While applications appear hand in hand with a better degree of platform control and new experimental and theoretical protocols are developed, so will the opportunities to address either new problems or old ones with new perspectives, pushing the boundaries of what these simulators are able to achieve. The development of quantum technologies naturally aims at a large-scale fault-tolerant universal quantum computer in the future: in any event, in this review we have shown how the high degree of current technological development of the different platforms, the combination of classical and quantum methods, the use of a resource-based approach, the application of mappings and of cross-disciplinary toolboxes, have already provided us with a large amount of new knowledge and applications. Thus, should this ultimate goal of a universal quantum computer not be fully met eventually, that would not change the already transformative effects that quantum technologies have had in the last decades and will have in a next future.

One last and equally important message we wish to convey follows on the potential impacts that the tremendous progress of quantum science and technology will have on society in the next decade. Examples include the environmentally safe and sustainable development of batteries and materials, fertilizers for food-supply chains, the engineering of precision medicine, quantum communications and criptography, and the mitigation or solution of complex-network problems important for logistics, finance, artificial intelligence. With such a list, hard and human sciences, including philosophical, economical, juridical, social, and policy-making sciences, sought to get entangled in new forms of scientific humanism. Along these lines, quantum technologies are expected to produce a shift to entirely novel economies and job markets, what is already occurring also with huge amounts of financial investments operated by both public and private bodies. While this happens, as history shows that occurs in correspondence of any technology revolution, the adoption of a responsible research and innovation approach to quantum technologies is mandatory, across all its six dimensions of open science, public engagement, gender equality actions, science education and outreach and ethics. The development of public awareness through education and outreach is a necessary step in this process. Considering that quantum science and technologies pose additional challenges in overcoming the more limited creativity, formal, and experimental literacies that are pillars of scientific thinking, renovated efforts and resources need to be invested to develop research-based educational and outreach tools. This would turn a potential limitation, such as the difficulty of educating to the counterintuitive concepts of quantum science, into the fantastic opportunity of providing citizens with culturo-scientific storytelling tools for future scaffolding.

All in all, atom-based quantum technologies represent one of the most relevant playgrounds in the future of transversal experimental, and even theoretical, physics. Moreover, the tools developed for their control and description do not only advance our understanding of quantum matter but open the door to applications of these models and protocols to other systems. While there are limitations in what we can achieve today, there is still a fruitful path ahead of us already in the NISQ era.

**Contribution by the authors** – M.L.C.: writeup, conceptualization, coordination. J.Y.M.: writeup and contribution to conceptualization. L.L. and L.G.: writeup. All authors have read and agreed to the published version of the manuscript.

**Acknowledgments**

The authors are grateful to Fausto Borgonovi, Stefano Carretta, Luca Celardo, Alessandro Chiesa, Elena Garlatti, Stefano Liberati, Massimo Mannarelli, Marco Nardecchia, G. Nerina, Filippo Simoni, and Augusto Smerzi for many enlightening discussions. M.L.C. is grateful to Simon Goorney, the AEDGE community, the STEQUEST core Team, and all the co-Authors of works that have seeded this review. M.L.C. and L.G. would like to thank the support of the Regione Toscana through the WQubit funds and the University of Pisa through the BIHO funding.

J.Y.M. and M.L.C. were supported by the European Social Fund REACT EU through the Italian national program PON 2014-2020, DM MUR 1062/2021. L. L. acknowledges financial support by a project funded under the National Recovery and Resilience Plan (NRRP), Mission 4 Component 2 Investment 1.3—Call for tender No. 341 of 15/03/2022 of Italian Ministry of University and Research funded by the European Union NextGenerationEU, award number PE0000023, Concession Decree No. 1564 of 11/10/2022 adopted by the Italian Ministry of University and Research, CUP D93C22000940001, Project title National Quantum Science and Technology Institute (NQSTI), spoke 2. This manuscript reflects only the authors' views and opinions, neither the European Union nor the European Commission can be considered responsible for them. M.L.C. acknowledges support from the National Centre on HPC, Big Data and Quantum Computing—SPOKE 10 (Quantum Computing) and received funding from the European Union Next-GenerationEU—National Recovery and Resilience Plan (NRRP)—MISSION 4 COMPONENT 2, INVESTMENT N. 1.4—CUP N. I53C22000690001. This research has received funding from the European Union's Digital Europe Programme DIGIQ under grant agreement no. 101084035. M.L.C. also acknowledges support from the project PRA_2022_2023_98 "IMAGINATION", from the MIT-UNIPI program, and in part by grants NSF PHY-1748958 and PHY-2309135 to the Kavli Institute for Theoretical Physics (KITP).

---